\documentclass[acmtog]{acmart}
\setcopyright{none}
\acmDOI{}
\acmISBN{}
\acmConference{}{}{}
\acmYear{2026}
\copyrightyear{2026}
\usepackage{algorithm}
\usepackage[noend]{algpseudocode}

\usepackage{xcolor}
\usepackage{amsmath}
\usepackage{booktabs}
\usepackage{tikz}
\usepackage{placeins}
\usepackage{multirow}
\usepackage{xspace}
\usepackage{float}
\usepackage{mathtools}
\usepackage{wrapfig}
\floatstyle{ruled}
\newif\ifdraftgraphics
\draftgraphicsfalse
\ifdraftgraphics
  \setkeys{Gin}{draft}
\fi

\newcommand{\FLIP}{\protect\reflectbox{F}LIP\xspace}

\providecommand{\Description}[1]{}

\usepackage{pifont}

\newcommand*{\trans}{^{\mkern-1.5mu\mathsf{T}}}
\newcommand*{\invtrans}{^{\mkern-1.5mu\mathsf{-T}}}

\newcommand{\duv}{\,\mathrm{d}u\,\mathrm{d}v}

\newcommand{\appropto}{\mathrel{\vcenter{
  \offinterlineskip\halign{\hfil$##$\hfil\cr
    \propto\cr\noalign{\kern2pt}\sim\cr\noalign{\kern-2pt}}}}}

\newcommand{\Pbase}{\mathbf{P}}          %
\newcommand{\Nbase}{\mathbf{N}}          %
\newcommand{\uvdom}{\Omega}              %

\newcommand{\nd}{n}                      %
\newcommand{\ndom}{\uvdom_{\nd}}         %

\newcommand{\raytmax}{t_{\max}}

\newcommand{\Unum}{U}                    %
\newcommand{\Vnum}{V}

\newcommand{\rtCases}{228}
\newcommand{\rtFrameSpeedup}{3.404}

\AtBeginDocument{%
}

\newcommand{\PaperTitle}{Geometric Queries on Tessellation-Free Displaced Surfaces}
\newcommand{\PaperShortTitle}{Geometric queries on tessellation-free displaced surfaces}

\graphicspath{{figures/acm/}{figures/}}
\newcommand{\ParentBoundsClearance}{\vspace{14pt}}
\newcommand{\SegmentCountsSkip}{5.5pt}

\author{Bing Xu}
\authornote{Equal contribution.}
\affiliation{%
	\institution{NVIDIA}
	\country{USA}
}
\email{xubinggl@gmail.com}

\author{Yang Zhou}
\authornotemark[1]
\affiliation{%
	\institution{Independent Researcher}
	\country{USA}
}
\email{yyp0502@gmail.com}

\author{Brian Budge}
\affiliation{%
	\institution{Independent Researcher}
	\country{Germany}
}
\email{brian.budge@gmail.com}

\title[\PaperShortTitle]{\PaperTitle}

\begin{document}

\begin{abstract}

Displacement maps encode fine geometric detail in compact textures and simplify surface authoring, but using them typically
requires pre-tessellating surfaces into dense triangle meshes, incurring a heavy memory cost.  Recent tessellation-free methods enable
direct ray tracing. However, broader geometric queries such as surface sampling, specular manifold search, and closest-point queries remain less explored.
Supporting these operations requires surface enclosures, area estimates, and normal bounds,
which scalar height intervals alone do not fully provide. We present a first-order texture-space hierarchy that unlocks these queries directly without pre-tessellation.
Each node stores a local affine height plane and a
conservative residual bound, retaining spatial height variation that scalar minimum--maximum bounds discard.
The plane and residual provide surface enclosures for intersection and distance searches, while the plane's height and
slopes supply area estimates for sampling. Additional slope bounds support specular light-path search.
Building on this representation, we develop query-specific traversal, pruning, and sampling algorithms, and demonstrate them in
caustics rendering, sampling displaced emitters, and surface partial differential equation (PDE) solvers.
We demonstrate memory savings over pre-tessellation and performance
improvements over query-specific baselines, while retaining efficient ray tracing.

\end{abstract}

\begin{teaserfigure}
  \centering
  \begin{tikzpicture}
    \node[anchor=south west, inner sep=0] (main) at (0,0)
      {\includegraphics[width=\linewidth]{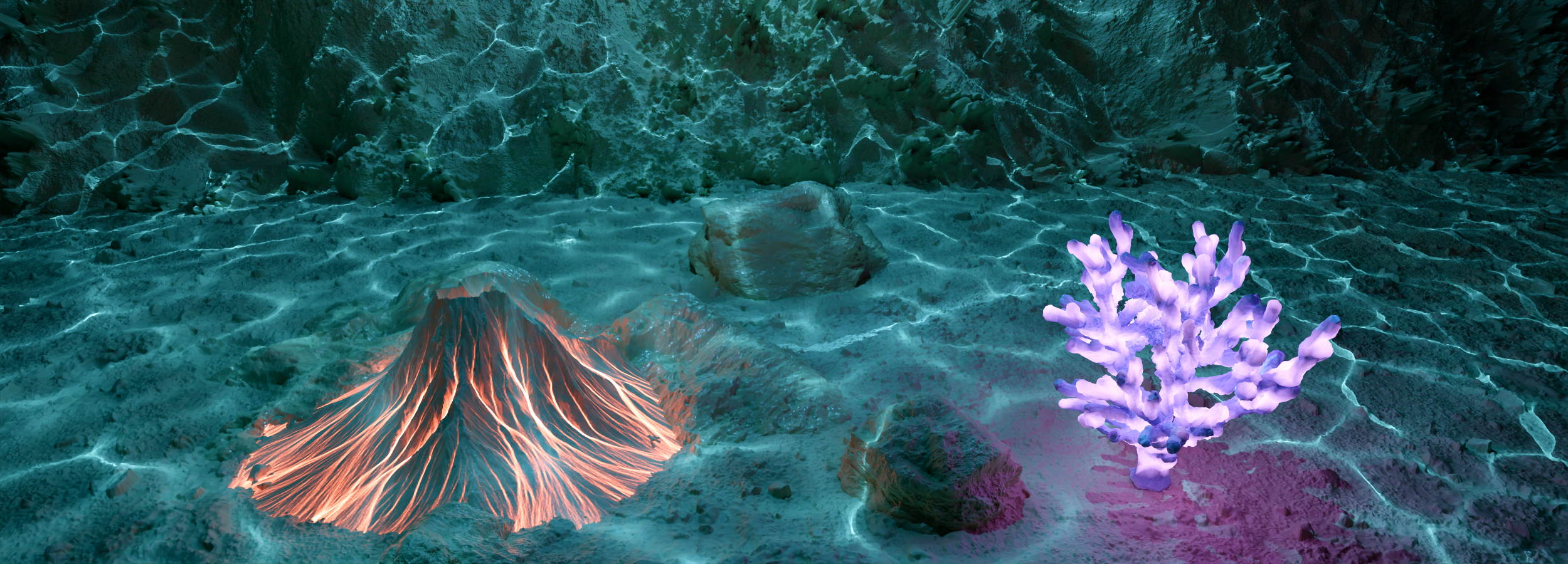}};
    \node[anchor=north east, inner sep=0] (inset) at (main.north east)
      {\includegraphics[width=0.16\linewidth]{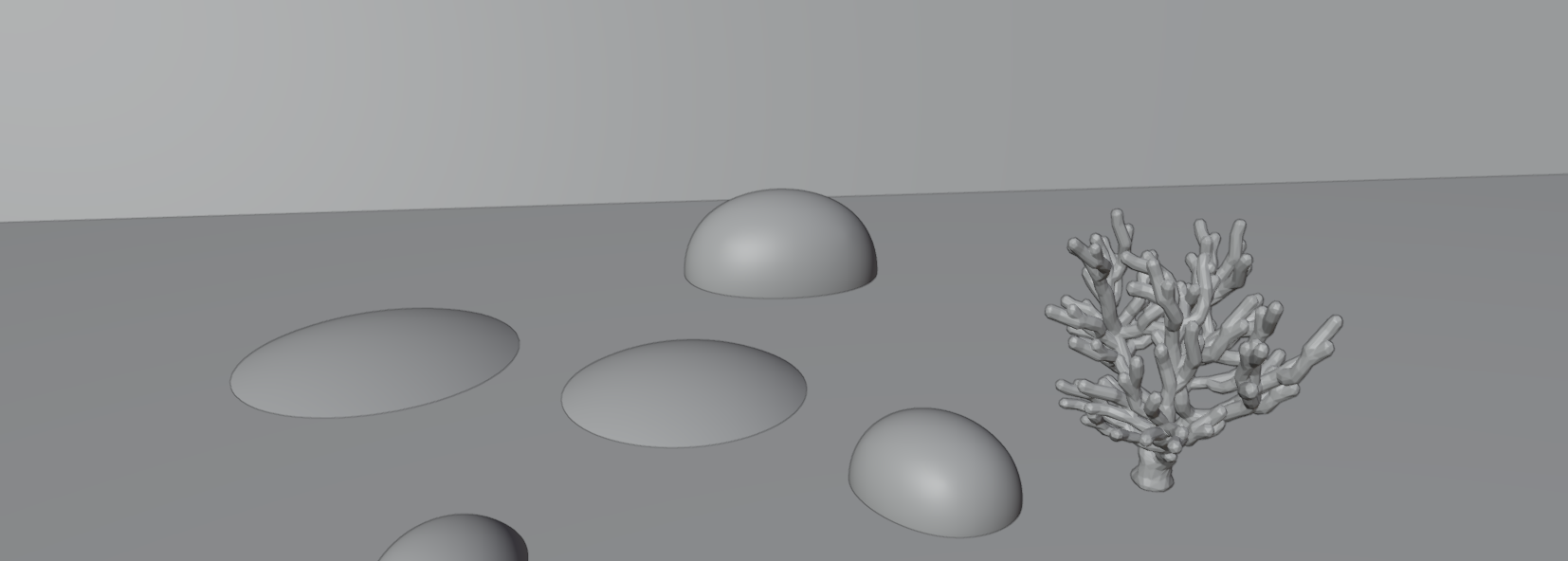}};
    \draw[white, line width=1pt] (inset.north west) |- (inset.south east);
  \end{tikzpicture}
  \caption{We enable versatile geometric queries on tessellation-free displaced surfaces for rendering and other
  graphics applications. In this \textsc{Underwater} scene, displaced surfaces serve not only as a realistic
  surface representation that greatly enhances geometric fidelity (base meshes shown in the inset), but also
  as a general-purpose graphics primitive: the entire scene is path traced by direct ray tracing; the caustics are
  cast by a displaced water surface; and the volcano and coral are complex light-emissive displaced surfaces, with
  the coral's emission distribution given by the solution to a screened Poisson equation defined on its displaced surface.
  All of these effects are made possible by the efficient geometric queries our method provides, with no tessellation
  ever required. }
  \label{fig:teaser}
\Description{An underwater scene rendered using displacement maps for ray intersection, caustics, area-light sampling, and a surface PDE field controlling coral emission.}

\end{teaserfigure}
\maketitle

\section{Introduction}
Displacement mapping is a widely used computer graphics technique that allows artists to create rich 
visual detail without modeling and storing fine features as explicit geometry. However, tessellation 
can negate this memory advantage by expanding the surface into a dense triangle mesh. Recent 
tessellation-free methods avoid this cost for ray tracing~\cite{thonat2021tessellation,ogaki2023nonlinear,thonat2023rmip}, 
but using displaced surfaces across computer graphics applications requires much more than ray intersection queries.
Our goal is to make tessellation-free displaced surfaces a general-purpose graphics primitive by 
equipping them with the operations needed to support a wide range of downstream applications.

These operations determine how detailed geometry participates in rendering and broader geometric computation.
Realistic rendering usually requires challenging light transport effects caused by difficult light paths or complex light 
source distribution. Beyond rendering, surface sampling supports point-cloud generation and
numerical integration~\cite{ling2025uniform}, while solving partial differential equations (PDEs) on a surface requires querying 
the distance to the surface. Although a displacement map makes it easy to evaluate 
the height at any single texture coordinate, answering these queries efficiently requires bounds or estimates over entire
surface regions. The challenge is to obtain this regional information without instantiating the detailed geometry.

The scalar height bounds typically used for direct ray tracing provide only part of this information. A 
minimum--maximum interval records the range of heights in a texture region but not how height varies with 
position~\cite{thonat2021tessellation}. Surface sampling requires estimates of area, while searching for challenging 
light paths usually asks about surface normals and their variation; both require information about slopes. Therefore, we propose to 
approximate height within each region by a local plane and conservatively bound the remaining error. The plane
supplies local slope estimates, while the residual encloses the height function throughout the region. We call this a 
first-order height representation because the plane is affine in texture coordinates. 

\begin{figure}[!htb]
\vspace{-5pt}
\centering
\includegraphics[width=\columnwidth]{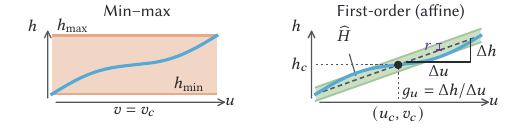}
\caption{Height bounds along a texture-space slice $v=v_c$. Min--max bounds retain only $[h_{\min},h_{\max}]$. Our first-order model is an affine height plane $\widehat H$, shown here as a line: $h_c$ is its height at the region center $(u_c,v_c)$, $g_u=\Delta h/\Delta u$ is its slope along $u$, and $r$ bounds the height error. See \autoref{sec:hierarchy} for the full construction.}
\label{fig:intro-height-bounds}
\vspace{-8pt}
\end{figure}

We build a texture-space hierarchy of height planes and residual bounds that is shared by all triangles 
using the same displacement map. Together with the base geometry, each node's plane and residual define 
a surface enclosure for intersection and distance queries, while the plane's height and slopes provide 
estimates for surface area sampling. Additional slope bounds enable pruning tests during specular path 
search. The hierarchy therefore provides both geometric bounds and sampling estimates, with 
query-specific data added only where needed.

Building on this representation, we develop traversal, pruning, and sampling algorithms for four geometric queries.
For ray intersection, a piecewise-linear approximation of the nonlinear texture-space ray enables inexpensive grid
traversal and tests against the stored height bands.
For specular light path search, a first-order bound over height and slope prunes regions and bootstraps local
constraint solvers, allowing them to find more admissible paths faster.
For surface sampling, our hierarchy estimates regional areas to guide hierarchical sampling, with weighting strategies
tailored to specific applications such as emitter sampling.
For closest-point queries, we derive lower bounds on world-space distances that guide the search and refinement
to a requested distance tolerance.
Together, these algorithms greatly expand the application landscape of compact displaced surfaces beyond ray
intersection while retaining efficient ray tracing.

\section{Related work}

\paragraph{Displacement mapping and tessellation-free ray tracing.}
Displacement maps allow content authors to drive geometric detail using textures instead of manually sculpting every 
surface vertex~\cite{cook1984shade}. Tessellation makes these displaced surfaces accessible to standard mesh processing 
or rendering, but storing the tessellated geometry and its acceleration structure can be costly. Early work addresses 
this problem by generating and caching displaced geometry during rendering~\cite{pharr1996geometry}. 
Smits et al.~\shortcite{smits2000direct} avoid storing the full tessellation by traversing an implicit barycentric grid and
generating displaced micro-triangles on demand. Maximum mipmaps~\cite{tevs2008maximum} accelerate height-field intersection 
through hierarchical height bounds. Modern GPUs support hardware tessellation that splits coarse geometry into dense, 
high-resolution geometry on the fly using dedicated hardware stages in the rasterization pipeline~\cite{niessner2016real}.

More recent work develops bounds and traversal methods for direct ray tracing of displaced triangle meshes. 
Thonat et al.~\shortcite{thonat2021tessellation}(TFDM) builds a min--max texture hierarchy and constructs world-space bounds
during traversal and ray intersection tests. The subsequent work~\cite{thonat2023rmip} (RMIP) alternates between tightening 
texture-space regions and world-space bounds, using elongated rectangular regions to follow the ray more closely. 
Ogaki~\shortcite{ogaki2023nonlinear} instead parameterizes the nonlinear texture-space ray by rational functions of height,
allowing direct intersection with texture-space boxes. Projective displacement mapping~\cite{hoetzlein2025pdm} uses 
fixed-step world-space ray marching with projective texture lookup, supporting interactive editing without a texture-space 
hierarchy.  In parallel, micro-meshes reduce storage by compressing finely subdivided surfaces~\cite{maggiordomo2023micro,gruen2026ray} 
and supports hardware acceleration. Our focus is a texture-space hierarchy for queries including and beyond ray intersection. 
It retains local height variation through height planes and residual bounds and, with query-specific enhancements where needed, 
supports specular-path search, surface sampling, and closest-point queries.

\paragraph{Bounding methods for geometric queries.}
Bounding methods are widely used in graphics to accelerate ray intersection, collision detection, and distance queries on various surface representations. On explicit meshes, these operations use bounding volume hierarchies~\cite{gottschalk1996obbtree,larsen1999proximity}. For continuous surfaces, bounds can be derived from the functions that define the geometry. Affine arithmetic guides subdivision for intersecting parametric surfaces~\cite{defigueiredo1996intersection}, while Taylor models combine polynomial approximations with bounded remainders for culling before tessellation~\cite{hasselgren2009pretessellation}. Sharp and
Jacobson~\shortcite{sharp2022spelunking} use range analysis for various geometric queries on neural implicit surfaces. In displacement rendering, Heidrich and Seidel~\shortcite{heidrich1998displacement} bound procedural displacement shaders using affine arithmetic, and Munkberg et al.~\shortcite{munkberg2010bounding} combine normal bounds and min--max mipmaps to enclose displaced B\'ezier patches. Recent work and ours use bounds to query displaced surfaces without pre-tessellation. TFDM and RMIP construct bounding volumes from the base geometry and scalar height intervals over texture regions. Queries involving surface area and normals also require derivative information. We store an affine height model and a bounded residual in each texture region, with additional slope bounds where needed. From this representation, we develop query-specific bounds, traversal strategies, and sampling algorithms to support queries beyond ray intersection.

\paragraph{Specular light transport.}
In Monte Carlo rendering, sampling specular paths that connect a receiver to a light through specular reflection or refraction 
is challenging for standard path tracing because the intermediate surface points must satisfy the geometric constraints 
imposed by optical laws. 
This is traditionally handled by photo mapping~\cite{jensen1996global, hachisuka2009stochastic} or path regularization~\cite{kaplanyan2013path} 
that introduce bias (blurring).
Manifold-based approaches~\cite{jakob2012manifold, hanika2015manifold, zeltner2020specular} explore or solve for 
admissible paths on a low-dimensional path space manifold. These approaches tend to fail on highly complex displaced 
surfaces due to poor initialization of equation-solving.
Path Cuts~\cite{wang2020path} use a hierarchical search to prune surface regions 
that cannot form valid specular paths. 
Fan et al.~\shortcite{fan2024specular} propose a polynomial formulation of specular constraints that can be solved efficiently.
Subsequently, Fan et al.~\shortcite{fan2025bernstein} derive position and irradiance bounds for caustics cast by
triangle tuples based on Bernstein bounds for rational functions, which reduces the search domain for specular light transport.
These methods all expect input geometry as ordinary triangle meshes (some need precomputation on them). They are incompatible 
with tessellation-free displaced surfaces or at least require significant extensions.
With a conservative test derived from height and slope bounds, our hierarchy supports efficient specular next-event 
estimation queries without pre-tessellation.

\paragraph{Surface and light sampling.}
Surface sampling is a fundamental operation in computer graphics~\cite{shirley1996monte, keller1997instant, alliez2003anisotropic, li2010anisotropic}, geometric machine learning~\cite{qi2017pointnet, hermosilla2018monte}, and point-based neural rendering~\cite{xu2026generalizable}.
Sampling a triangle mesh by area can be done by selecting triangles proportional to areas and 
sampling uniformly within each triangle~\cite{pharr2023pbrt}. This first step requires enumerating over all triangles. 
Ling et al.~\shortcite{ling2025uniform} show hot to generate uniform
surface samples for arbitrary types of surface with a all-hit ray-intersection routine. 
For rendering, sampling 
should also account for the lighting integrand. Hierarchical product sampling combines multiple factors during sample 
generation~\cite{clarberg2005wavelet,clarberg2008practical}, and light trees use spatial, directional, and emission 
information to select lights according to their estimated contribution~\cite{conty2018manylights}. Our hierarchy instead 
guides sampling within a displaced surface, using the stored height and slopes to estimate regional area. Combining 
these estimates with emission and receiver-dependent weights supports importance sampling of displaced emitters without 
pre-tessellation.

\paragraph{Closest-point queries and surface PDEs.}
Bounding volume hierarchies accelerate closest-point queries on triangle meshes, as implemented in FCPW~\cite{FCPW}. When the mesh approximates a displaced surface, query accuracy depends on the tessellation resolution. Methods for continuous representations query the surface without first converting it to a triangle mesh: Marschner et al.~\shortcite{marschner2021sos} formulate closest-point queries on polynomial patches through sum-of-squares optimization, while Sharp and Jacobson~\shortcite{sharp2022spelunking} use range analysis on neural implicit surfaces. We derive distance lower bounds from the height planes and residuals in our hierarchy, enabling closest-point queries directly on displaced surfaces. These queries provide surface projections for Projected Walk on Spheres~\cite{sugimoto2024pwos}, which solves surface PDEs through Monte Carlo walks with repeated closest-point projections.

\section{Overview and background}
\label{sec:overview}
We first define the displaced surface and its shell coordinates, then describe the regional information needed for ray intersection, specular-path search, surface sampling, and closest-point queries.

\subsection{Displaced surface}
\label{sec:displace_surface}
A displaced triangle mesh consists of a base mesh and a scalar height map. Each base triangle carries three vertex positions $\mathbf P_i$, displacement directions $\mathbf N_i$, and texture coordinates $\mathbf T_i=(U_i,V_i)$, where $i\in{0,1,2}$ indexes the vertices. Displacement moves points on the base triangle along their interpolated directions by an amount determined by the scalar texture.

\begin{figure}[!ht]
    \vspace{-4pt}
    \centering
    \includegraphics[width=\columnwidth,trim=0 19bp 0 4bp,clip]{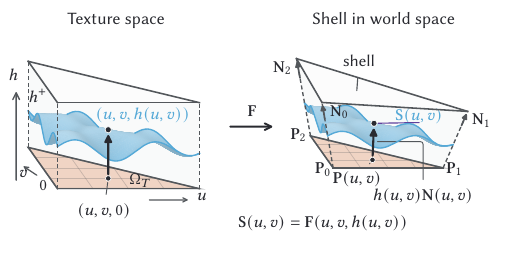}
 \vspace{-3pt}
    \caption{The shell map takes the height graph over the texture footprint $\Omega_T$ (left) to the displaced surface (blue) inside the world-space shell (right). Arrows connect corresponding base and surface points. The example uses $0\le h\le h_{\max}$ with unnormalized displacement directions $\mathbf N$ (see \autoref{sec:displace_surface}).}
    \label{fig:shell-space}
    \vspace{-4pt}
  \end{figure}

To bound and query the displaced surface, we describe the surrounding volume using two surface coordinates and an independent height coordinate. Following Ogaki~\shortcite{ogaki2023nonlinear}, we first consider a constant height $h$. Moving each vertex along its displacement direction gives $\mathbf S_i(h)=\mathbf P_i+h\mathbf N_i$. We can then interpolate these positions using barycentric coordinates $(1-\alpha-\beta,\alpha,\beta)$:
\begin{equation}
\begin{aligned}
\mathbf S(\alpha,\beta,h)&=(1-\alpha-\beta)\mathbf S_0(h)\\
&\quad+\alpha\mathbf S_1(h)+\beta\mathbf S_2(h).
\end{aligned}
\label{eq:shell-interpolation}
\end{equation}
For each fixed $h$, this describes a triangle (note this might not be parallel to the base triangle). Varying $h$ over an interval containing the displacement heights sweeps out a volume around the base triangle, called the \textbf{shell} (see \autoref{fig:shell-space} right). We refer to this volume in world coordinates as shell space, and to $(\alpha,\beta,h)$ as canonical-space coordinates.

The scalar displacement map is defined in texture coordinates, which can be obtained from the same barycentric interpolation:
\begin{equation}
(u,v)=(1-\alpha-\beta)\mathbf T_0+\alpha\mathbf T_1+\beta\mathbf T_2.
\label{eq:texture-coordinates}
\end{equation}
We denote the triangle's texture footprint by $\Omega_T$ and assume that its texture mapping is nonsingular, so each $(u,v)\in\Omega_T$ determines a unique pair $(\alpha,\beta)$. We can therefore express the interpolated base
position and direction as functions of texture position:
\begin{equation}
\begin{aligned}
\mathbf P(u,v)&=(1-\alpha-\beta)\mathbf P_0+\alpha\mathbf P_1+\beta\mathbf P_2,\\
\mathbf N(u,v)&=(1-\alpha-\beta)\mathbf N_0+\alpha\mathbf N_1+\beta\mathbf N_2.
\end{aligned}
\label{eq:interpolated-fields}
\end{equation}
We use $\mathbf N(u,v)$ without normalization after interpolation, following Ogaki~\shortcite{ogaki2023nonlinear}. Together with $h$, the coordinates $(u,v)$ form texture-space coordinates $(u,v,h)$. We now rewrite
\autoref{eq:shell-interpolation} using these coordinates, as the shell map
\begin{equation}
\mathbf F(u,v,h)=\mathbf P(u,v)+h\mathbf N(u,v).
\label{eq:shell-map}
\end{equation}

Let $h(u,v)$ denote the displacement height at texture coordinates $(u,v)$.  Within a texel cell, heights are interpolated bilinearly from the samples at the four texel centers. Evaluating the shell map at this height gives the displaced face (\autoref{fig:shell-space}):
  \begin{equation}
  \begin{aligned}
  \mathbf S(u,v)&=\mathbf F(u,v,h(u,v))\\
  &=\mathbf P(u,v)+h(u,v)\mathbf N(u,v),
  \qquad (u,v)\in\Omega_T.
  \end{aligned}
  \label{eq:displaced-surface}
  \end{equation}
We call the portion of the displaced surface corresponding to a texture region a surface patch. The shell map also relates height bounds to surface enclosures: a height interval over a texture region maps to a world-space region containing the corresponding displaced surface.

Surface area and geometric normals depend on how the displaced surface changes with texture position. Wherever $h$ is differentiable, its surface derivatives are
\begin{equation}
\begin{aligned}
\partial_u\mathbf S
&=\partial_u\mathbf P+(\partial_u h)\mathbf N+h\,\partial_u\mathbf N,\\
\partial_v\mathbf S
&=\partial_v\mathbf P+(\partial_v h)\mathbf N+h\,\partial_v\mathbf N,
\end{aligned}
\label{eq:surface-derivatives}
\end{equation} 
where $\partial_u$ and $\partial_v$ denote differentiation with respect to $u$ and $v$.
These expressions account for variation in both height and displacement direction. Note that even when $h$ is affine, a varying direction field can produce a nonplanar displaced patch.

\subsection{Geometric queries and their requirements}
\label{sec:overview_different_queries}
\begin{figure*}[!t]
  \centering
  \includegraphics[width=\textwidth]{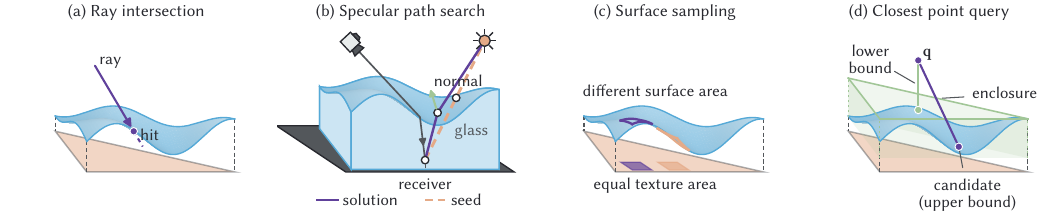}
  \caption{Different geometric queries on displaced surfaces. (a) A ray finds an intersection with the surface. (b) A specular manifold search connects a receiver to a light through a displaced glass interface: the orange dashed seed path is adjusted via manifold walk to satisfy refraction, yielding the purple path. The green arrow marks the surface normal at the specular interaction. (c) Equal texture areas can map to different surface areas. (d) The distance to a surface candidate gives an upper bound on the closest distance; the distance to an enclosing volume gives a lower bound. Any enclosure with a lower bound bigger than our current best distance can be directly rejected.}
  \label{fig:query-overview}
  \vspace{-8pt}
\end{figure*}

The surface definition provides positions and derivatives at individual texture coordinates. Efficient queries also need information over entire regions. A texture-space hierarchy groups neighboring cells into progressively larger regions, allowing queries to reject a region or assign it a sampling weight without examining every texel. Tessellation-free ray tracers use regional height bounds to guide traversal~\cite{thonat2021tessellation,ogaki2023nonlinear}. Here we describe the requirements of the four queries illustrated in \autoref{fig:query-overview}.

\paragraph{Ray intersection.}
Ray intersection determines visibility and locate surface interactions along light paths. Given a ray $\mathbf I(t)=\mathbf o+t\boldsymbol\omega$, the closest-hit query seeks the smallest admissible $t$ for which $\mathbf I(t)=\mathbf S(u,v)$. 
Regional height bounds allow traversal to reject portions of the surface that cannot intersect the ray before performing local intersection tests. Both scalar height intervals and our first-order height bounds
support this rejection; efficient evaluation also greatly depends on the traversal strategy (\autoref{sec:ray-query}).

\paragraph{Specular light path search.}
To render caustics, we seek points on a displaced surface that connect a receiver to a light source via specular
reflection or refraction. Admissible light paths must obey optical laws such as the law of reflection or Snell's law,
and arbitrarily sampled surface points generally do not satisfy the geometric constraints these laws impose. Finding valid
specular connections can therefore be viewed as a higher-order generalization of ray intersection: rather than a
single line segment, each solution is a path made up of multiple segments.

As in accelerated ray intersection, our general strategy is to use a hierarchical data structure to speed up the
search. Unlike a standard acceleration structure, however, ours must bound not only surface positions but also
normals, so that invalid regions can be pruned before the constraint is solved locally. These bounds must account for
both the base geometry and the slope field that defines the normals. \autoref{sec:specular-query} develops the pruning
test based on a first-order bound and the corresponding query algorithm.

\paragraph{Surface sampling.}
Sampling a displaced surface requires accounting for how area changes between texture space and the space the 
instantiated surface lives in. For a regular surface patch, the area element is
\begin{equation}
\mathrm dA=J(u,v)\,\mathrm du\,\mathrm dv,\qquad
J(u,v)=\|\partial_u\mathbf S\times\partial_v\mathbf S\|,
\label{eq:query-area-element}
\end{equation}
where the Jacobian $J$ is evaluated at the sampled point.
A sample drawn with texture-space density $p_{uv}$ therefore has area density $p_A=p_{uv}/J$ on a one-to-one patch. 
Therefore, we can expect uniform sampling in texture space is generally not uniform in surface area. 
Regional area estimates can guide sample allocation, while the pointwise Jacobian converts the resulting density to 
surface-area measure. Since $J$ depends on height and its derivatives, height intervals alone do not supply these estimates. 
\autoref{sec:measure-query} uses these area estimates to guide hierarchical sampling, with additional weighting strategies 
to support use cases such as displaced emitters.

\paragraph{Closest-point query.}
Given a point $\mathbf q$, the closest-point query finds a surface point at the minimum Euclidean distance
\begin{equation}
d^*(\mathbf q)=\min_T\min_{(u,v)\in\Omega_T}
\|\mathbf q-\mathbf S(u,v)\|.
\label{eq:query-closest-distance}
\end{equation}
Here the outer minimum ranges over the base triangles, and $\mathbf S$ stands for points on each triangle's displaced surface. 
The distance to a surface point already found gives an upper bound on $d^*(\mathbf q)$. A surface enclosure gives a lower bound on the distance to an unexplored region; if this exceeds the best distance found so far, the region can be discarded. Closest-point search therefore requires regional distance lower bounds and candidate points on the surface. These queries support surface projection in geometry processing; we demonstrate their use in Projected Walk on Spheres~\cite{sugimoto2024pwos}.

These queries use regional information in different ways: searches need bounds that can exclude solutions, while sampling needs estimates that guide sample placement. We therefore seek a representation that retains local height and slope information together with bounds on approximation error. \autoref{sec:hierarchy} introduces this representation and its construction.

\section{First-order hierarchy}
\label{sec:hierarchy}
We represent height over each texture region by an affine function and a conservative residual bound. The affine function retains local height variation and provides slope estimates, while the residual encloses the height function throughout the region. We construct these records hierarchically so that queries can use them at different spatial scales.

\subsection{Node representation}
\label{sec:node-representation}
We organize the height map into a regular texture-space quad-tree hierarchy. We use $n$ to index its nodes and denote the rectangular texture region covered by node $n$ as $\Omega_n$. At the finest level, a leaf covers one texture cell where the height should be bilinearly interpolated from four neighboring samples
Then each parent covers four children, as in a min--max mipmap used in~\cite{thonat2021tessellation, ogaki2023nonlinear}. Notice that the hierarchy is shared by triangles using the same height map. For a node covering $\Omega_n$ and a triangle with texture footprint $\Omega_T$, the query uses the clipped texture region $\Omega_n\cap\Omega_T$ and its corresponding surface patch.

Previous methods~\cite{thonat2021tessellation,thonat2023rmip} use scalar interval $[h_{\min},h_{\max}]$ to bound the heights within a region but does not describe how they vary with texture position. We instead approximate the height function by a local plane. This first-order (affine) model stores a height offset and two slopes. Higher-order models could also capture curvature, but would require more coefficients and more involved bound calculations. 
As shown in \autoref{fig:intro-height-bounds}, over each texture region, we write
\begin{equation}
\widehat h_n(u,v)
=h_n+g_{u,n}(u-u_c)+g_{v,n}(v-v_c).
\label{eq:overview-height-predictor}
\end{equation}
Here, $h_n$ is the plane's height at the region center ($u_c$ and $v_c$ denote this center's texture coordinates), and $g_{u,n},g_{v,n}$ are its slopes. We store a residual bound $r_n$ such that %
\begin{equation}
|h(u,v)-\widehat h_n(u,v)|\le r_n,
\qquad (u,v)\in\Omega_n.
\label{eq:overview-height-bound}
\end{equation}
The resulting height band follows the plane's slope and can remain narrow even when the region spans a large height range. The stored slopes describe this plane; they need not equal the height function's derivatives at the region center. Likewise, the residual bounds height error, but does not bound derivative error. Queries combine the height band with the base geometry and displacement directions through the shell map. Although the height plane is affine in texture space, its image in world space need not be planar when displacement directions vary.

In summary, each node's height representation contains four values $(h_n,g_{u,n},g_{v,n},r_n)$ and we compute the texture region and center from its grid coordinates and hierarchy level on the fly. 
For ray intersection, subtracting the height plane from a linear ray segment reduces rejection to a test against the residual band. For closest-point queries, the plane and residual combine with the base geometry to give a world-space distance lower bound. The plane's height and slopes also provide a regional area estimate for sampling. Specular-path search additionally requires bounds on the slope field used to define its normals. The following query sections derive these operations and the additional data they require.

\subsection{Hierarchy construction}
\label{sec:hierarchy-construction}
Here we describe how we build the hierarchy from the leaves upward. At each leaf, we compute a height plane and residual bound from the height map's interpolation rule. We then combine the records of each group of four children to form their parent, repeating this process up to the coarsest level.

\paragraph{Leaf initialization.}
Within a texture cell, bilinear interpolation gives
\begin{equation}
h(u,v)=h(u_c,v_c)+a_u\Delta u+a_v\Delta v
    +a_{uv}\Delta u\Delta v,
\label{eq:bilinear-leaf}
\end{equation}
where $\Delta u=u-u_c$ and $\Delta v=v-v_c$ are offsets from the cell center. The coefficients are the derivatives of the bilinear height function at that center:
\begin{equation}
a_u=\partial_u h(u_c,v_c),\qquad
a_v=\partial_v h(u_c,v_c),\qquad
a_{uv}=\partial_u\partial_v h(u_c,v_c).
\end{equation}
We use the constant and linear terms as the leaf's height plane, setting $h_n=h(u_c,v_c)$, $g_{u,n}=a_u$, and $g_{v,n}=a_v$. The remaining cross term $a_{uv}\Delta u\Delta v$ is the difference between the height function and this plane. For cell half-widths $s_u,s_v$, the offsets satisfy $|\Delta u|\le s_u$ and $|\Delta v|\le s_v$, which gives us the residual bound
\begin{equation}
r_n=|a_{uv}|s_us_v.
\label{eq:leaf-node}
\end{equation}
The residual in \autoref{eq:leaf-node} bounds the deviation of the cell's four
corner samples from the leaf plane. Both bilinear interpolation and piecewise-linear interpolation over the cell's two triangles preserve this bound, since they interpolate the corner heights and the affine plane using the same weights. The latter provides the height interpolant used for ray intersection (\autoref{par:ray-leaf-tests}). We refer to Thonat et al.~\shortcite{thonat2021tessellation} for a discussion of other interpolation rules.

\paragraph{Parent node construction.}
Starting from the leaves, we construct each parent from its four child nodes without reading the height map again. Let $p$ denote the parent and $(u_c,v_c)$ denote its center. We first choose its slopes $g_{u,p},g_{v,p}$: our choice is by
averaging the child slopes. We then determine the height offset and residual needed to enclose all four
The construction is valid for any choice of parent slopes.
\begingroup
\setlength{\intextsep}{0pt}
\setlength{\columnsep}{3pt}
\begin{wrapfigure}{r}{0.50\columnwidth}
\centering
\includegraphics[width=\linewidth]{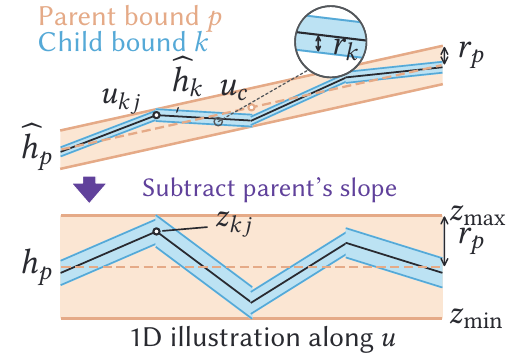}
\end{wrapfigure}

For each child $k$, we evaluate its four corners, indexed by $j\in\{0,1,2,3\}$, with texture coordinates $(u_{kj},v_{kj})$. Subtracting the parent slope contribution from the child's plane height gives
\par
\ParentBoundsClearance
\endgroup
\begin{equation}
z_{kj}=\widehat h_k(u_{kj},v_{kj})
    -g_{u,p}(u_{kj}-u_c)-g_{v,p}(v_{kj}-v_c).
\end{equation}
Each $z_{kj}$ is the parent height offset that would make its plane pass through that point on the child plane. The child residual expands it to the interval $[z_{kj}-r_k,z_{kj}+r_k]$. We take the range over all four children and
their corners, then use its midpoint and half-width as the parent's height offset and residual:
\begin{equation}
\begin{aligned}
z_{\min}&=\min_{k,j}(z_{kj}-r_k),&
z_{\max}&=\max_{k,j}(z_{kj}+r_k),\\
h_p&=\tfrac12(z_{\min}+z_{\max}),&
r_p&=\tfrac12(z_{\max}-z_{\min}).
\end{aligned}
\label{eq:parent-height}
\end{equation}
Since our representation is affine, evaluating the corners suffices. Including the child residual encloses its entire height band. Repeating this construction up to the coarsest level preserves the height bounds throughout the hierarchy. 

Our first-order hierarchy provides tighter bounds and extra information needed for more geometric queries that follow. In the following sections, we will describe how we use these quantities, and derive query specific hierarchy traversal, pruning, and sampling strategies. Although our main goal is to support queries beyond ray intersection, we begin by showing how the representation also supports our efficient tessellation-free ray traversal strategies, then turn to specular path search, surface sampling, and closest-point queries.

\section{Ray intersection}
\label{sec:ray-query}
 
We begin with ray intersection, using the height planes and residuals from \autoref{sec:hierarchy} to reject surface regions during traversal. In texture space, each node encloses the height function between $\widehat h_n-r_n$ and $\widehat h_n+r_n$. Testing whether a ray remains outside this band therefore requires evaluating its height relative to the stored plane. Following Ogaki~\shortcite{ogaki2023nonlinear}, we perform these tests in texture space, where the bands are defined directly, rather than constructing world-space enclosures that may overlap between neighboring texture regions.
  
\begin{figure}[!htb]
 \vspace{-4pt}
 \centering
 \includegraphics[width=\columnwidth,trim=0 2bp 0 3bp,clip]{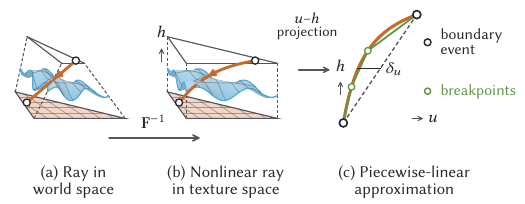}
 \vspace{-7pt}
 \caption{(a) A straight world-space ray maps to (b) a nonlinear texture-space ray. \textbf{Shell distortion is exaggerated} to expose the curvature. (c) Uniform subdivision in the $u$--$h$ projection: the midpoint deviation $\delta_u$ of the single dashed segment sets the piece count, 
 here three equal steps in height (green). Black circles mark boundary events and green circles mark breakpoints.}
 \label{fig:ray-overview}
 \vspace{-11pt}
\end{figure}

The difficulty is that a straight world-space ray generally becomes a curve in texture space (\autoref{fig:ray-overview}(a, b)). Ogaki represents this curve by rational functions of height and intersects it with texture-space boxes through quadratic root solves. As traversal moves between nodes, their box boundaries change, requiring new solves to determine where the curve enters and leaves each box. We instead approximate the curve by linear segments before traversing the hierarchy. For each segment, successive grid crossings can be computed incrementally, avoiding repeated nonlinear intersection tests. This also simplifies evaluation of our first-order bounds: subtracting a node's affine height plane from the segment's height gives an affine function whose extrema occur at the endpoints. We can therefore test the sloped height band directly without expanding it into a scalar height interval. Padding accounts for the segment approximation, with its limitations described in \autoref{sec:ray-traversal}.

Our choice of linear segments is motivated by the observation that many mapped rays deviate little from straight segments over the traversed intervals. A small number of segments can therefore represent these curves closely while enabling inexpensive traversal and height-band tests. Across \rtCases{} mesh--displacement map pairs, 98.94\% of 396.9 million ray--prism builds that produce segments use at most four segments, with an approximation threshold of 0.06 texel. \autoref{sec:ray-results} evaluates rendering speed and agreement with NRT.
\par\vspace{\SegmentCountsSkip}
\noindent\includegraphics[width=\columnwidth]{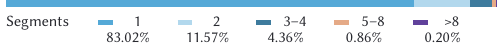}
\par\vspace{\SegmentCountsSkip}

Our scene bounding volume hierarchy (BVH) first identifies candidate displaced triangles by intersecting the ray with prism-based shell bounds. For each candidate, we approximate the mapped ray by linear segments and traverse them in increasing world-ray parameter $t$, using incremental grid steps and endpoint tests against the node bounds.
\autoref{alg:ray-intersection} summarizes this per-triangle query; the following subsections describe segment construction and traversal.

\begin{algorithm}[!htb]
\caption{Ray intersection within a displaced triangle}
\label{alg:ray-intersection}
\small
\algrenewcommand{\algorithmicindent}{1em}
\definecolor{raycomment}{HTML}{55A9D7}
\algrenewcommand{\algorithmiccomment}[1]{\hfill\textcolor{raycomment}{$\triangleright$\,\textit{#1}}}
\begin{algorithmic}[1]
\Require Ray $\mathbf I$ with interval $[t_{\min},t_{\max}]$, base triangle $T$, height hierarchy
\State $\mathrm{best}\gets\mathrm{miss}$
\State $\mathcal C\gets\Call{BuildSegments}{\mathbf I,T,[t_{\min},t_{\max}]}$ \Comment{\autoref{sec:ray-chords}}
\For{each segment $c\in\mathcal C$ in increasing ray parameter $t$}
  \State $n\gets\Call{FirstNode}{c,T,t_{\max}}$ \Comment{\autoref{sec:ray-traversal}}
  \While{$n$ exists}
      \State $J\gets\Call{ClipSegment}{c,\Omega_n,t_{\max}}$ \Comment{Interval inside node}
      \If{$J=\varnothing$ or $\Call{Reject}{c,n,J}$} \Comment{\autoref{eq:first-order-ray-test}}
          \State $n\gets\Call{AdvanceDDA}{c,n,t_{\max}}$ \Comment{Skip node and descendants}
      \ElsIf{$n$ is above the terminal level}
          \State $n\gets\Call{Descend}{c,n,J}$ \Comment{First crossed child}
      \Else
          \State $\mathrm{hit}\gets\Call{IntersectLeaves}{\mathbf I,T,c,n,J}$ \Comment{\autoref{par:ray-leaf-tests}}
          \If{$\mathrm{hit}$ is valid and closer than $\mathrm{best}$}
              \State $\mathrm{best}\gets\mathrm{hit}$;
                     $t_{\max}\gets t_{\mathrm{hit}}$
          \EndIf
          \State $n\gets\Call{AdvanceDDA}{c,n,t_{\max}}$
      \EndIf
  \EndWhile
\EndFor
\State \Return $\mathrm{best}$
\end{algorithmic}
\end{algorithm}

\subsection{Piecewise-linear ray approximation}
\label{sec:ray-chords}
We approximate the mapped ray with linear segments by partitioning at geometric events, then subdividing each interval uniformly according to its midpoint deviation (\textsc{BuildSegments} in \autoref{alg:ray-intersection}). \autoref{fig:ray-overview} illustrates the construction.

\paragraph{Partitioning at geometric events.} Following Ogaki~\shortcite{ogaki2023nonlinear}, we parameterize the mapped ray by height.
For a world-space ray $\mathbf I(t)=\mathbf o+t\boldsymbol\omega$, eliminating $t$ from the shell intersection equation $\mathbf F(u,v,h)=\mathbf I(t)$ gives
\begin{equation}
\mathbf x(h)=\left(\frac{\Unum(h)}{d(h)},\frac{\Vnum(h)}{d(h)}\right)\trans.
\label{eq:rational-ray}
\end{equation}
Here, $\mathbf x(h)=(u(h),v(h))\trans$ is the ray's texture position at height $h$. The coordinate numerators $U(h),V(h)$ and common denominator $d(h)$ are obtained from the ray--shell intersection equation following Ogaki~\shortcite{ogaki2023nonlinear}. All three have degree at most two. This construction relies on the affine displacement-direction field in \autoref{eq:interpolated-fields};
  
Before approximating the curve, we partition its height range at three types of events. Singularities separate branches where the rational parameterization is undefined. Turning points of the barycentric functions $\alpha(h)$ and $\beta(h)$ split each branch into intervals over which both coordinates are monotonic.  Triangle-boundary crossings delimit the portions inside the triangle. We retain only interior intervals without singularities, then approximate each interval by linear segments. 

\paragraph{Uniform subdivision.}
Within each retained interval, we estimate how many segments are needed by measuring the curve's deviation at the midpoint in height. We compare the texture position of this point with the midpoint of the straight segment joining the interval's endpoints (\autoref{fig:ray-overview}c). A small deviation allows fewer segments; a larger deviation produces more. We then sample the curve at equal height increments and connect consecutive samples. Thus, spacing is uniform within each event interval, while the segment count adapts to its estimated deviation.

We use an approximation threshold of 0.06 texel.
The midpoint estimate determines the subdivision count but does not bound the maximum deviation between samples. Traversal therefore uses additional padding to account for the difference between the segments and the curve, as described in \autoref{sec:ray-traversal}.

We parameterize each segment by $s\in[0,1]$, with texture coordinates $\mathbf x(s)=(u(s),v(s))^{\mathsf T}$ and height $h(s)$ linear in $s$. This permits incremental grid traversal and endpoint tests against a node's affine height plane.

\subsection{Hierarchical DDA traversal}
\label{sec:ray-traversal}
Approximating the nonlinear texture-space ray by linear segments enables efficient traversal with a two-dimensional digital differential analyzer (DDA). Grid crossings can be updated incrementally, avoiding the repeated nonlinear ray--box intersection tests used by Ogaki~\shortcite{ogaki2023nonlinear}. We process segments in increasing world-ray parameter $t$. For each segment, we clip it to the triangle's texture footprint and current ray interval, choose a starting hierarchy level based on its texture-space extent, and locate the cell containing its starting point (\textsc{FirstNode} in \autoref{alg:ray-intersection}). The DDA tracks the next horizontal and vertical grid crossings and advances to whichever occurs first.

At each visited node, we test the segment against the height enclosure defined in \autoref{sec:node-representation}. We clip it to the node's texture region (\textsc{ClipSegment}) and measure its height relative to the stored plane:
\begin{equation}
q_n(s)=h(s)-\widehat h_n(\mathbf x(s)).
\label{eq:plane-compensated-ray}
\end{equation}
Subtracting the plane turns the node's height band into the constant interval $[-r_n,r_n]$, where $r_n$ is the stored height residual. Since $q_n$ is affine in $s$, its extrema occur at the interval endpoints. To account for the difference between the segment and the curved ray, we extend the tested interval and widen the height band by $\eta_n$. Let $q_a$ and $q_b$ be the values of $q_n$ at the extended interval's endpoints. We reject when both values lie below or both lie above the widened band (\textsc{Reject}):
\begin{equation}
\max(q_a,q_b)<-r_n-\eta_n\quad\text{or}\quad\min(q_a,q_b)>r_n+\eta_n.
\label{eq:first-order-ray-test}
\end{equation}
The padding $\eta_n$ accounts for the change in plane height caused by the curve's sideways deviation from the segment. The supplementary material gives the padding calculations and their limitations.
\par\vspace{2pt}
{\centering
\includegraphics[width=.72\columnwidth]{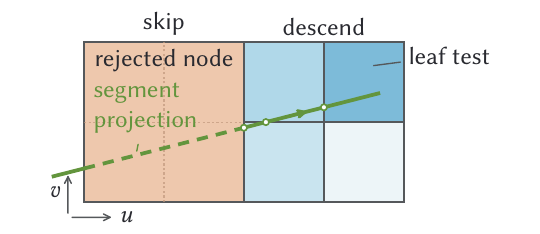}\par}
\vspace{-3pt}
\noindent A rejected interval skips the node and its descendants. Otherwise, we narrow the interval to where the segment overlaps the padded height band. Since $q_n$ is affine, this interval is obtained by solving linear inequalities. We descend into the children crossed within this interval (\textsc{Descend}). After processing a node's interval, traversal advances to the next crossed region, ascending the hierarchy as needed (\textsc{AdvanceDDA}). Because a segment only approximates the curved ray, the ray may pass through cells beside the segment. Enlarging the height band does not make traversal visit these cells; additional cell tests are described in the supplementary material. Whenever we find a closer intersection, we shorten the remaining ray interval by updating $\raytmax$. Visibility queries stop as soon as an intersection is found.

\phantomsection\label{par:ray-leaf-tests}
The stored ray-tracing hierarchy ends at $4\times4$-texel blocks, omitting the two finest hierarchy levels. Within each surviving block, we traverse individual texel cells and construct local surface triangles directly from the height-map samples. Each cell requiring a local test supplies four displaced corner samples, which form two planar triangles for intersection with the world-space ray. We follow the local-triangulation approach of Thonat et al.~\shortcite{thonat2021tessellation} and refer to their study for comparisons of leaf-intersection methods. Shading normals follow \autoref{eq:surface-derivatives}. The planar leaf tests and piecewise-linear traversal make the complete renderer approximate.

\subsection{Results}
\label{sec:ray-results}

\begin{table}[t]
 \centering\small
 \setlength{\tabcolsep}{1.25pt}
  \caption{Gallery rendering times (ms) at 1 spp and $1280\times1024$ resolution (Skull helmet: $960\times960$). Times are medians over 30 measured frames.
   Bold marks the fastest method. LPIPS~\cite{zhang2018lpips} is measured on 256-spp images where lower is better. The final row reports geometric-mean speedups over NRT and mean LPIPS. PDM settings are described in \autoref{sec:ray-results}.}
 \label{tab:ray-results}
 \fontsize{7.5}{9}\selectfont
 \begin{tabular}{@{}lrrrrr@{\hspace{5pt}}r@{\hspace{4.5pt}}r@{}}
 \toprule
 & \multicolumn{5}{c}{Time (ms)} & \multicolumn{2}{c@{}}{LPIPS}\\
 \cmidrule(lr){2-6}\cmidrule(l{3pt}){7-7}\cmidrule(l{3pt}){8-8}
+ Scene & TFDM & NRT & PDM$_{0.005}$ & PDM$_{0.4\%}$ & Ours & \shortstack{PDM$_{0.005}$ vs\\PDM$_{0.4\%}$} & \shortstack{Ours vs\\NRT}\\
 \midrule
Reactor & 25.36 & 34.90 & \textbf{6.94} & 37.91 & 9.45 & 0.0466 & 0.0013\\
Wood block & 21.48 & 25.72 & 8.75 & 38.11 & \textbf{7.89} & 0.0157 & 0.0006\\
Skull helmet & 71.52 & 37.87 & 16.38 & 70.14 & \textbf{8.17} & 0.0421 & 0.0038\\
Quilted cushion & 18.99 & 23.54 & \textbf{7.49} & 24.54 & 7.81 & 0.0048 & 0.0002\\
Dry branches & 47.72 & 64.54 & 12.33 & 46.17 & \textbf{9.75} & 0.0340 & 0.0003\\
Motherboard & 19.19 & 21.28 & \textbf{4.03} & 44.14 & 6.65 & 0.0130 & 0.0003\\
Woven knot & 31.51 & 42.50 & \textbf{8.95} & 39.09 & 10.78 & 0.0025 & 0.0002\\
Stone spiral & 19.93 & 26.94 & 8.28 & 29.68 & \textbf{7.22} & 0.0183 & 0.0002\\
Stepped torus & 49.60 & 73.47 & 36.55 & 75.98 & \textbf{17.51} & 0.0378 & 0.0009\\
 \midrule
Speedup / mean & 1.18$\times$ & $1\times$ & 3.56$\times$ & 0.84$\times$ & 3.92$\times$ & 0.0239 & 0.0009\\
 \bottomrule
 \end{tabular}
\end{table}

\definecolor{EGRayTileBackground}{RGB}{0,0,0}
\newcommand{\EGRayGalleryTile}[5]{%
  \begin{tikzpicture}[baseline=(frame.south)]
    \node[inner sep=0pt, outer sep=0pt, minimum width=#1,
      minimum height=\dimexpr#1*4/5\relax,
      fill=EGRayTileBackground] (frame) {};
    \node[inner sep=0pt] (beauty) at (frame.center)
      {\includegraphics[width=#1,height=\dimexpr#1*4/5\relax,
        keepaspectratio]{figs/beauty_gallery/#3}};
    \node[anchor=north west, outer sep=0pt, fill=black, text=white, inner sep=1.1mm,
      font=\scriptsize]
      at (frame.north west) {\textsc{#5}};
    \node[anchor=south east, inner sep=0pt, outer sep=0pt] (base)
      at (frame.south east)
      {\includegraphics[width=#2]{beauty_gallery/#4}};
    \draw[white, line width=1.2pt, line cap=butt, line join=miter]
      ([xshift=.6pt]base.south west) --
      ([shift={(.6pt,-.6pt)}]base.north west) --
      ([yshift=-.6pt]base.north east);
  \end{tikzpicture}%
}

\begin{figure*}[tp]
 \centering
 \makebox[\textwidth][c]{%
  \EGRayGalleryTile{.320\textwidth}{.078\textwidth}
   {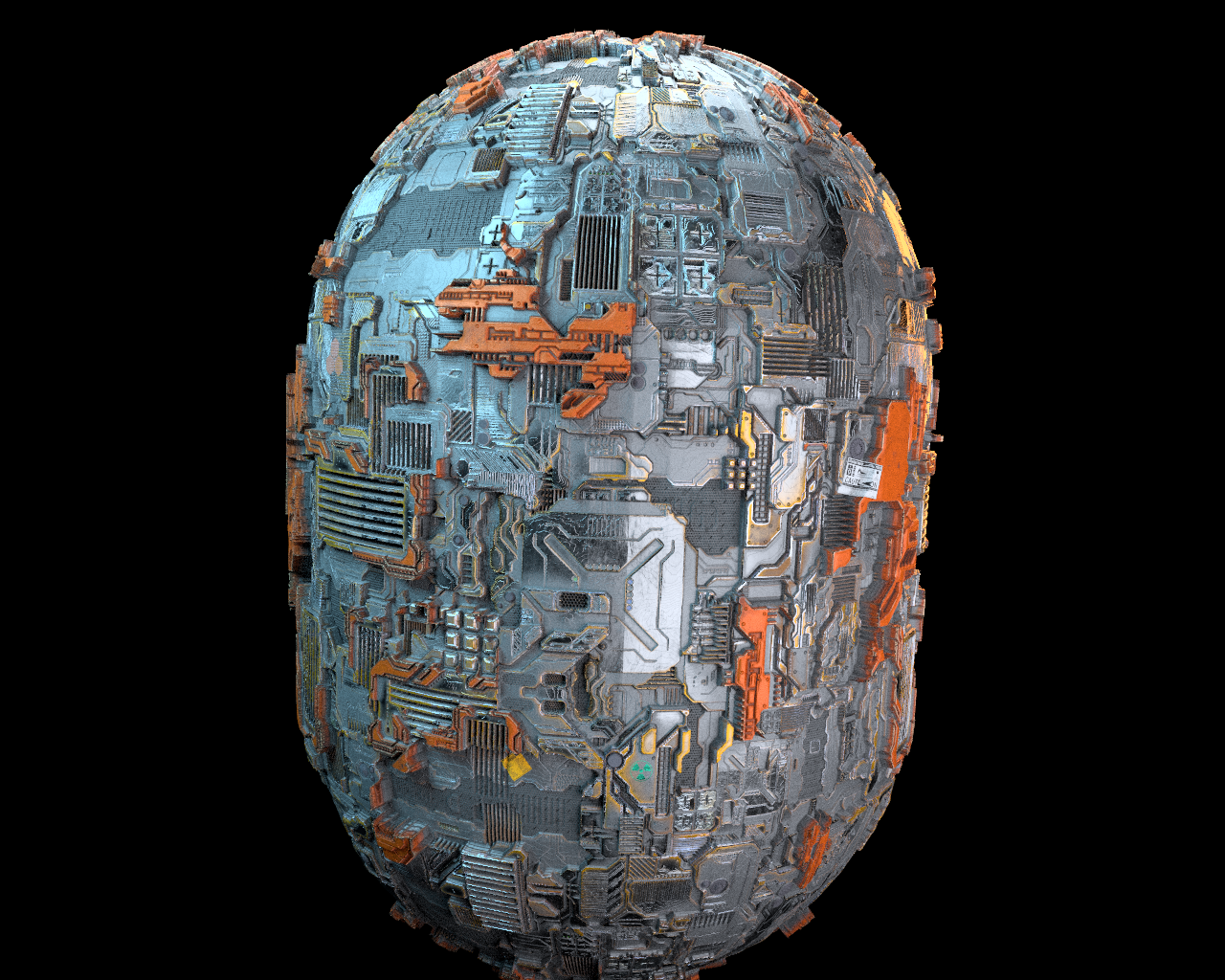}{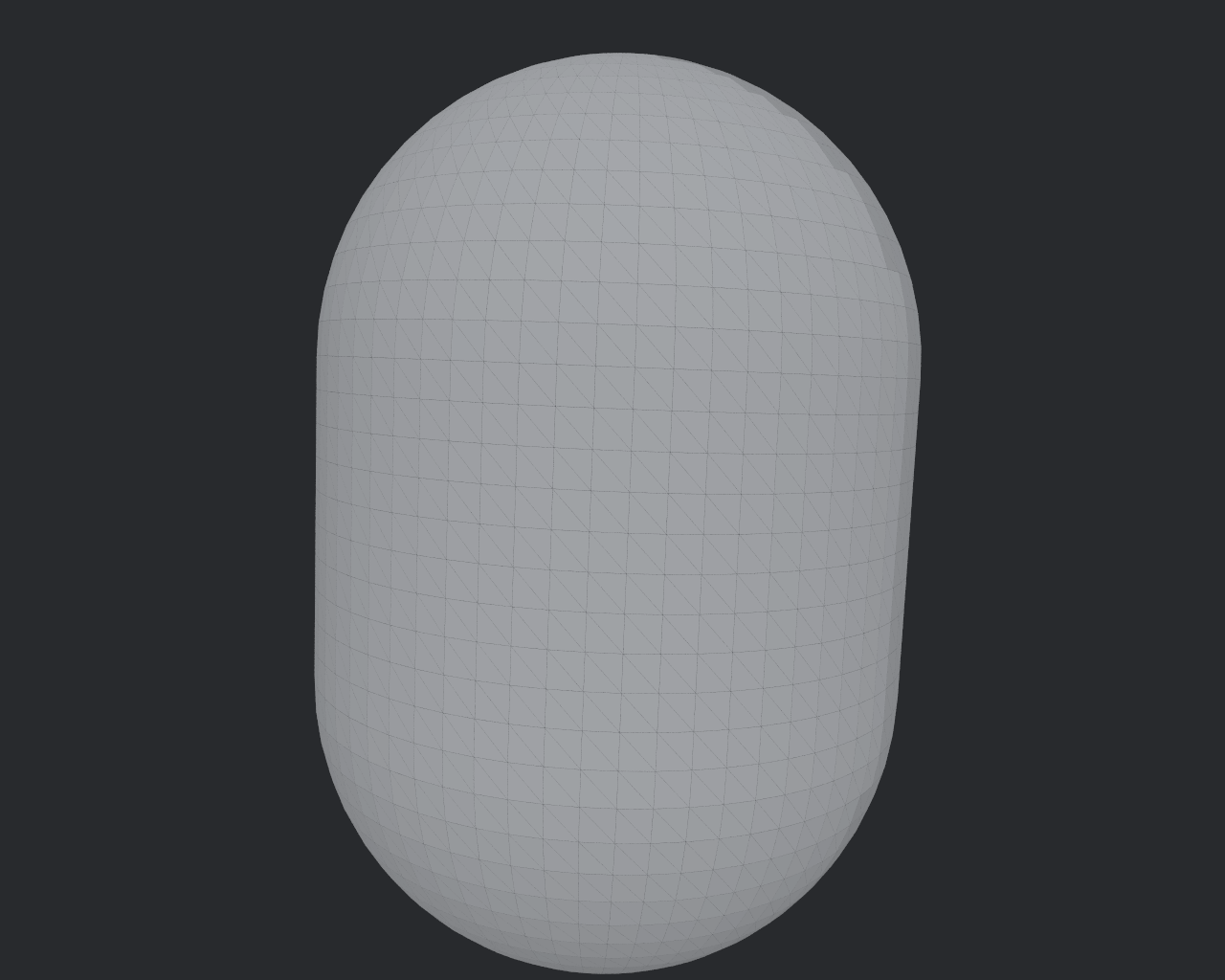}
   {Reactor}\hfill
  \EGRayGalleryTile{.320\textwidth}{.078\textwidth}
   {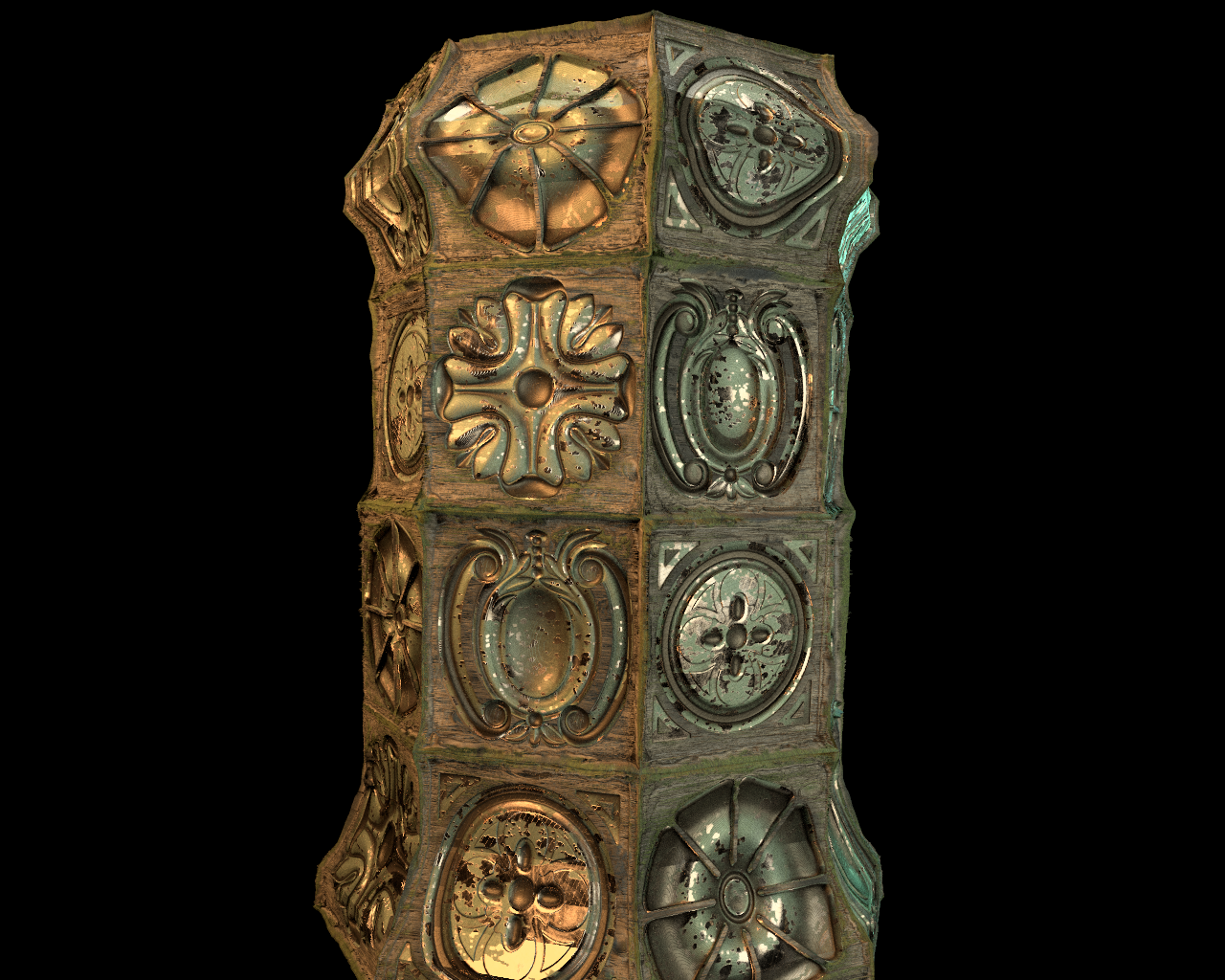}{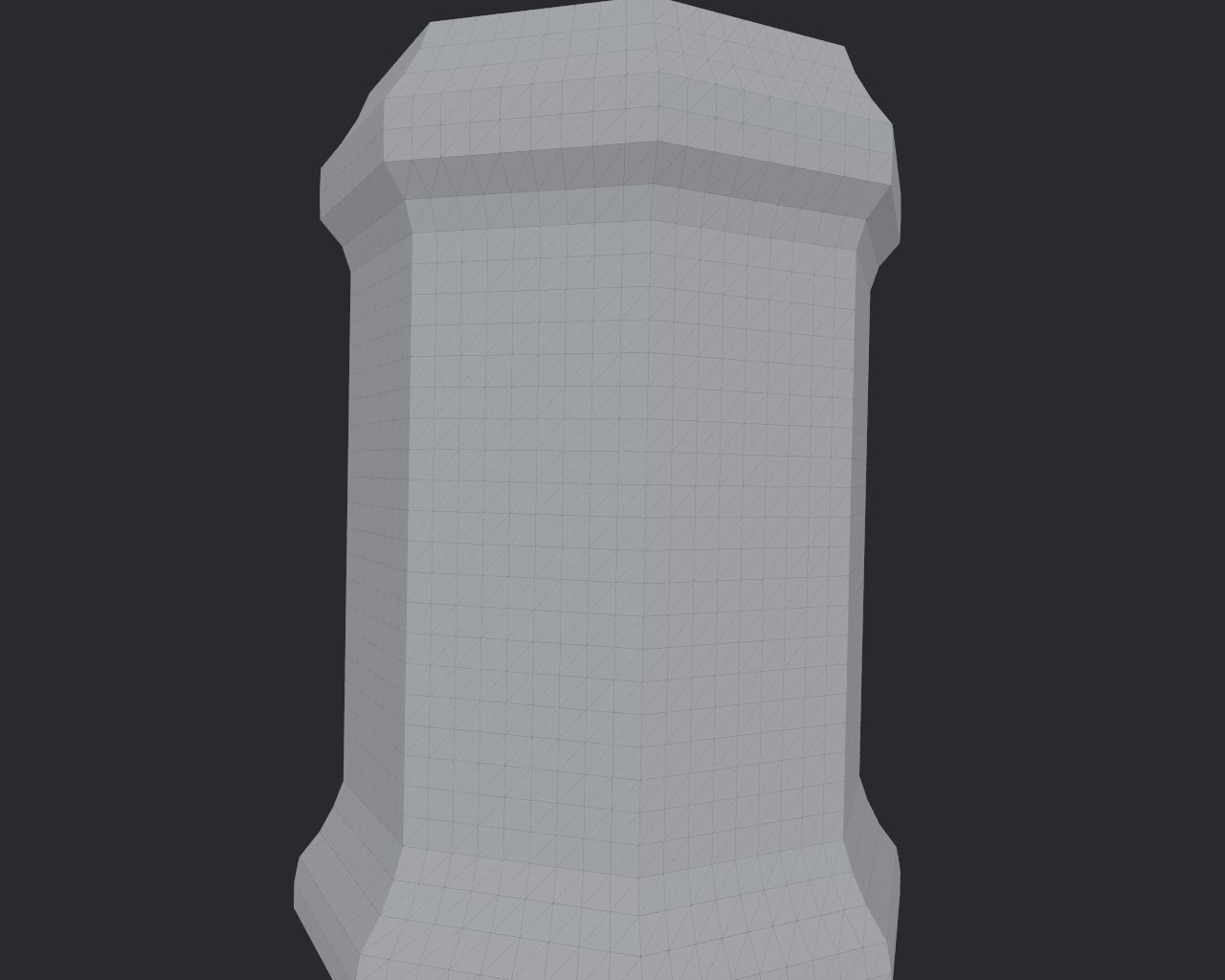}
   {Wood block}\hfill
  \EGRayGalleryTile{.320\textwidth}{.078\textwidth}
   {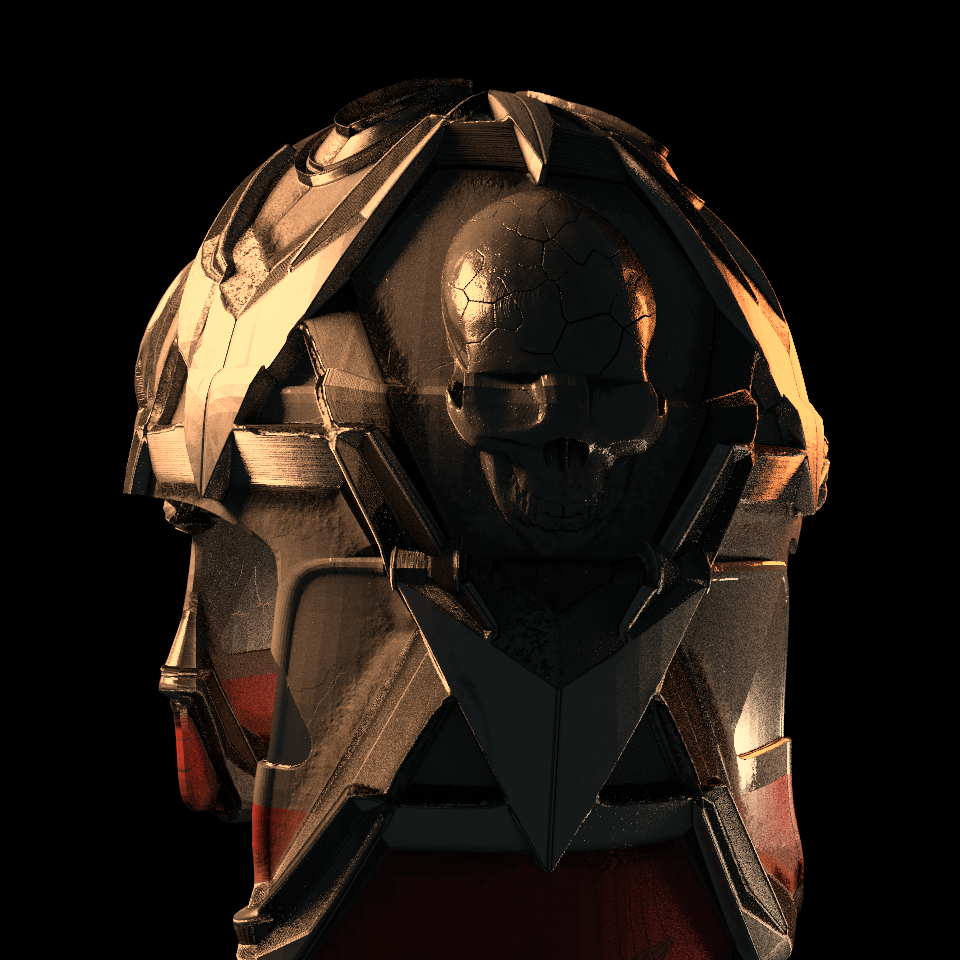}{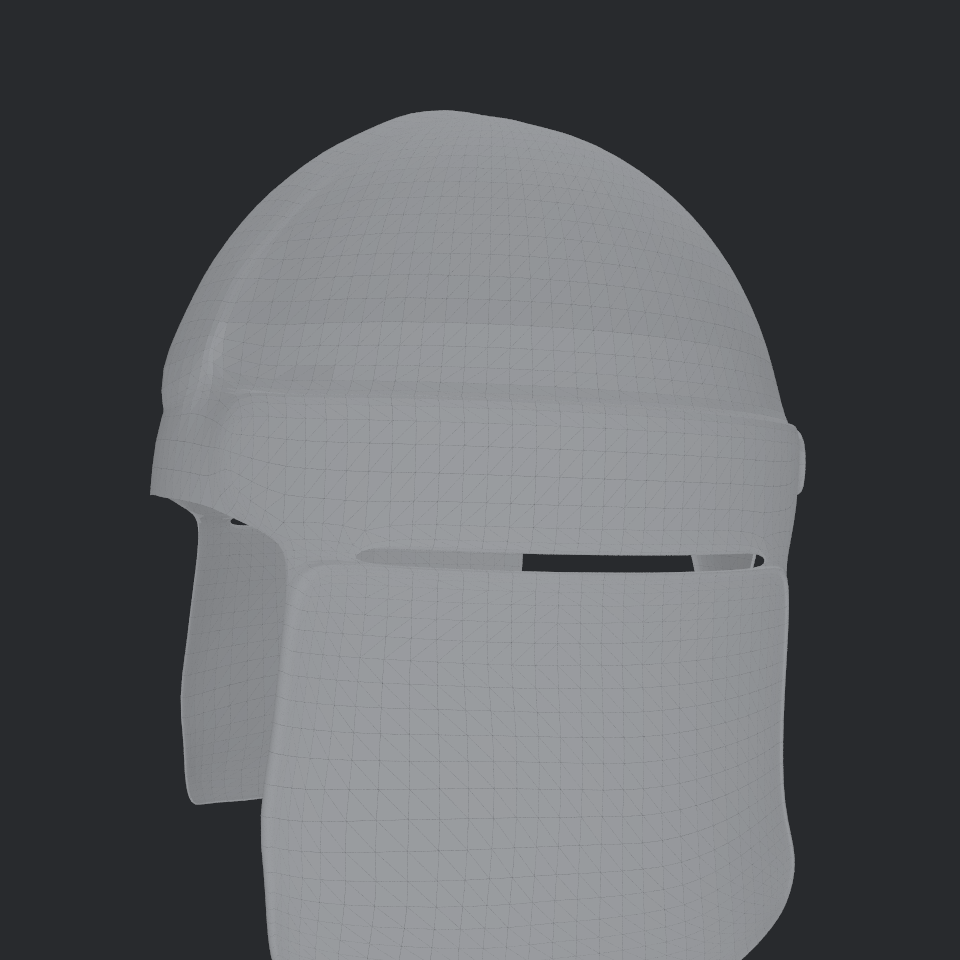}
   {Skull helmet}}

 \vspace{1.1mm}
 \makebox[\textwidth][c]{%
  \EGRayGalleryTile{.320\textwidth}{.078\textwidth}
   {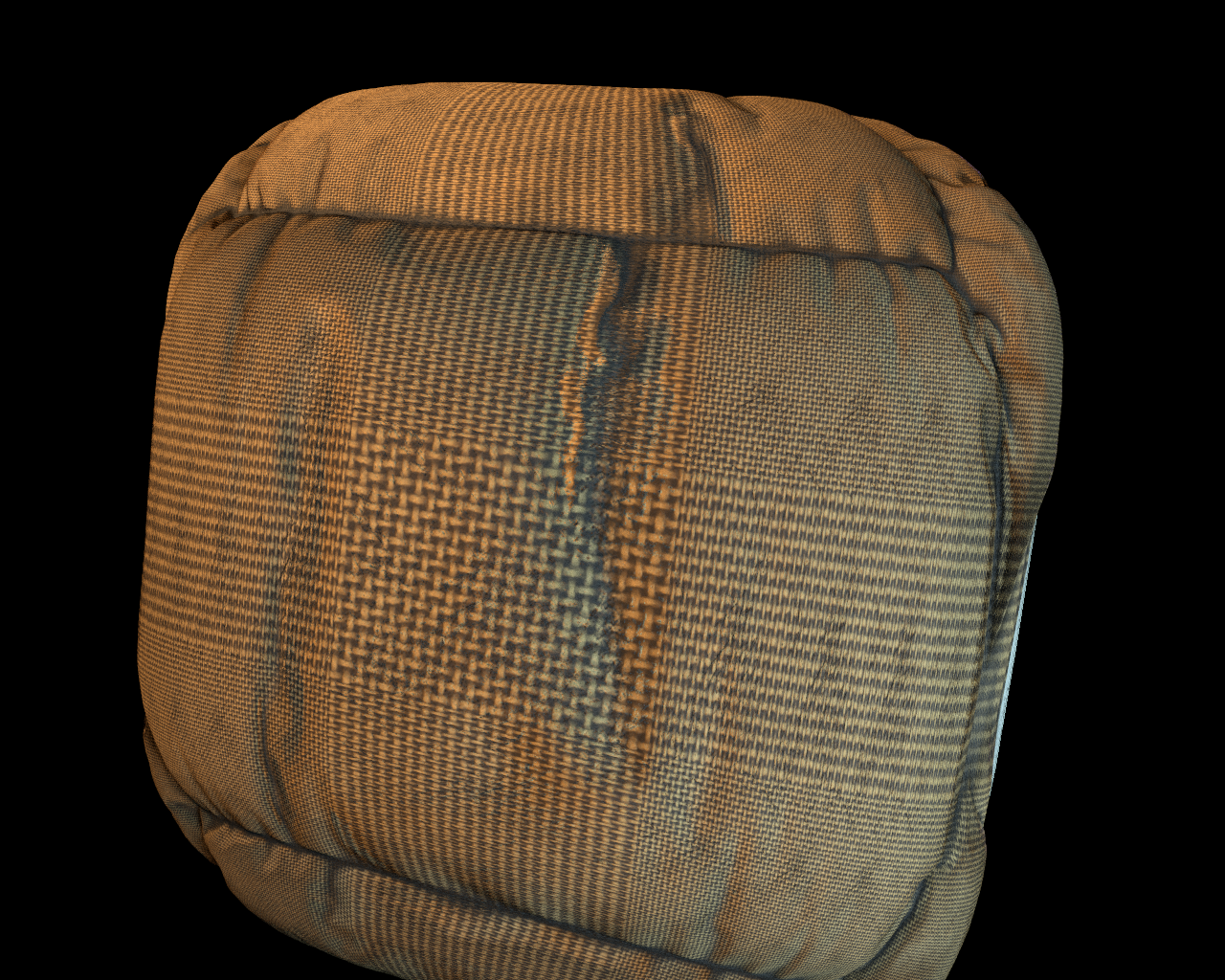}{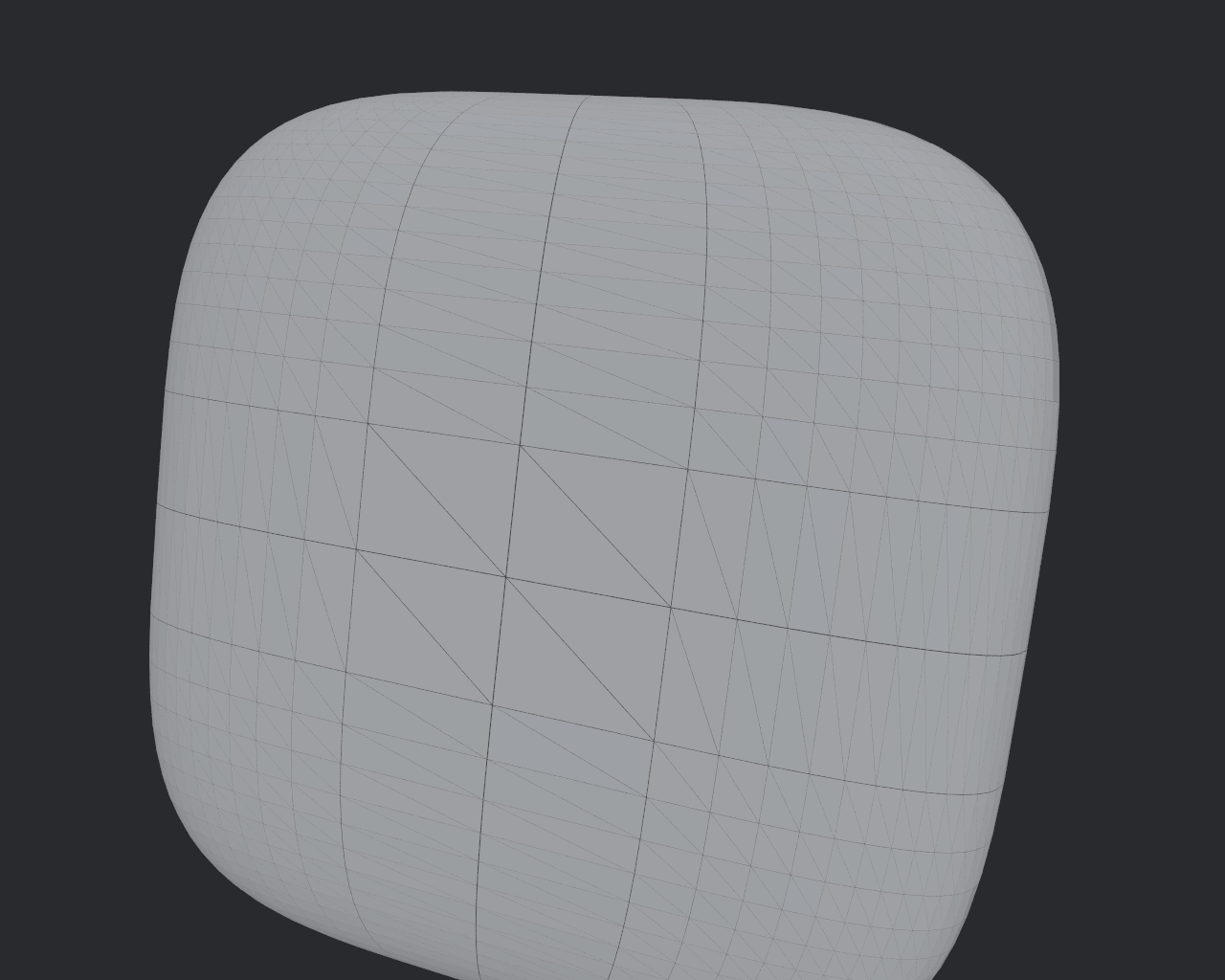}
   {Quilted cushion}\hfill
  \EGRayGalleryTile{.320\textwidth}{.078\textwidth}
   {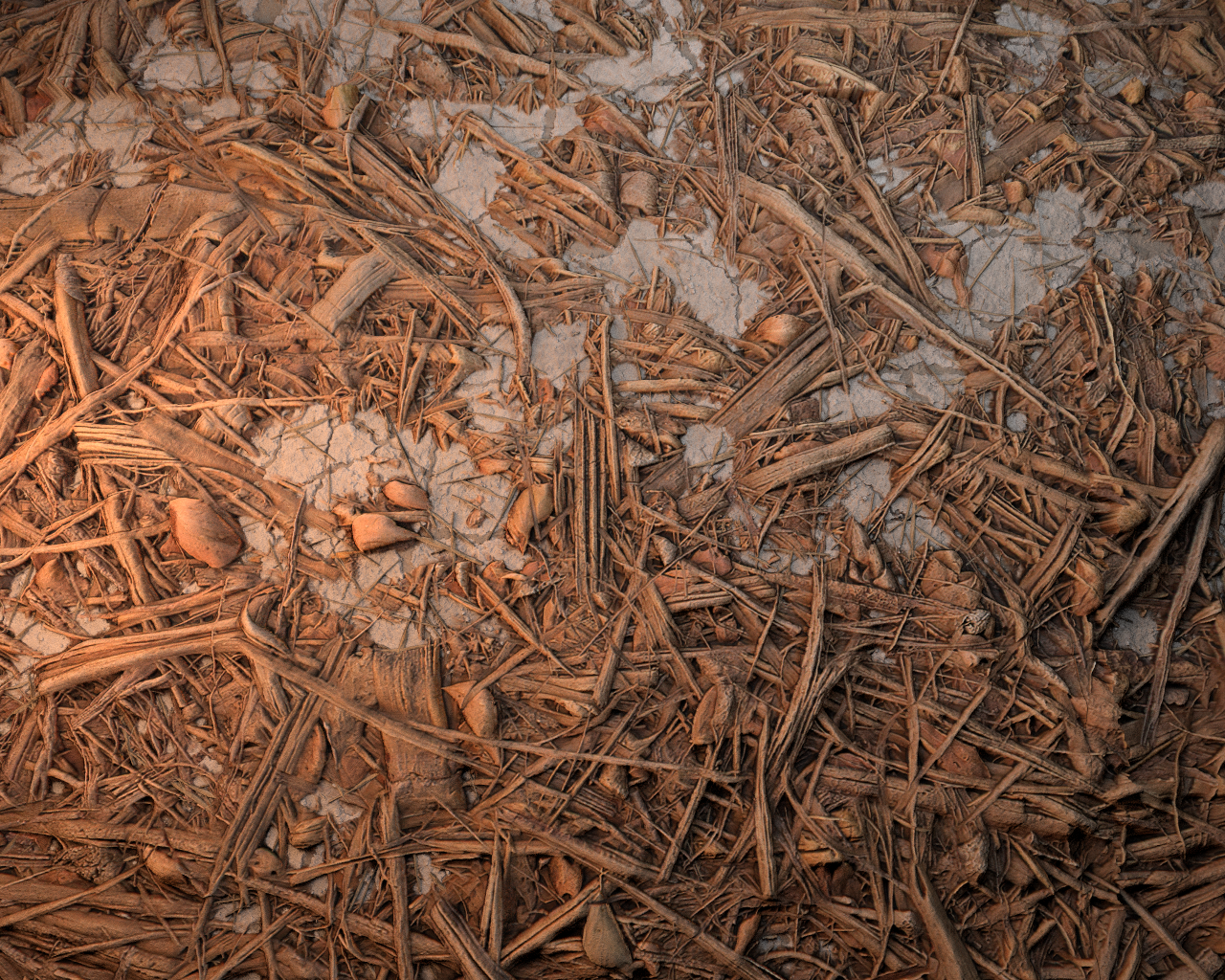}{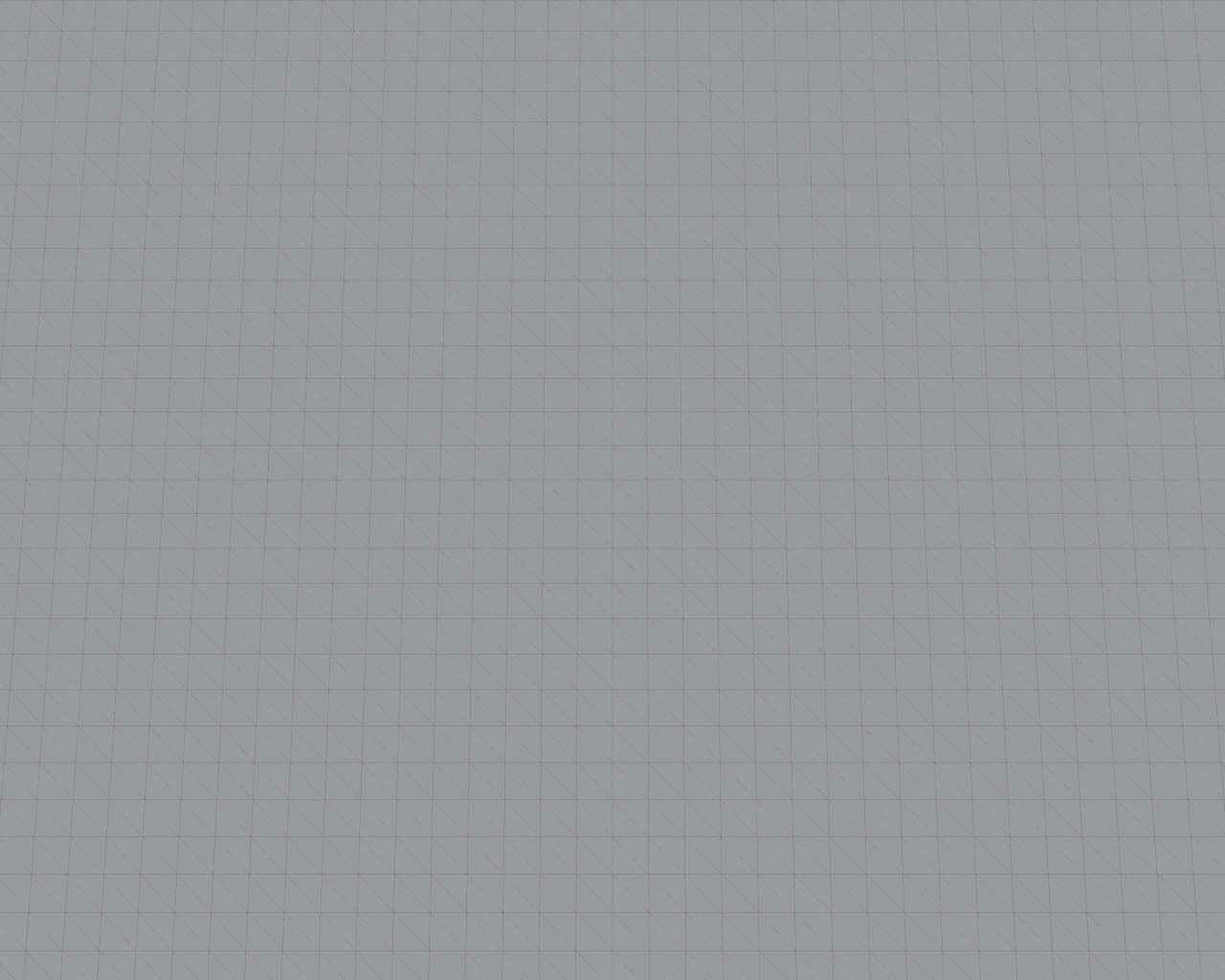}
   {Dry branches}\hfill
  \EGRayGalleryTile{.320\textwidth}{.078\textwidth}
   {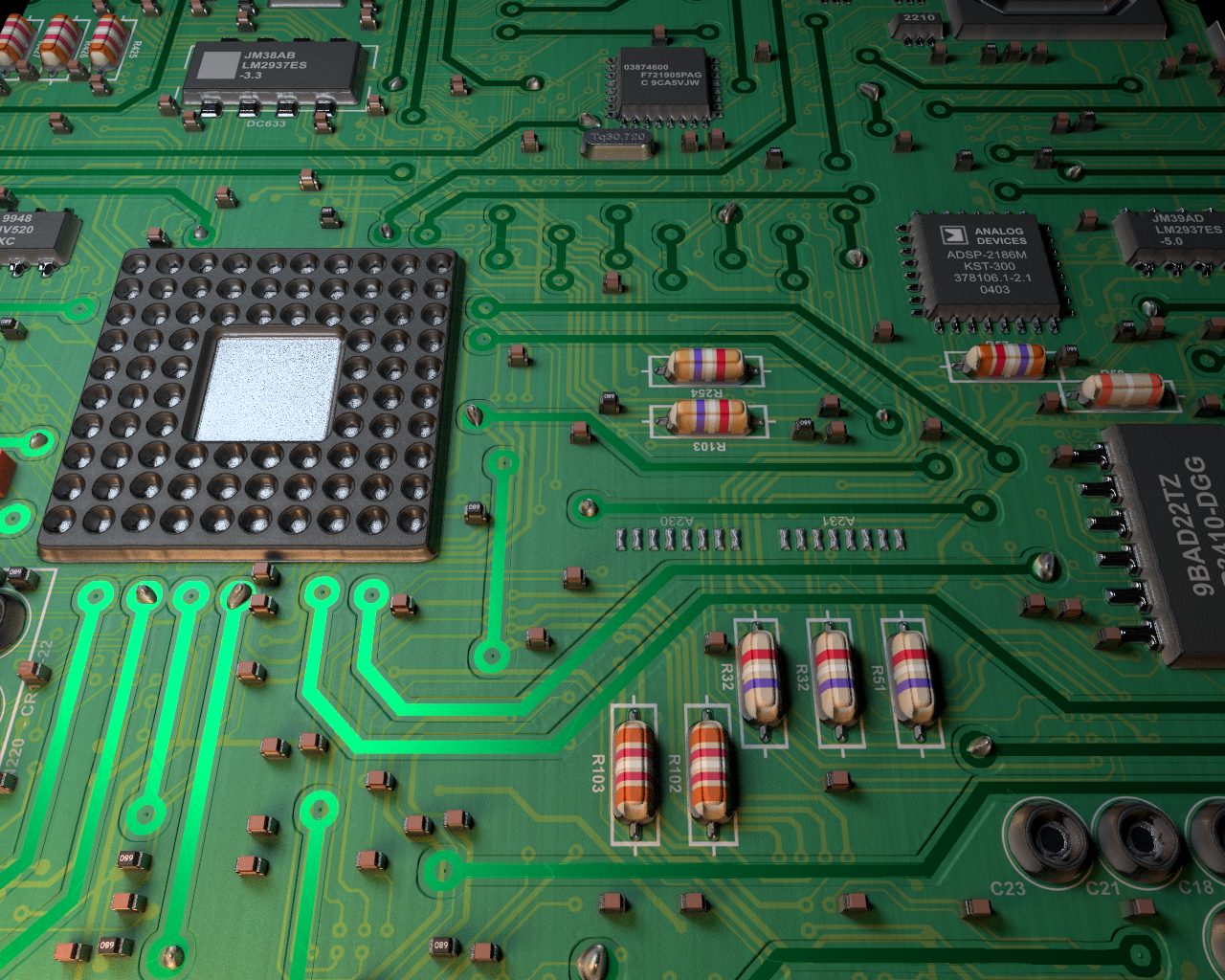}{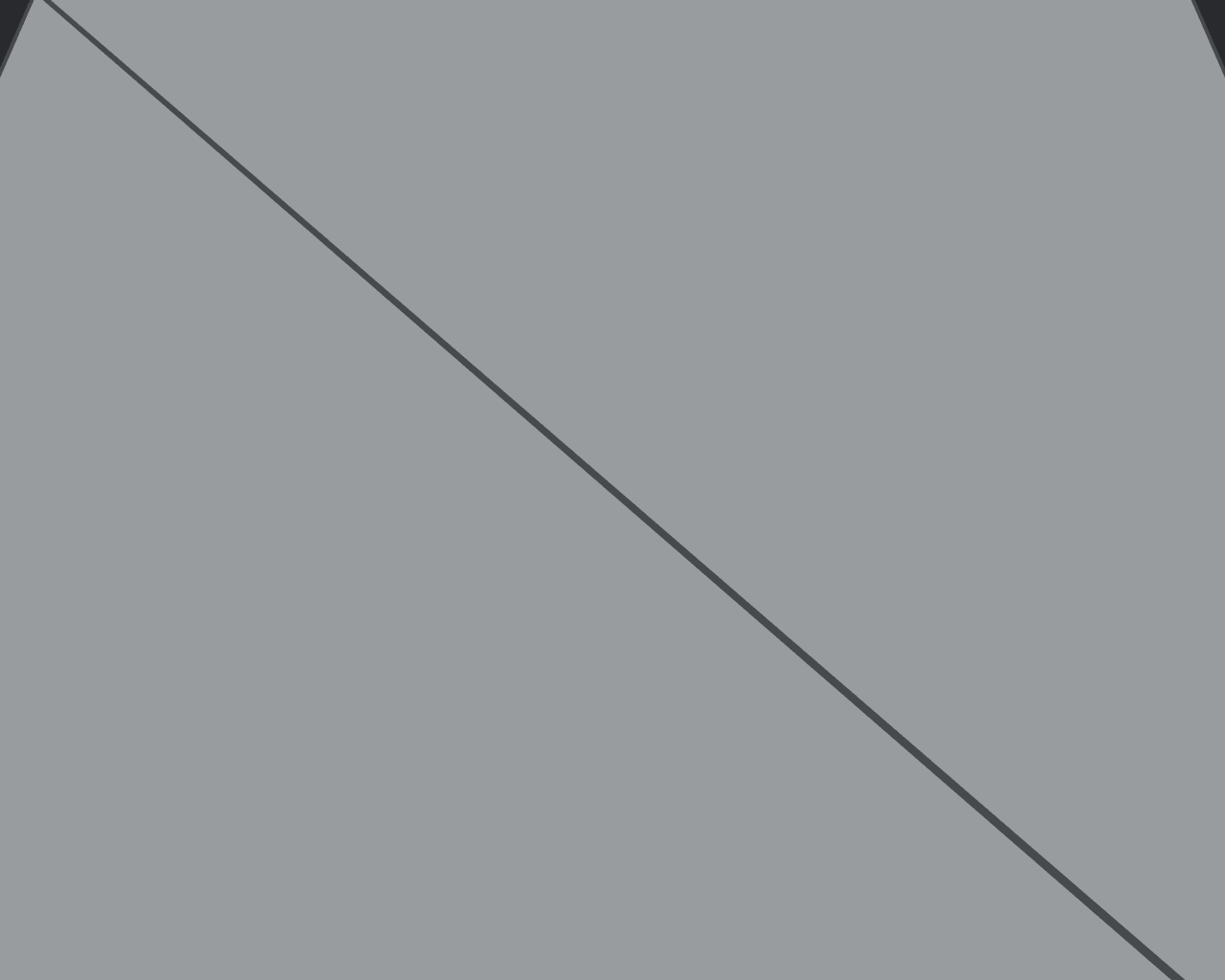}
   {Motherboard}}

 \vspace{1.1mm}
 \makebox[\textwidth][c]{%
  \EGRayGalleryTile{.320\textwidth}{.078\textwidth}
   {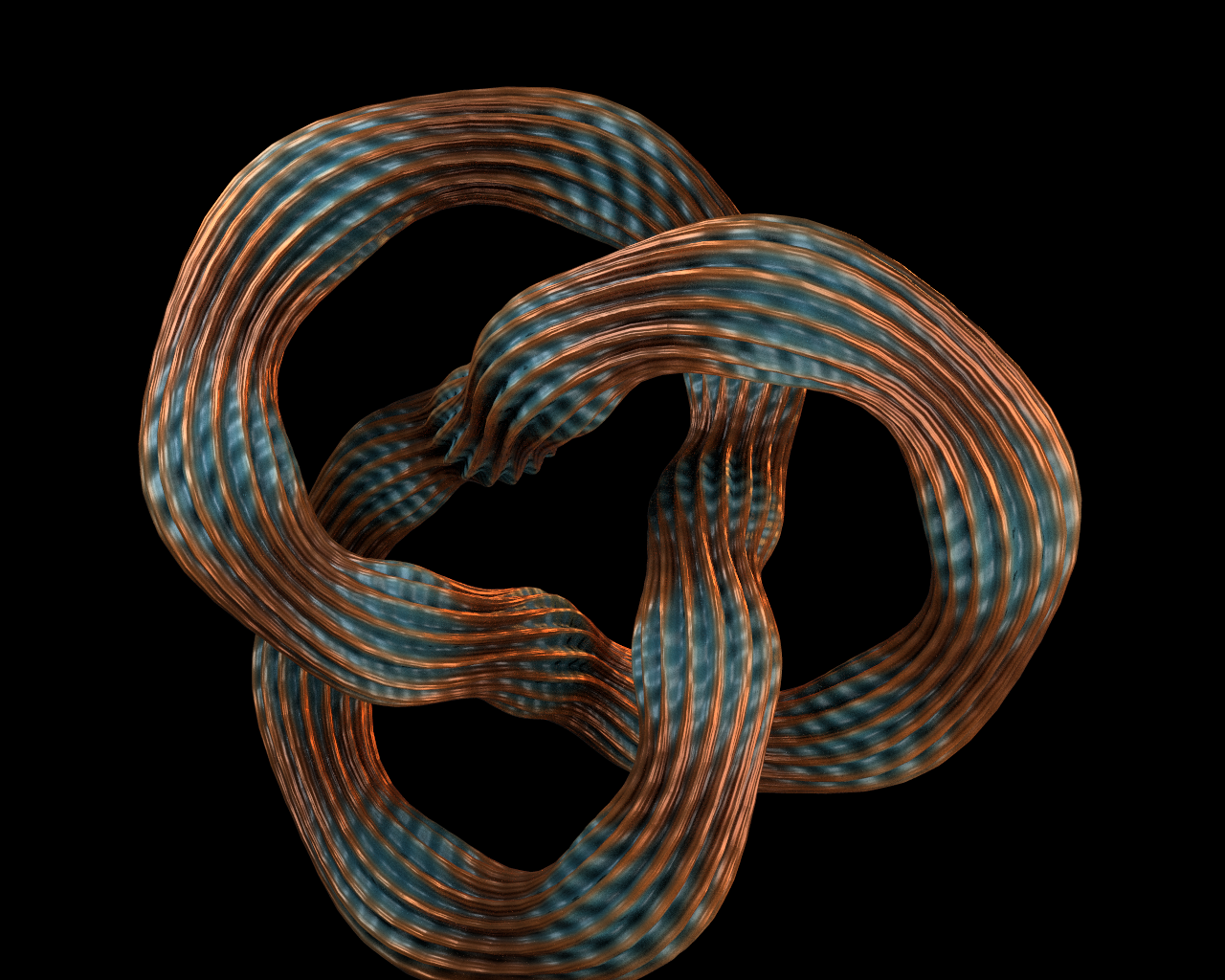}{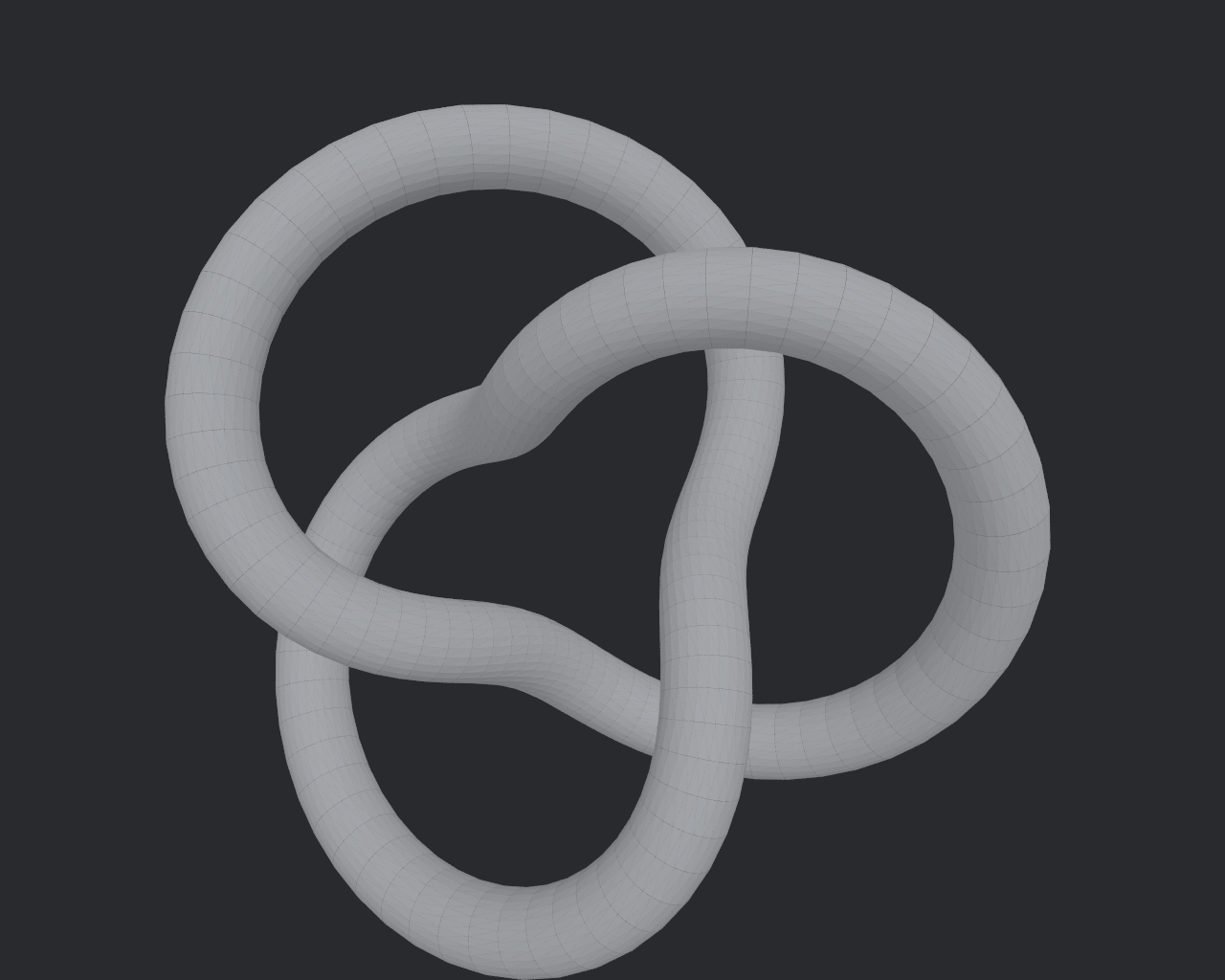}
   {Woven knot}\hfill
  \EGRayGalleryTile{.320\textwidth}{.078\textwidth}
   {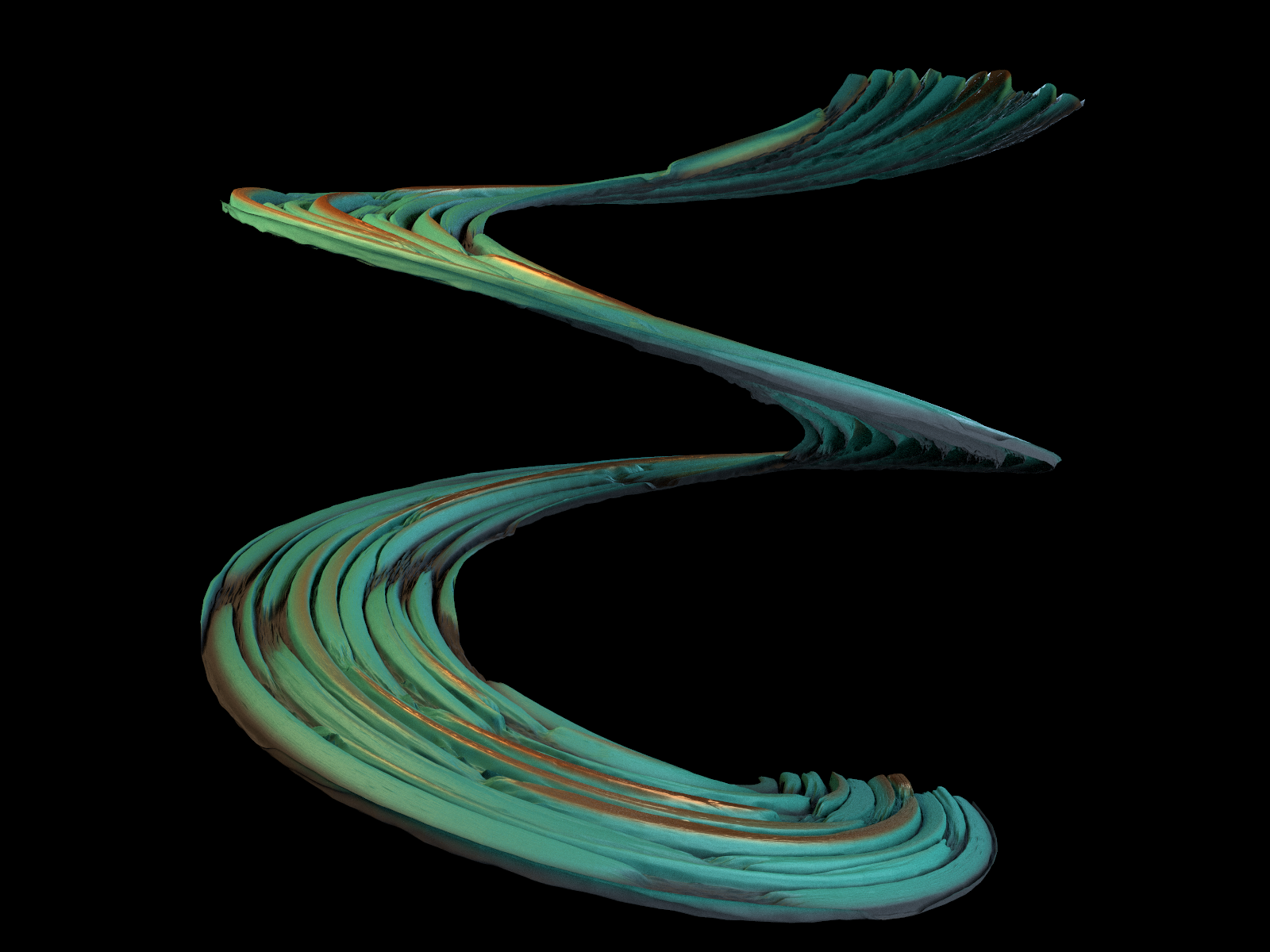}{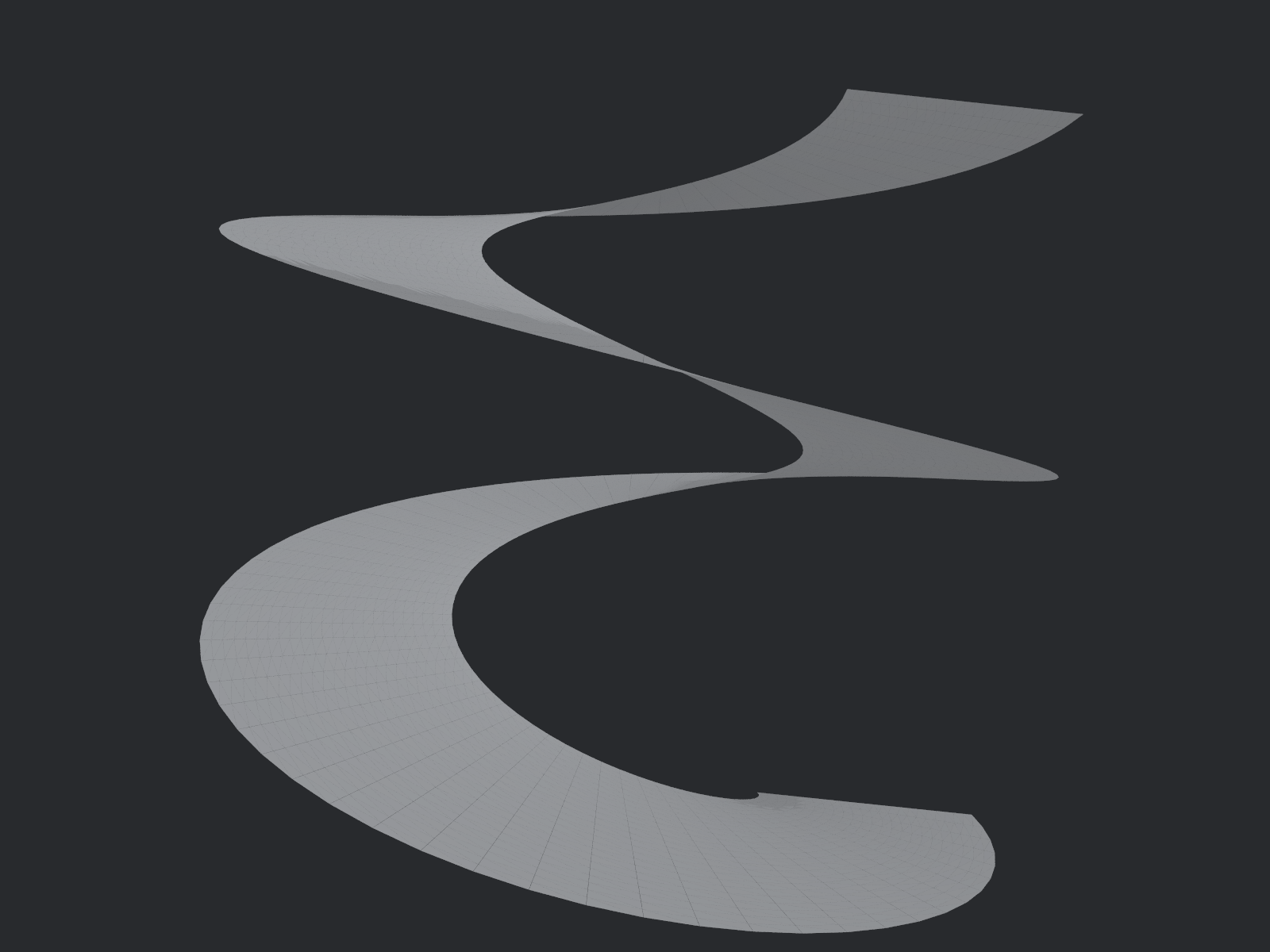}
   {Stone spiral}\hfill
  \EGRayGalleryTile{.320\textwidth}{.078\textwidth}
   {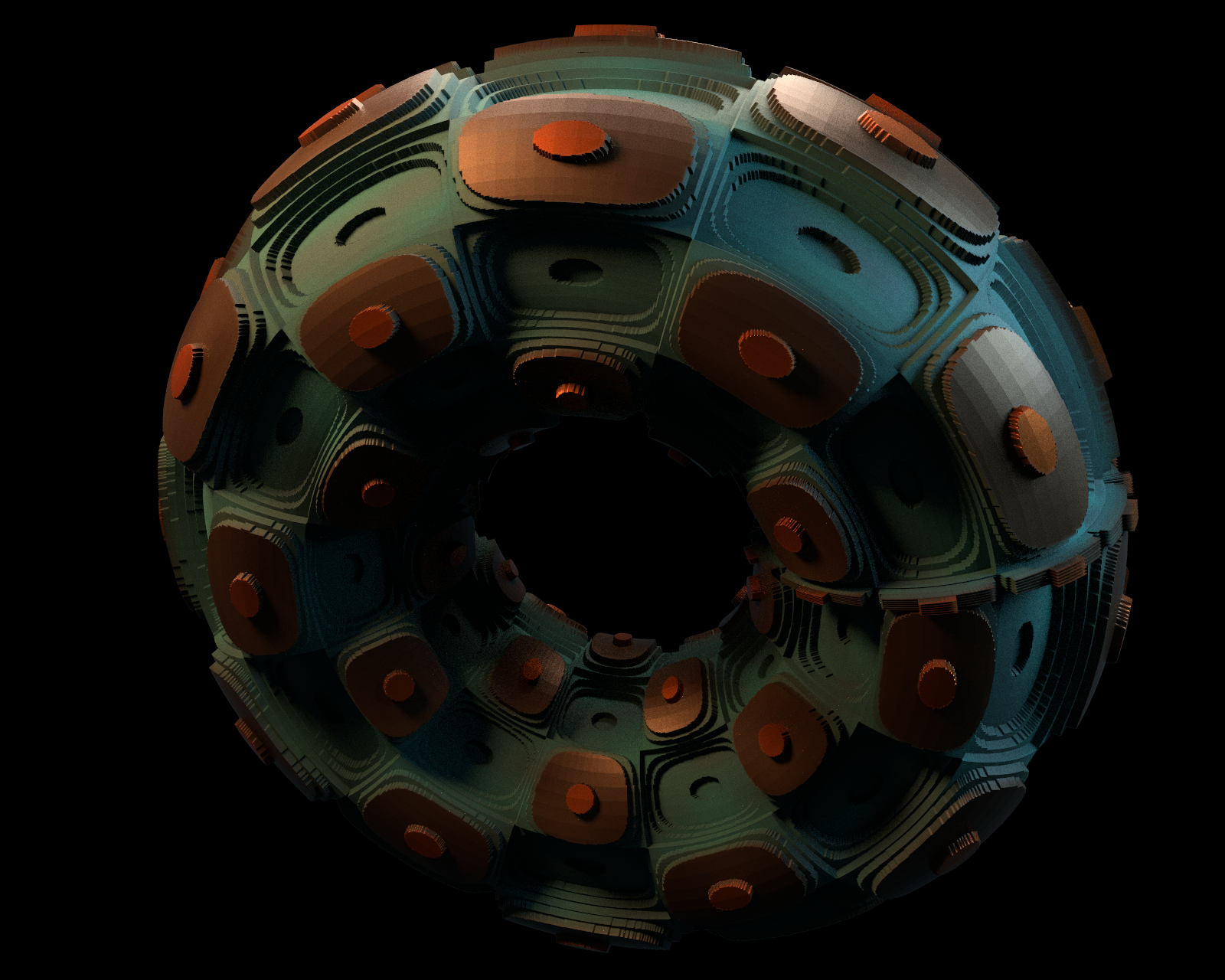}{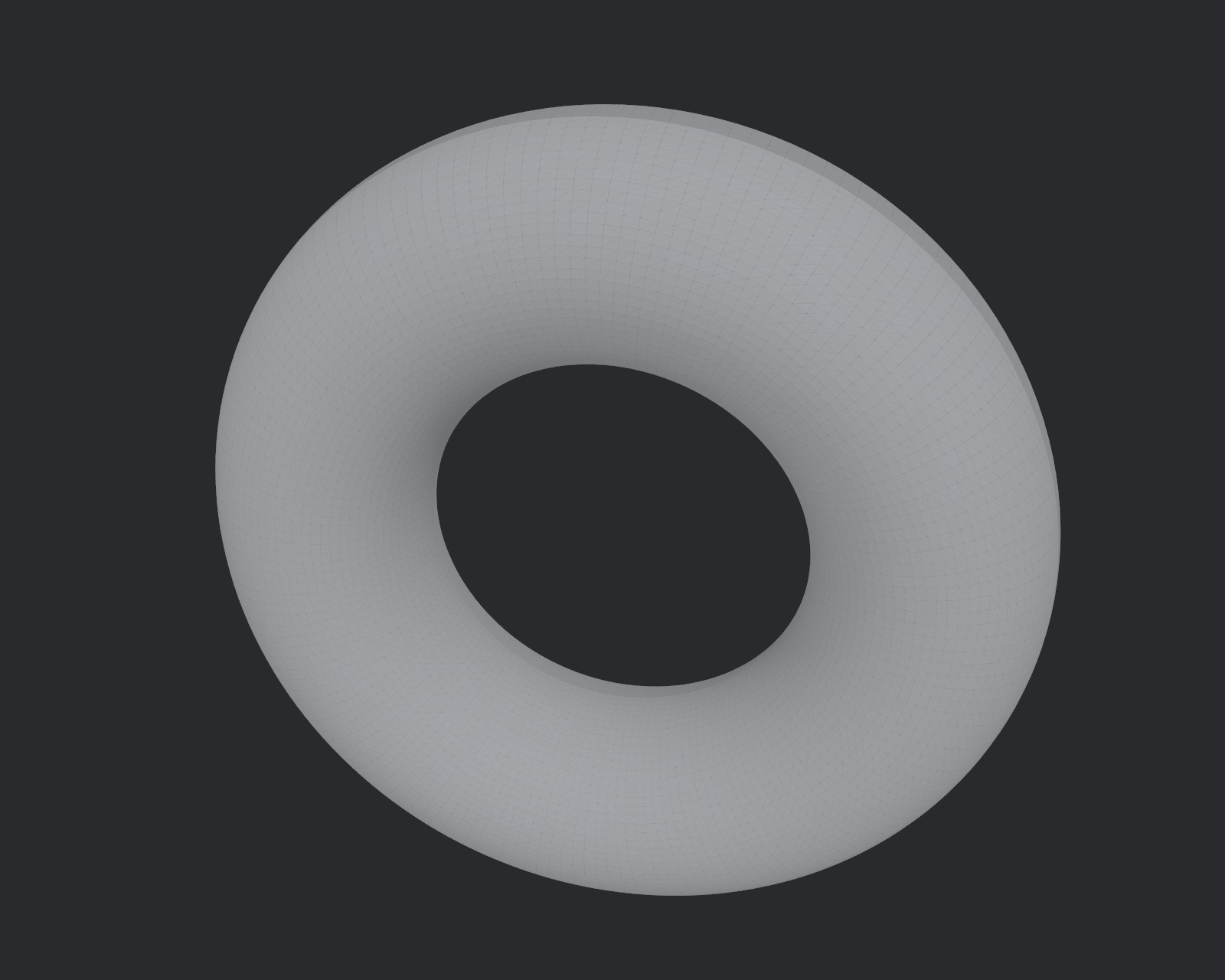}
   {Stepped torus}}
 \caption{Gallery displaced-surface renders across diverse mesh geometry and displacement maps. We show the base meshes in insets under neutral diffuse shading with topology edges. The motherboard uses a two-triangle plane; the other inputs include flat, curved, closed, and open surfaces. }
 \label{fig:ray-gallery}
\end{figure*}

We implement our method in a custom OptiX renderer~\cite{parker2010optix} and compare with unofficial implementations of NRT~\cite{ogaki2023nonlinear} and TFDM~\cite{thonat2021tessellation}, and our GPU adaptation of the released PDM code~\cite{hoetzlein2025pdm}, on an NVIDIA RTX 6000 Ada. PDM$_{0.005}$ uses the released default step and PDM$_{0.4\%}$ uses our displacement-relative interpretation of the paper's setting. All methods share inputs and rendering settings, but differ in surface reconstruction, so the comparisons do not imply equal accuracy. The supplementary material gives the configurations.

Our method accelerates ray tracing across the test scenes with diverse geometry and displacement maps (\autoref{fig:ray-gallery}, \autoref{tab:ray-results}). The speedups mainly come from our traversal’s replacement of repeated nonlinear ray--box tests with incremental grid steps and linear height-band tests. PDM$_{0.005}$ is faster on four scenes, but shows visible artifacts, particularly around thin features. Although our method uses a linear segment approximation, it maintains high accuracy with no visible discrepancies. Across nine scenes, it passes our quantitative checks for hit positions, normals, and image differences against NRT, with LPIPS values of at most 0.004. Differences between PDM's two settings, reported separately in the table, measure sensitivity to step size rather than error against a reference. We also evaluate on a broader dataset containing \rtCases{} mesh--displacement map pairs and obtain a \rtFrameSpeedup$\times$ average speedup over NRT.

\paragraph{Memory.}
Stopping the stored hierarchy at $4\times4$-texel blocks reduces hierarchy storage while retaining texel-level intersection tests. Across the broader testset (228 pairs), our geometry and acceleration data occupy a geometric mean 10.3\,MB, including the 8-bit height map, retained hierarchy levels, per-triangle records, and prism BVH. NRT and TFDM retain full min--max pyramids and use 32.3\,MB, while PDM uses 4.8\,MB without a texture hierarchy. Pre-tessellation at one micro-quad per texel is estimated to require 544\,MB including its BVH, about $53\times$ our footprint.

\paragraph{Compatibility with ray tracing.}
Our first-order representation supports queries beyond ray intersection while retaining efficient ray traversal. With traversal and leaf tests fixed, it performs similarly to scalar bounds on the tested curved base meshes and improves rendering speed by about 14\% on two planar-base scenes. The overall renderer speedups also include traversal and leaf-intersection changes. These results show that supporting broader queries need not sacrifice ray-tracing performance on the tested scenes.

\section{Specular next-event estimation}
\label{sec:specular-query}

So far, we have used our hierarchy for direct ray tracing of displaced surfaces, an application
shared by previous techniques. Although direct ray intersection is important, it is by no means
the only application. Many rendering and graphics tasks require more challenging types of
geometric queries on displaced surfaces. These queries have largely been overlooked and still
lack efficient, tessellation-free solutions.

One important example arises in rendering caustics cast by (displaced) specular surfaces. Caustics are light transport 
effects frequently encountered in realistic scenes that form intricate, visually striking patterns. They are challenging to 
render because they are contributed by specular paths (\autoref{fig:query-overview}), a subset of light paths that are hard 
or even impossible to sample by standard path tracing~\cite{veach1997robust}. In particular, these specular light paths lie on a low-dimensional 
manifold of the path space~\cite{jakob2012manifold} because their construction is governed by geometric constraints imposed by optical laws. 
This geometric nature of the problem motivates us to utilize and enhance our hierarchy to improve the sampling efficiency.

\begin{figure}[t]
  \centering
  \includegraphics[width=0.8\columnwidth]{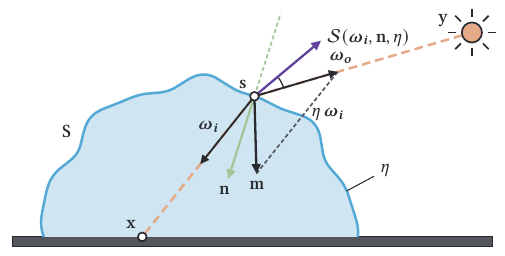}
  \caption{The 3-vertex specular setup in a 2D cross-section. A candidate vertex $\mathbf{s}$ on the displaced
  surface $\mathbf{S}$ (normal $\mathbf{n}$ pointing inwards) connects the shading point $\mathbf{x}$ to the point light $\mathbf{y}$. This path is 
  not valid as the scattered direction $\mathcal{S}(\boldsymbol{\omega_i}, \mathbf{n}, \eta)$ deviates from $\boldsymbol{\omega_o}$.}
  \label{fig:specular-setup}
\end{figure}

Formally, we consider a 3-vertex setup as illustrated in \autoref{fig:specular-setup}. Let $\mathbf{x}$ be a shading point, $\mathbf{y}$ be a 
point light source, and $\mathbf{s}$ on a displaced surface $\mathbf{S}$ that acts as a smooth reflecting or refracting 
interface. In order for the path $\overline{\mathbf{x}\mathbf{s}\mathbf{y}}$ to be physically valid, it needs to either 
obey the law of reflection and Snell's law. Following the notation of Zeltner et al.~\shortcite{zeltner2020specular}, we define
a generalized scattering function to unify both cases. Given $\mathbf{\omega_i}$, the unit direction from $\mathbf{s}$ to $\mathbf{x}$,
The scattered direction is
\begin{align} \label{eq:specular_scattering_function}
\mathcal{S}(\boldsymbol{\omega_i}, \mathbf{n}, \eta) &=
\begin{cases}
  \mathcal{S}^{\mathrm{r}}(\boldsymbol{\omega_i}, \mathbf{n}), & \text{if reflection,} \\
  \mathcal{S}^{\mathrm{t}}(\boldsymbol{\omega_i}, \mathbf{n}, \eta), & \text{if transmission,}
\end{cases}
\qquad \text{where} \\ \nonumber
\mathcal{S}^{\mathrm{r}}(\boldsymbol{\omega_i}, \mathbf{n}) &= 2 (\boldsymbol{\omega_i} \cdot \mathbf{n}) \mathbf{n} - \boldsymbol{\omega_i}, \\ \nonumber
\mathcal{S}^{\mathrm{t}}(\boldsymbol{\omega_i}, \mathbf{n}, \eta) &= -\eta \, \bigl(\boldsymbol{\omega_i} - (\boldsymbol{\omega_i} \cdot \mathbf{n}) \mathbf{n}\bigr) - \mathbf{n} \sqrt{1 - \eta^2 \bigl(1 - \| \boldsymbol{\omega_i} \cdot \mathbf{n} \|^2\bigr)}, \nonumber
\end{align}
where $\mathbf{n}$ is the surface normal at $\mathbf{s}$ and $\eta$ is the relative index of refraction.
We can measure how much an arbitrary path deviates from satisfying \autoref{eq:specular_scattering_function} by the 
angle difference objective~\cite{zeltner2020specular}
\begin{equation} \label{eq:angle_difference_objective}
    \mathbf{C} =
\begin{pmatrix}
  \theta\bigl(\mathcal{S}(\mathbf{\omega_i}, \mathbf{n}, \eta)\bigr) - \theta(\mathbf{\omega_o}) \\
  \phi\bigl(\mathcal{S}(\mathbf{\omega_i}, \mathbf{n}, \eta)\bigr) - \phi(\mathbf{\omega_o}),
\end{pmatrix}
\end{equation}
where $\mathbf{\omega}_o$ is the unit direction from $\mathbf{s}$ to $\mathbf{y}$, $\theta = \arccos(\mathbf{\omega}_z)$, 
and $\phi = \arctan(\mathbf{\omega}_y, \mathbf{\omega}_x)$. With \autoref{eq:angle_difference_objective}, the caustic 
rendering problem can be transformed to a root finding problem to find admissible paths on $\mathbf{S}$ that satisfy 
$\mathbf{C} = 0$.

A common approach to this root-finding problem is Newton's method~\cite{hanika2015manifold,zeltner2020specular,wang2020path}.
However, its convergence depends on an initialization close to the solution. The problem is further complicated by the
geometry of displaced surfaces, which can admit many solutions, whereas each Newton solve finds at most one. 
Our key idea is to extend our texture-space hierarchy to quickly prune the many regions containing no solution.
The remaining regions then serve as a complete seed set for the Newton solver, allowing it to find \emph{almost all}
\footnote{See \autoref{subsec:hierachical_specular_nee_query}.}
solutions.

\subsection{Conservative first-order pruning of specular paths} \label{subsec:first_order_pruning}

\begin{figure}[t]
  \centering
  \includegraphics[width=\columnwidth]{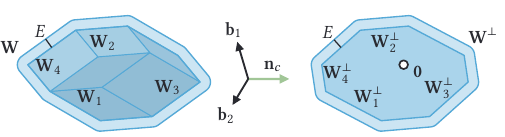}
  \caption{Conservative node test. 
  \textbf{Left}: over a node, all possible $\mathbf{W}$
  lie in a 3D convex polytope with edges along $\mathbf{W}_1, \dots, \mathbf{W}_4$,
  widened by a ball of radius $E$.
  \textbf{Right}: projecting along $\mathbf{n}_c$ onto $\mathbf{b}_1, \mathbf{b}_2$ gives
  $\mathbf{W}^{\perp}$, a 2D polygon with at most 4 pairs of parallel edges along
  $\mathbf{W}^{\perp}_1, \dots, \mathbf{W}^{\perp}_4$, widened by a disc of radius $E$.
  Here the node cannot be pruned because it contains the origin.
  }
  \label{fig:specular-pruning}
\end{figure}

The core of our approach is a conservative test of whether any point on the portion of the displaced surface 
contained within a node can form a valid connection. If this test fails, the node and its entire subtree can be safely 
pruned. Given $\mathbf{\omega_i}$ and $\mathbf{\omega_o}$, a specular vertex $\mathbf{s}$ is valid iff its normal 
$\mathbf{n}$ is parallel to the \emph{generalized half vector} $\mathbf{m} = \eta \omega_i + \omega_o$
\begin{equation} \label{eq:cross_product_test}
    \mathbf{n} \times \mathbf{m} = \mathbf{0}.
\end{equation}
Note that both vectors do not need to be normalized for this test. Here $\mathbf{n}$ is the shading normal of the
displaced surface, formed from the two tangents of \autoref{eq:surface-derivatives} with the slope field used by the
renderer; the supplementary material states the exact model. Our strategy is to conservatively bound the range of
$\mathbf{n} \times \mathbf{m}$ and test whether it includes the zero vector.
To achieve this, we express any vector quantity $\mathbf{z}$ that varies within the footprint of a node in the following 
linear form, relative to its value at center:
\begin{equation*}
    \mathbf{z}(\Delta u, \Delta v) = \mathbf{z}_c + \Delta u \, \partial_u \mathbf{z}_{c} + \Delta v \, \partial_v \mathbf{z}_c + \dots,
\end{equation*}
where $(\Delta u, \Delta v)$ is the offset of UV coordinates, $\mathbf{z}_c$ is the value at the node center, $\partial_u \mathbf{z}_{c}, \partial_v \mathbf{z}_{c}$ are the partial 
derivatives evaluated at the center, and ``$\dots$'' includes the remainder error terms. 
The supplementary material gives the full derivation from \autoref{eq:linear_form_n} to \autoref{eq:proj_linear_form}.
We can now express $\mathbf{n}$ as
\begin{equation} \label{eq:linear_form_n}
     \mathbf{n}(\Delta u, \Delta v) = \mathbf{n}_c+ \Delta u\, \partial_u \mathbf{n}_{c} + \Delta v \, \partial_v \mathbf{n}_{c}+\xi_u\,\mathbf{e}_u+\xi_v\,\mathbf{e}_v +  \mathbf{e}_\mathbf{n},
\end{equation}
where $\mathbf{e}_u \propto r_{n,u}$ and $\mathbf{e}_v \propto r_{n,v}$ are terms associated with the slope-error bounds $(r_{n,u}, r_{n,v})$,
which bound the deviation of the slope field from a first-order model over the node. 
\textbf{These are computed and stored additionally alongside the hierarchy during its construction.}
Here, $\xi_u, \xi_v \in [-1, 1]$ are two unknown
symbols representing the errors along each axis, and $\mathbf{e}_\mathbf{n}$ encodes all remaining error. Encoding $\mathbf{e}_u$
and $\mathbf{e}_v$ with separate symbols yields a tighter bound, since the slope error along one axis can be
much larger than along the other. Similarly, $\mathbf{m}$ can be expanded into
\begin{equation} \label{eq:linear_form_m}
    \mathbf{m}(\Delta u, \Delta v) = \mathbf{m}_c+ \Delta u\, \partial_u \mathbf{m}_{c} + \Delta v \, \partial_v \mathbf{m}_{c} + \mathbf{e}_\mathbf{m},
\end{equation}
but this time without slope-specific error terms. Substituting \autoref{eq:linear_form_n} and \autoref{eq:linear_form_m} 
into \autoref{eq:cross_product_test} yields
\begin{align} \label{eq:linear_form_nxm}
    \mathbf{W} &= \mathbf{n} \times \mathbf{m} \nonumber \\
    &= \mathbf{W}_c + \Delta u\, \mathbf{W}_1 + \Delta v \, \mathbf{W}_2  +  \xi_u\, \mathbf{W}_3 + \xi_v\, \mathbf{W}_4 + \mathbf{e}_\mathbf{W}, \quad \| \mathbf{e}_{\mathbf{W}} \| \leq E.
\end{align}
Geometrically, the set of all possible $\mathbf{W}$ defines a 3D region in space that is a convex polytope widened by a
ball of radius $E$, as illustrated in \autoref{fig:specular-pruning}.
The test of \autoref{eq:cross_product_test} is equivalent to check whether the origin lies in this region.
However, as it is costly to perform this test in 3D, we project $\mathbf{W}$ to 2D along the direction of $\mathbf{n}_c$:
\begin{equation} \label{eq:proj_linear_form}
    \mathbf{W}^{\perp} = \mathbf{W}_c^{\perp} + \Delta u\, \mathbf{W}_1^{\perp} + \Delta v \, \mathbf{W}_2^{\perp}  +  \xi_u\, \mathbf{W}_3^{\perp} + \xi_v\, \mathbf{W}_4^{\perp} + \mathbf{e}_{\mathbf{W}^{\perp}},
\end{equation}
where $\mathbf{W}^{\perp}_{\ast} = (\mathbf{b}_1 \; \mathbf{b}_2)\trans \, \mathbf{W}_{\ast}$ are the coefficients along
the orthonormal pair $\mathbf{b}_1, \mathbf{b}_2$ of $\mathbf{n}_c$. The 2D test is not as tight as the full 3D test. 
However, it is conservative and much cheaper to evaluate. 
Note that $\| \mathbf{e}_{\mathbf{W}^{\perp}} \| \leq E$ still holds.
The choice of $\mathbf{n}_c$ is important to help $\mathbf{W}^{\perp}$ preserve projected area instead of becoming
degenerated. This is because $\mathbf{W}$ is perpendicular to $\mathbf{n}$ and $\mathbf{n}$
stays close to $\mathbf{n}_c$ within the extent of a node. $\mathbf{W}^{\perp}$ now defines a 2D polygon enlarged by a
disc of radius $E$, and we proceed to check if it contains the 2D origin (\autoref{fig:specular-pruning}).
This is much cheaper as $\mathbf{W}^\perp$ has
at most 4 pairs of parallel edges oriented along $\mathbf{W}_1^\perp$, $\mathbf{W}_2^\perp$, $\mathbf{W}_3^\perp$, and
$\mathbf{W}_4^\perp$. The test is done by using the Separating Axis Theorem (SAT) with all 4
unique edge normals as axes.

\subsection{Hierarchical specular next-event estimation query} \label{subsec:hierachical_specular_nee_query}
With the conservative node-level pruning in \autoref{subsec:first_order_pruning}, we are ready to develop a 
hierarchical query that finds all admissible specular paths. Given a base triangle $T$, we start the traversal from the 
smallest node that covers it. This is a standard top-down traversal. At each level, we pop the parent node and examine 
each of its child node. A child node is either pruned or pushed to the traversal stack. Once we find an accepted 
leaf node, we solve for $\mathbf{C} = \mathbf{0}$ (\autoref{eq:angle_difference_objective}) using Newton's method, starting at 
the texel center. The solver usually converges within a few steps because the initialization is very close, and the constraint 
$\mathbf{C}$ on a small smooth patch is nearly linear. We accumulate the contribution from all leaf node solutions to finish the query. The final specular incident radiance received by $\mathbf{x}$ is 
\begin{equation}
    L_i(\mathbf{x}) = \sum_j f(\mathbf{s}_j) G(\mathbf{x} \leftrightarrow \mathbf{s_j} \leftrightarrow \mathbf{y}) L_e(\mathbf{y}),
\end{equation}
where $\{\mathbf{s}_j\}$ is the set of all solutions, $f$ is the specular BSDF, $G$ encodes the geometry term between vertices, and $L_e$ is emission.

Our query assumes one solution inside one leaf node. Theoretically, this is not always true as there can be 2 solutions 
when $|\nabla \mathbf{C}|$ changes sign within the texel. This happens when a texel sits exactly on where a caustic ``folds'' onto itself.
Our leaf solver will only find one solution, causing a slight energy loss. However, we note that this case is rare and 
becomes less likely as the texel size shrinks. As will be shown next, we did not observe any noticeable darkening or 
artifact in practice.

\subsection{Results} \label{subsec:specular_results}
We implemented our specular next-event estimation query in a CPU renderer. 
We compare our method with Specular Manifold Sampling (SMS)~\cite{zeltner2020specular}, which to our 
knowledge is the only modern specular path sampling method that supports arbitrary types of surfaces, including 
tessellation-free displaced surfaces. Both methods \textbf{run on CPU and use a shared ray tracing backend based on our 
intersection query} (\autoref{sec:ray-query}) for parity. We refrain from comparing with methods
~\cite{wang2020path, fan2025bernstein} that require building data structures on top of tessellated 
triangle meshes and alternative solving techniques~\cite{fan2024specular} that are orthogonal to our leaf solver.
Comparisons and measurements were performed on an AMD Ryzen 9 9950X3D CPU with 32 GB of main memory.

\autoref{fig:teaser} showcases how our method enables tessellation-free displaced surfaces as caustic emitters to fulfill 
the artistic direction of the scene.
\autoref{fig:specular_nee} includes equal-time caustic rendering comparisons between our method and SMS (both unbiased 
and biased variants). The tested scenes feature both reflective and refractive caustics, all cast by displaced surfaces 
from point lights. Our method demonstrates significantly faster convergence as the caustics are effectively 
noise-free. The noise only comes from scattering from diffuse surfaces. Without good initialization, SMS struggles to 
find solution on complex displaced surfaces, resulting in either excessive noise (unbiased), or lost energy (biased).

\definecolor{SpecNeeOrange}{HTML}{FF0280}
\definecolor{SpecNeeSkyBlue}{HTML}{00A7E1}
\newlength{\SpecNeeGap}
\newlength{\SpecNeeInsetW}
\newlength{\SpecNeeStackH}
\newlength{\SpecNeeFrameW}
\newcommand{\SpecNeeScale}{0.9}
\newcommand{\SpecNeeInset}[2]{%
  \begin{tikzpicture}[baseline=(img.south)]
    \node[inner sep=0pt, outer sep=0pt] (img)
      {\includegraphics[width=\SpecNeeInsetW]{figures/specular_nee/#2}};
    \draw[draw=#1, line width=\SpecNeeFrameW]
      ([shift={(.5\SpecNeeFrameW,.5\SpecNeeFrameW)}]img.south west) rectangle
      ([shift={(-.5\SpecNeeFrameW,-.5\SpecNeeFrameW)}]img.north east);
  \end{tikzpicture}%
}
\newcommand{\SpecNeeStack}[2]{%
  \vbox{\offinterlineskip
    \hbox{\SpecNeeInset{SpecNeeOrange}{#1}}%
    \vskip\SpecNeeGap
    \hbox{\SpecNeeInset{SpecNeeSkyBlue}{#2}}}%
}

\newcommand{\SpecNeeDispFrac}{0.3}
\newcommand{\SpecNeeMain}[6]{%
  \begin{tikzpicture}[baseline=(img.south)]
    \node[inner sep=0pt, outer sep=0pt, anchor=south west] (img) at (0,0)
      {\includegraphics[height=\SpecNeeStackH]{figures/specular_nee/#1}};
    \begin{scope}[x={(img.south east)}, y={(img.north west)}, shift={(img.south west)}]
      \draw[SpecNeeOrange, line width=\SpecNeeFrameW]
        ({#2/1920},{1-#3/1080}) rectangle ({(#2+100)/1920},{1-(#3+150)/1080});
      \draw[SpecNeeSkyBlue, line width=\SpecNeeFrameW]
        ({#4/1920},{1-#5/1080}) rectangle ({(#4+100)/1920},{1-(#5+150)/1080});
    \end{scope}
    \node[inner sep=0pt, outer sep=0pt, anchor=north west] (disp) at (img.north west)
      {\includegraphics[width=\SpecNeeDispFrac\SpecNeeStackH]{figures/specular_nee/#6}};
    \draw[white, line width=\SpecNeeFrameW, line cap=butt, line join=miter]
      ([shift={(-.5\SpecNeeFrameW,0)}]disp.north east) --
      ([shift={(-.5\SpecNeeFrameW,.5\SpecNeeFrameW)}]disp.south east) --
      ([shift={(0,.5\SpecNeeFrameW)}]disp.south west);
  \end{tikzpicture}%
}

\begin{figure*}[tb]
\setlength{\SpecNeeGap}{2pt}
\setlength{\SpecNeeInsetW}{\dimexpr(9\textwidth-52\SpecNeeGap)/84\relax}
\setlength{\SpecNeeStackH}{\dimexpr3\SpecNeeInsetW+\SpecNeeGap\relax}
\setlength{\SpecNeeFrameW}{1.2pt}
\renewcommand{\arraystretch}{0.5}
 \centering

    \scalebox{\SpecNeeScale}{%
    \begin{tabular}{@{}c*{4}{@{\hspace{\SpecNeeGap}}c}@{}}
    \textsc{\textbf{Displaced Swimming Pool}} &
    \makebox[\SpecNeeInsetW]{\textbf{Reference}} &
    \makebox[\SpecNeeInsetW]{\textbf{Ours}} &
    \makebox[\SpecNeeInsetW]{\textbf{SMS unbiased}} &
    \makebox[\SpecNeeInsetW]{\textbf{SMS biased}} \\[1pt]
    $1920 \times 1080$, $5$ CPU minutes, Ours shown &
    \makebox[\SpecNeeInsetW]{$\approx 3$ hours} &
    \makebox[\SpecNeeInsetW]{$100$ spp} &
    \makebox[\SpecNeeInsetW]{$19$ spp} &
    \makebox[\SpecNeeInsetW]{$4$ spp, $8$ trials} \\ [1pt]
    &
    \makebox[\SpecNeeInsetW]{MSE$\downarrow$, \FLIP$\downarrow$} &
    \makebox[\SpecNeeInsetW]{$\mathbf{2.9\text{\textbf{e-}}4}$, $\mathbf{0.049}$} &
    \makebox[\SpecNeeInsetW]{$1.3\text{e-}2$, 0.135} &
    \makebox[\SpecNeeInsetW]{$6.3\text{e-}3$, 0.154} \\ [1pt]
    \SpecNeeMain{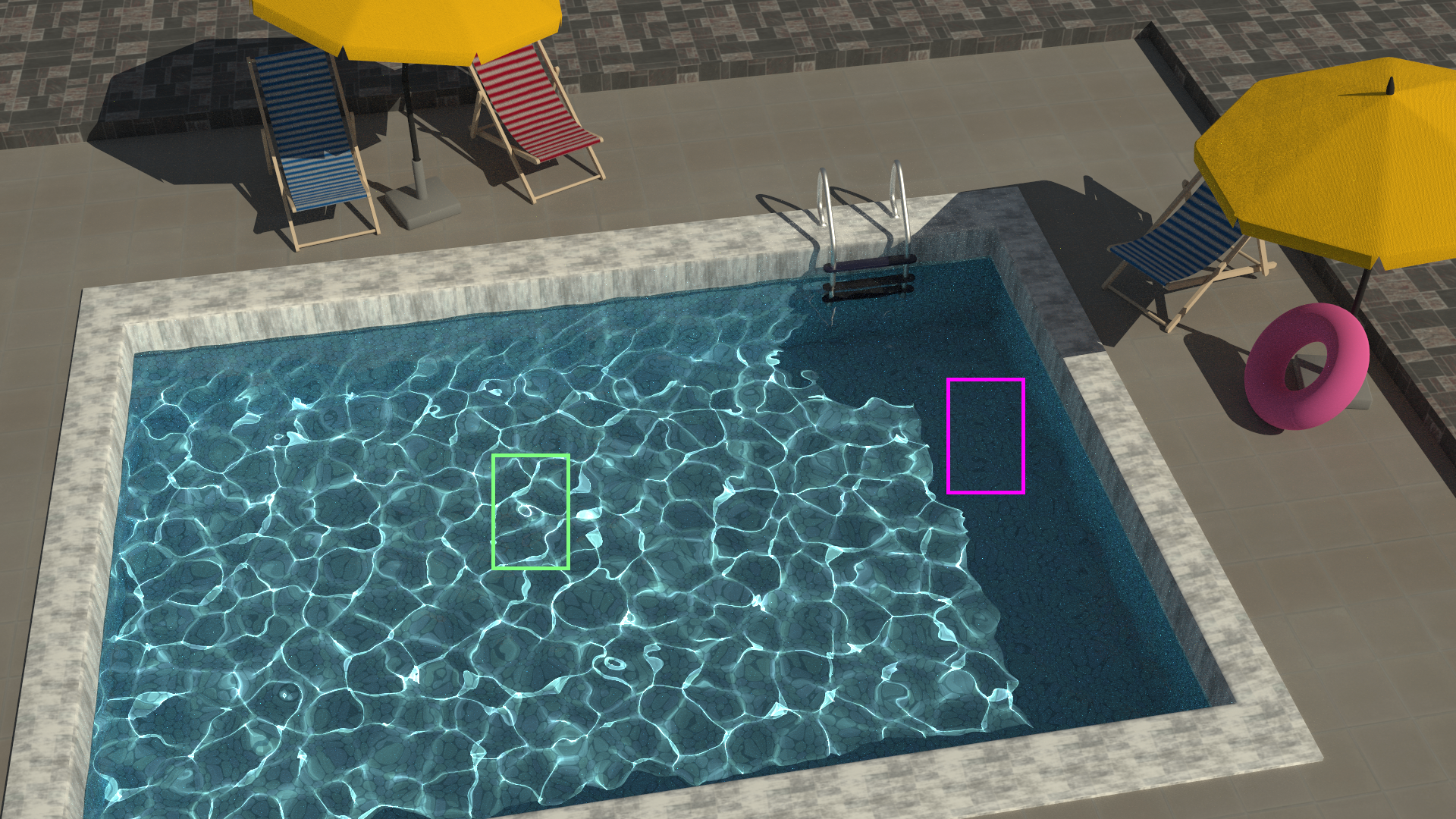}{650}{600}{1250}{500}{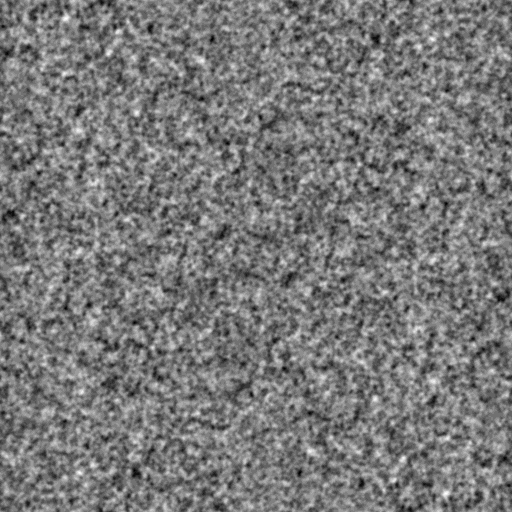} &
    \SpecNeeStack{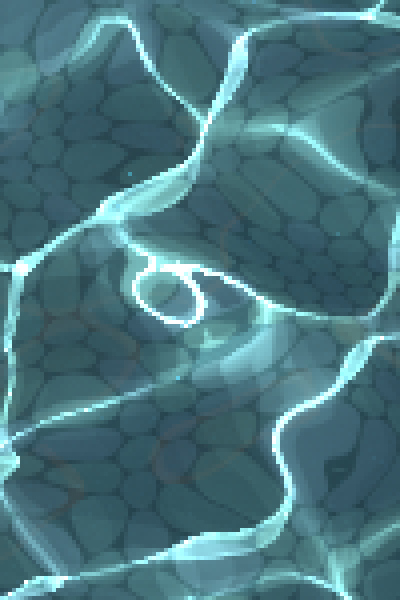}
                 {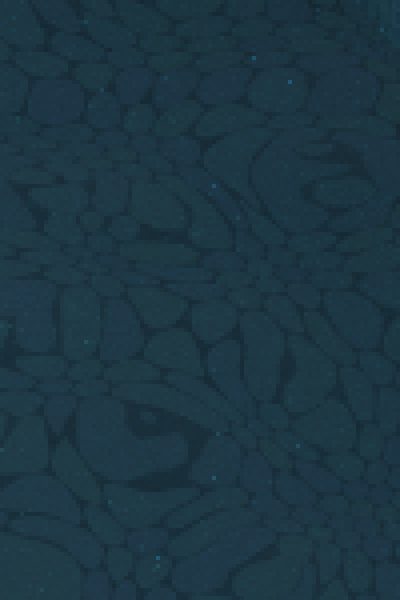} &
    \SpecNeeStack{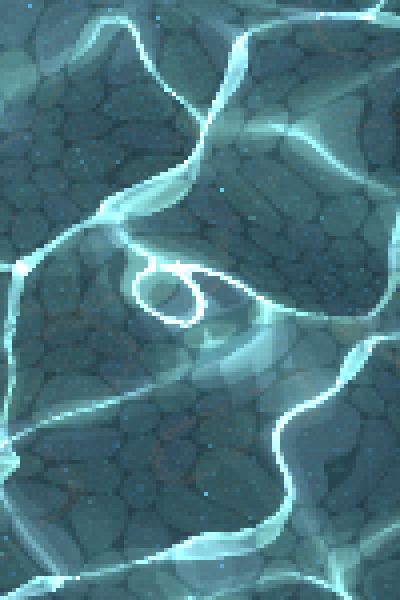}
                 {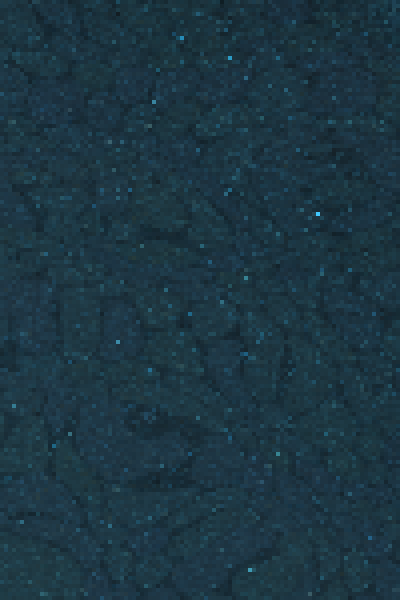} &
    \SpecNeeStack{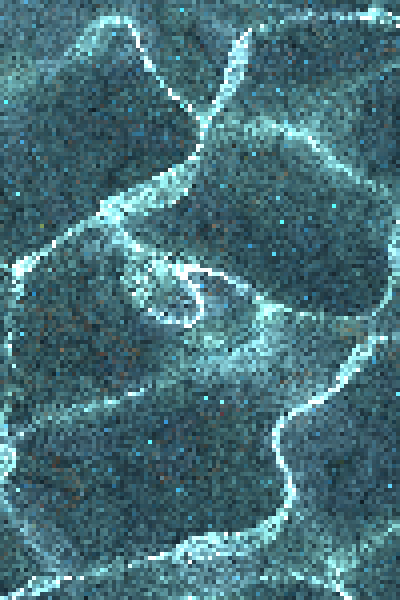}
                 {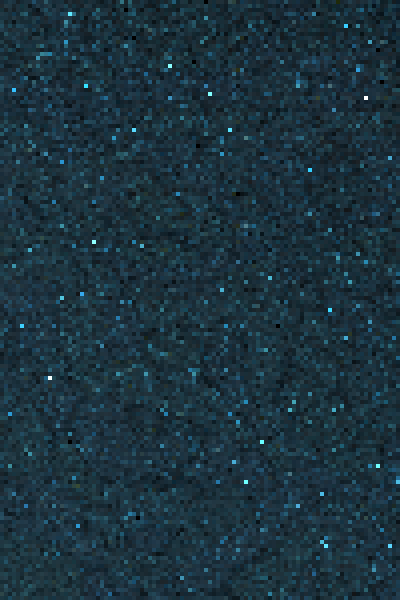} &
    \SpecNeeStack{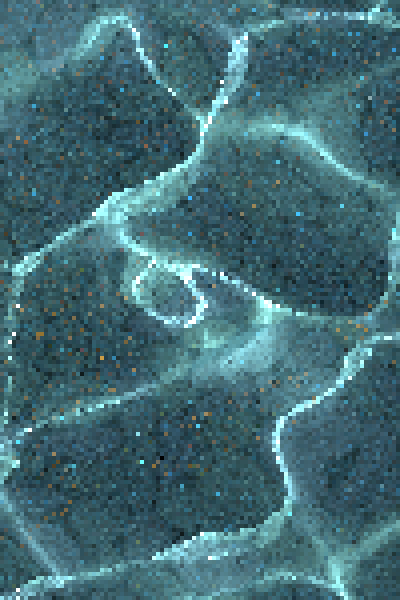}
                 {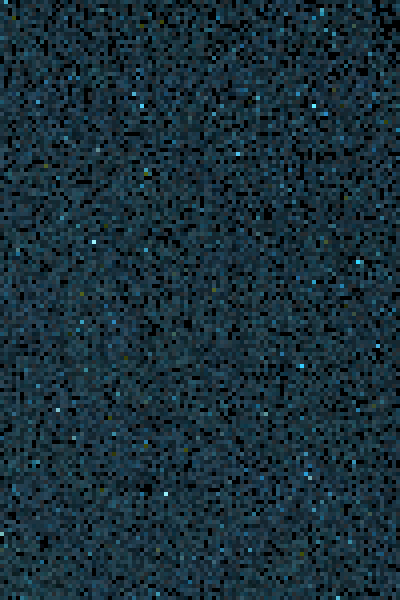} \\ [1pt]

    \textsc{\textbf{Copper Plane}} &
    \makebox[\SpecNeeInsetW]{$\approx 3$ hours} &
    \makebox[\SpecNeeInsetW]{$7$ spp} &
    \makebox[\SpecNeeInsetW]{$2$ spp} &
    \makebox[\SpecNeeInsetW]{$5$ spp, $8$ trials} \\ [1pt]
    $1920 \times 1080$, $5$ CPU minutes, Ours shown &
    &
    \makebox[\SpecNeeInsetW]{$\mathbf{1.3\text{\textbf{e-}}3}$, $\mathbf{0.045}$} &
    \makebox[\SpecNeeInsetW]{$1.8\text{e-}2$, $0.099$} &
    \makebox[\SpecNeeInsetW]{$5.8\text{e-}3$, $0.104$} \\ [1pt]
    \SpecNeeMain{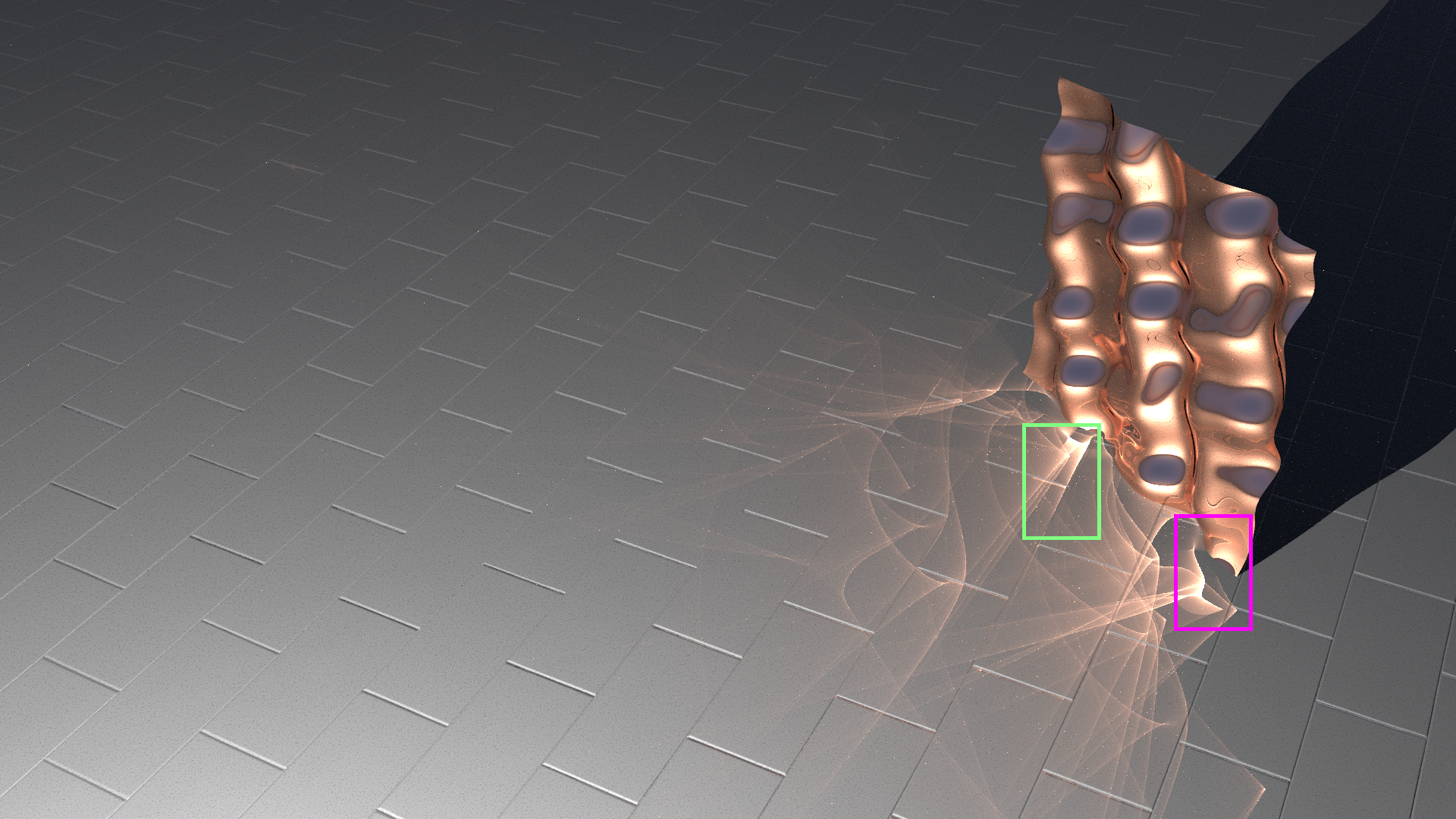}{1350}{560}{1550}{680}{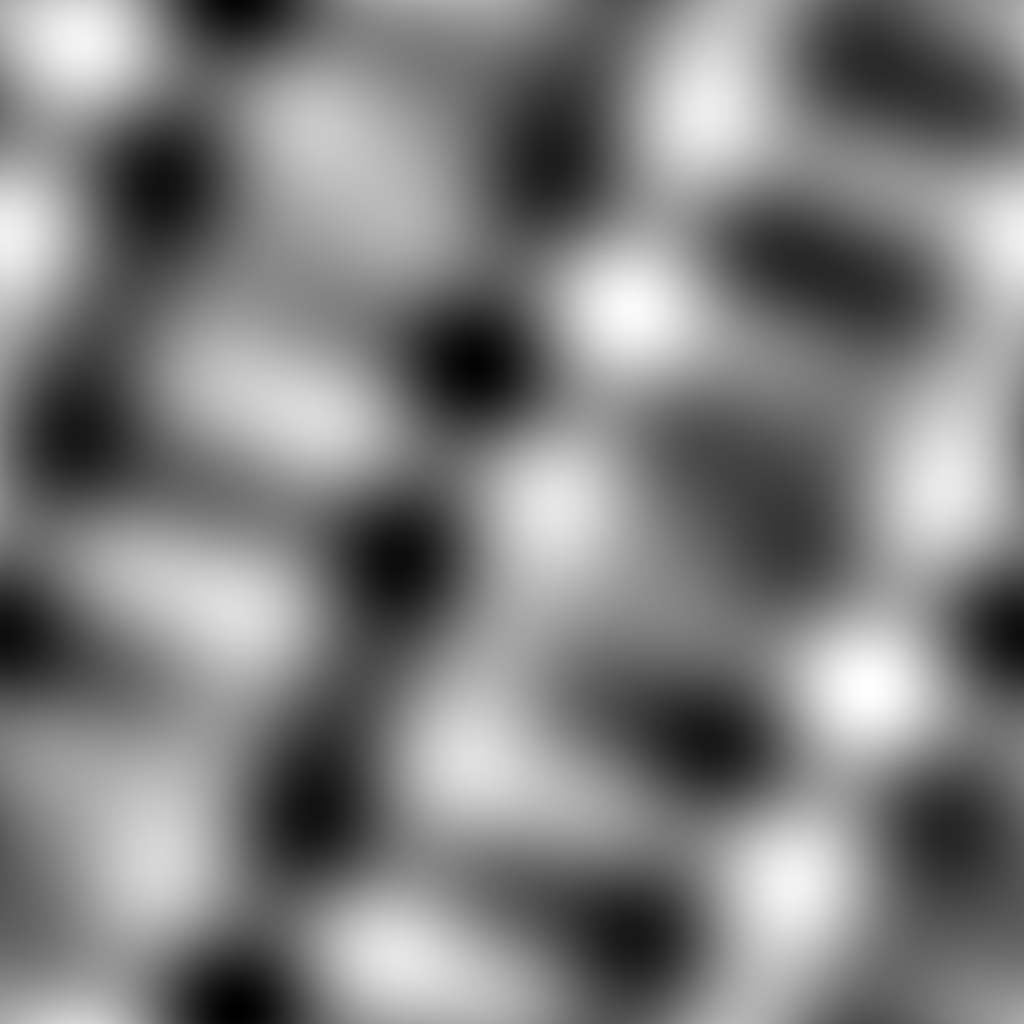} &
    \SpecNeeStack{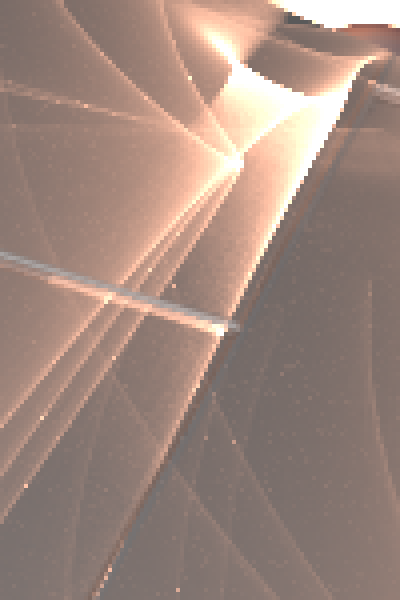}
                 {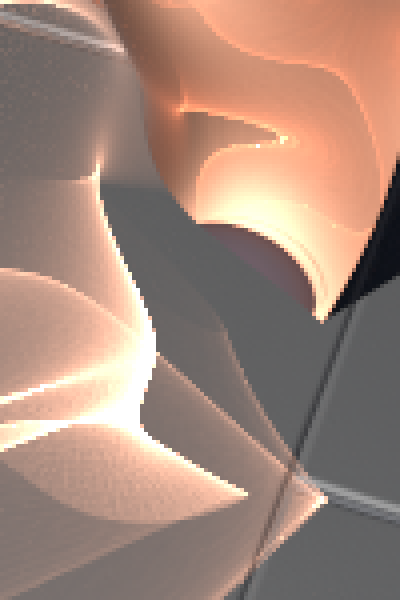} &
    \SpecNeeStack{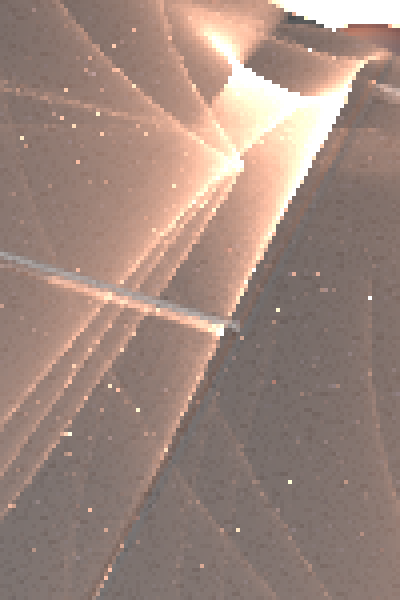}
                 {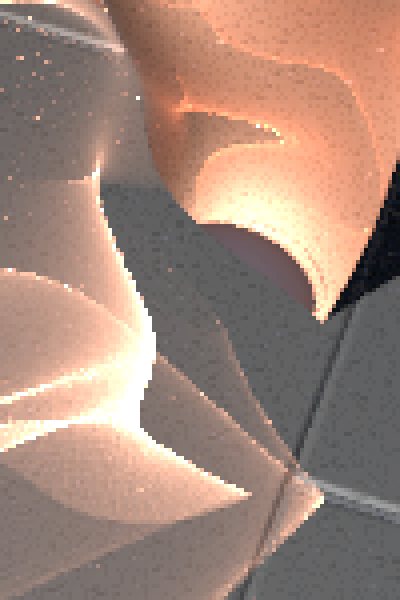} &
    \SpecNeeStack{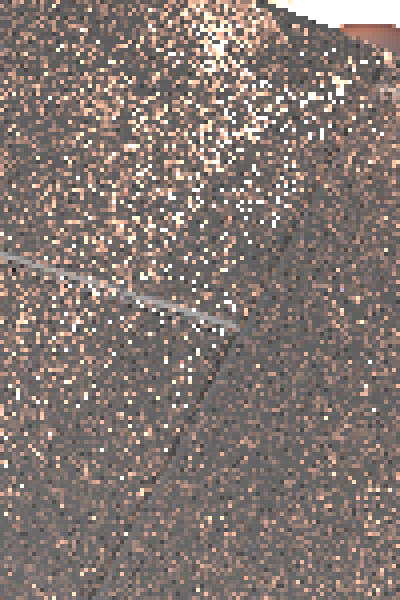}
                 {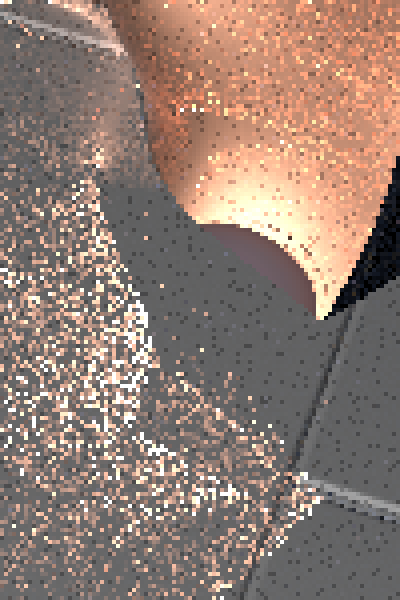} &
    \SpecNeeStack{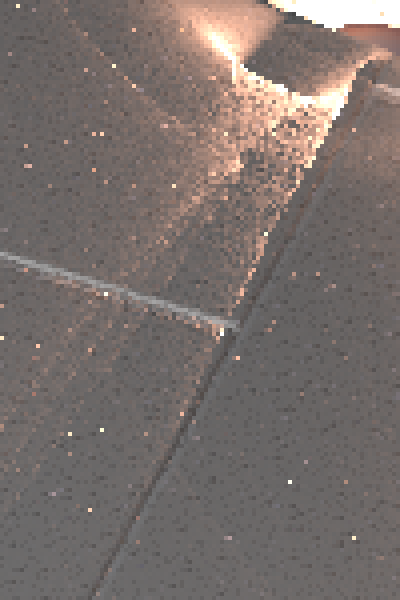}
                 {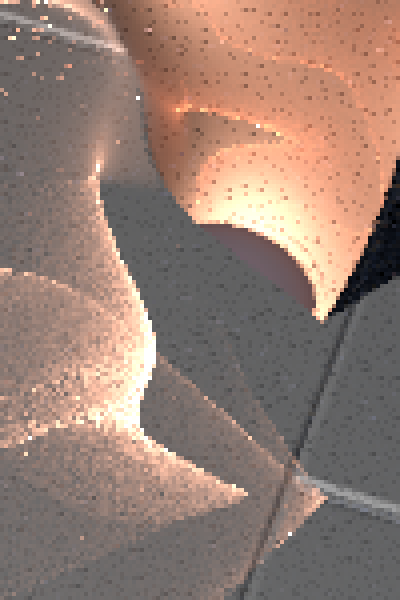} \\ [1pt]
                 
    \textsc{\textbf{Glass Sphere}} &
    \makebox[\SpecNeeInsetW]{$\approx 2.5$ hours} &
    \makebox[\SpecNeeInsetW]{$21$ spp} &
    \makebox[\SpecNeeInsetW]{$8$ spp} &
    \makebox[\SpecNeeInsetW]{$2$ spp, $8$ trials} \\ [1pt]
    $1920 \times 1080$, $5$ CPU minutes, Ours shown &
    &
    \makebox[\SpecNeeInsetW]{$\mathbf{1.6\text{\textbf{e-}}5}$, $\mathbf{0.022}$} &
    \makebox[\SpecNeeInsetW]{$5.8\text{e-}4$, $0.060$} &
    \makebox[\SpecNeeInsetW]{$2.7\text{e-}4$, $0.077$} \\ [1pt]
    \SpecNeeMain{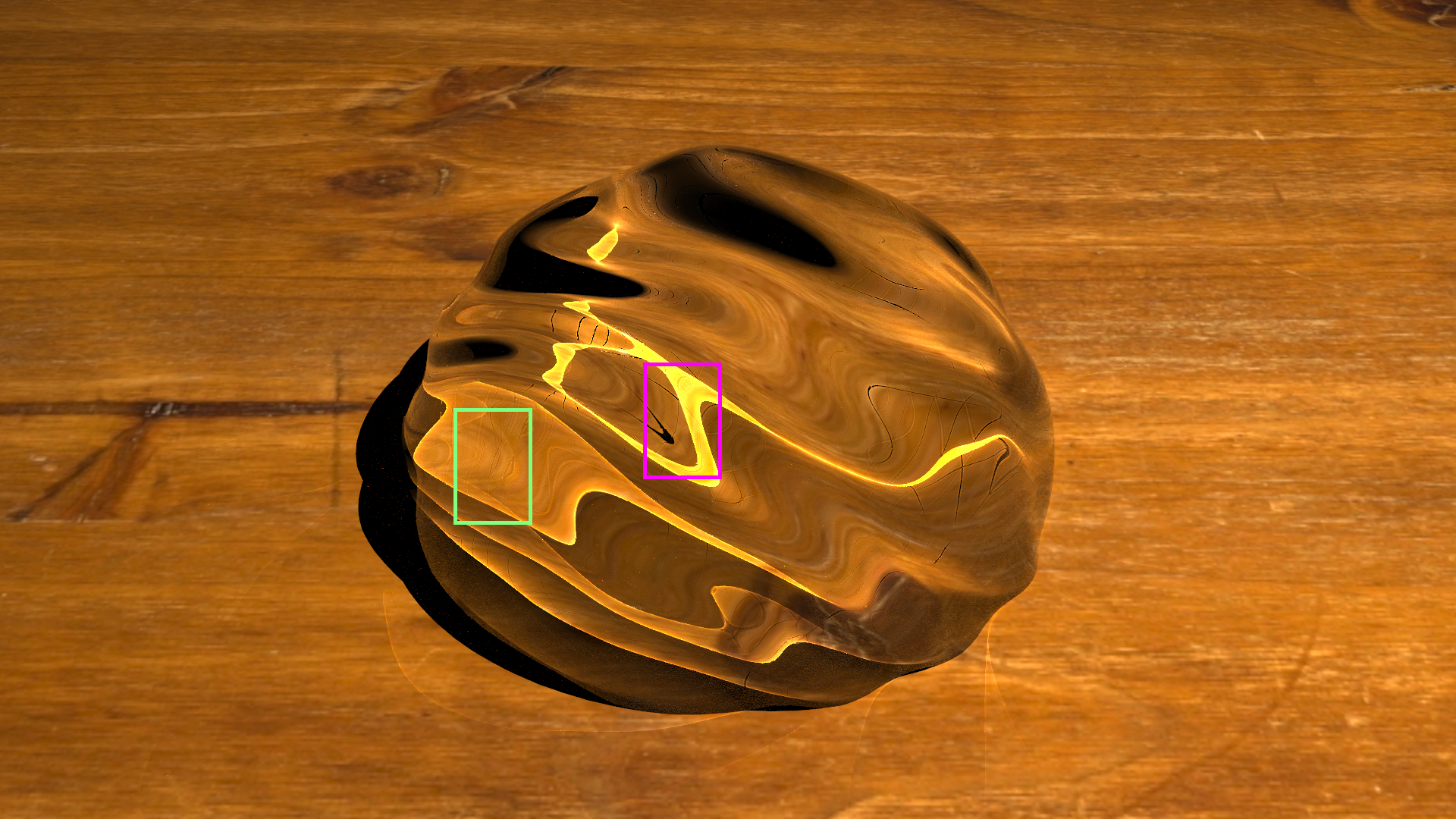}{600}{540}{850}{480}{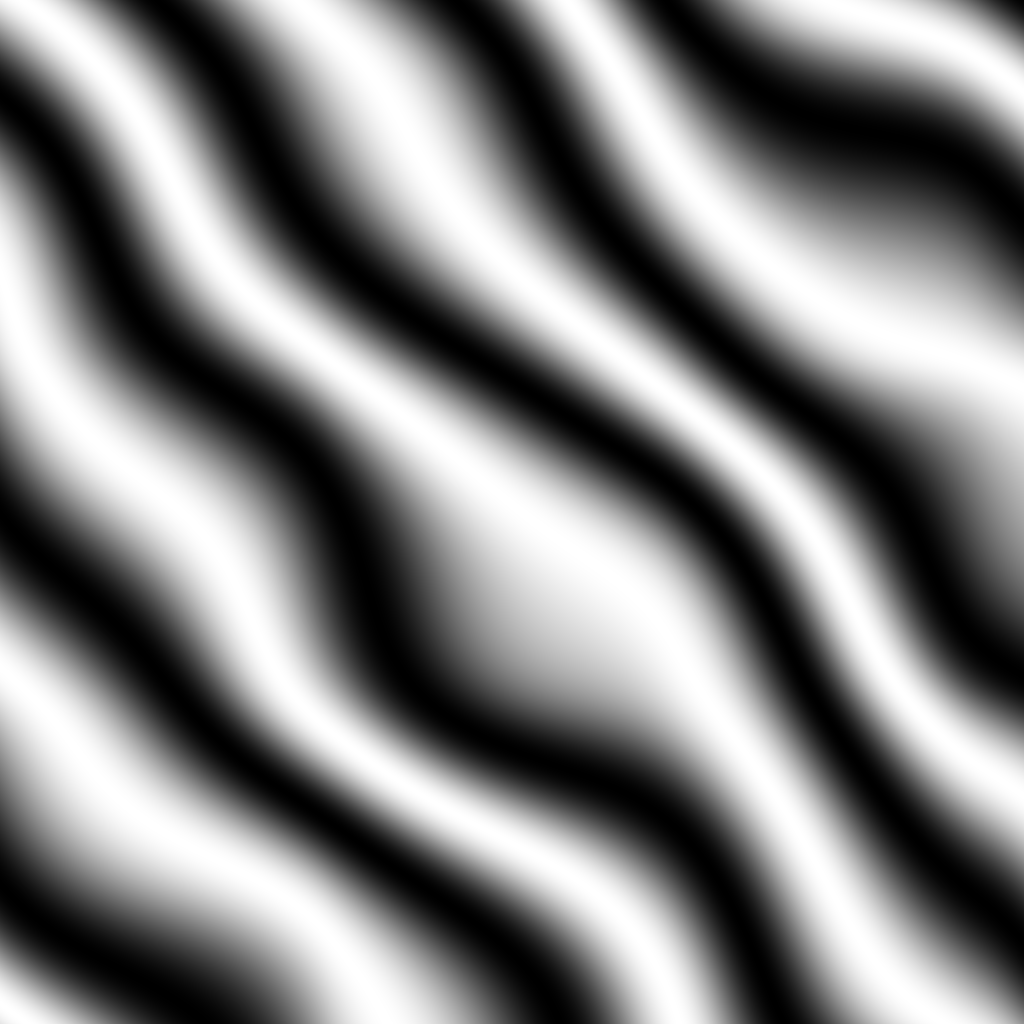} &
    \SpecNeeStack{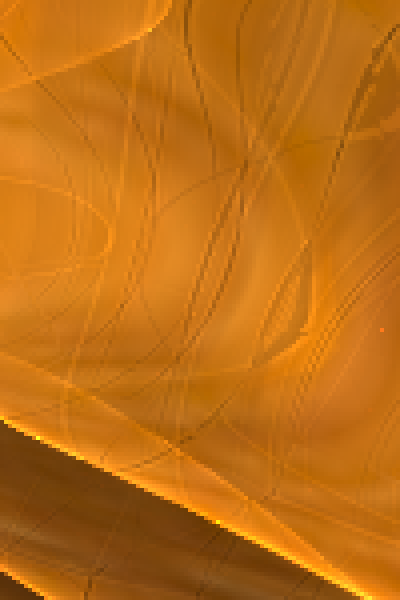}
                 {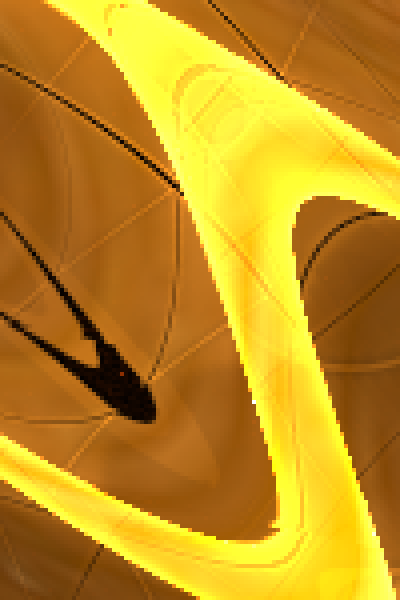} &
    \SpecNeeStack{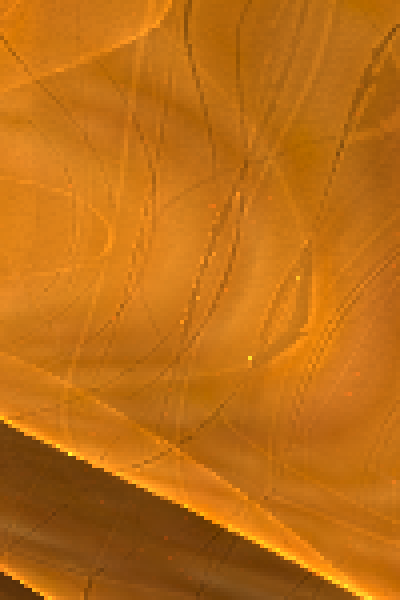}
                 {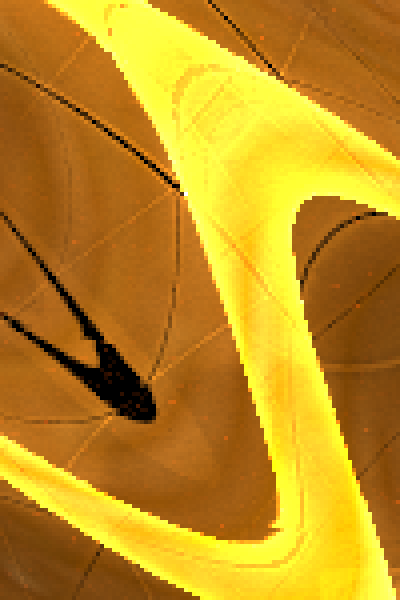} &
    \SpecNeeStack{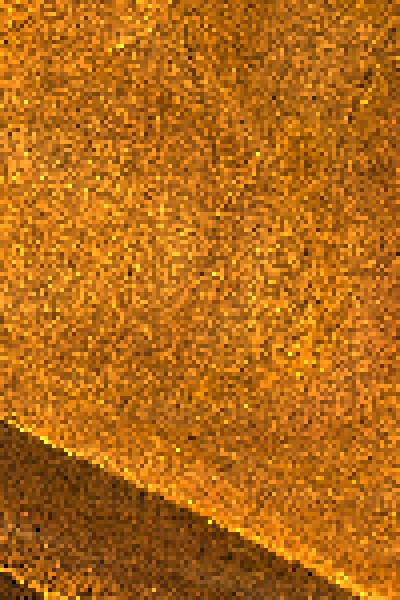}
                 {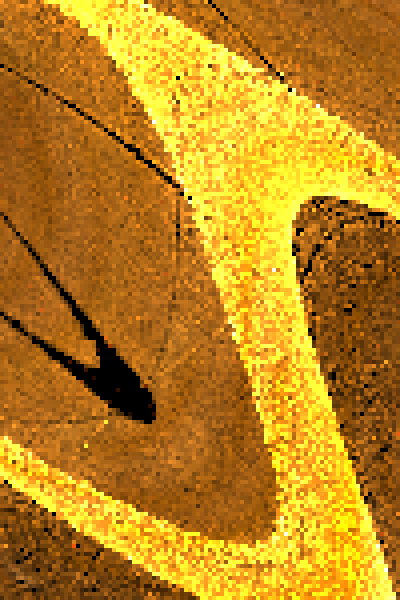} &
    \SpecNeeStack{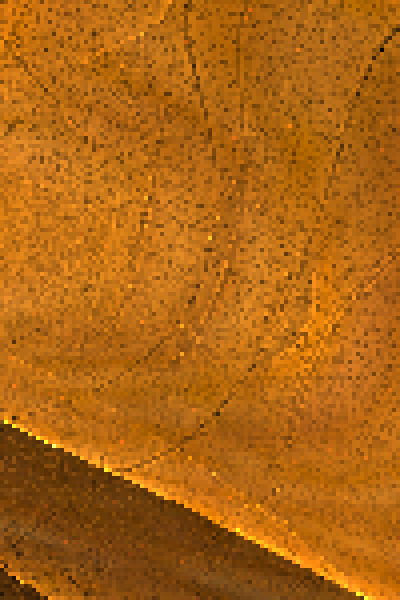}
                 {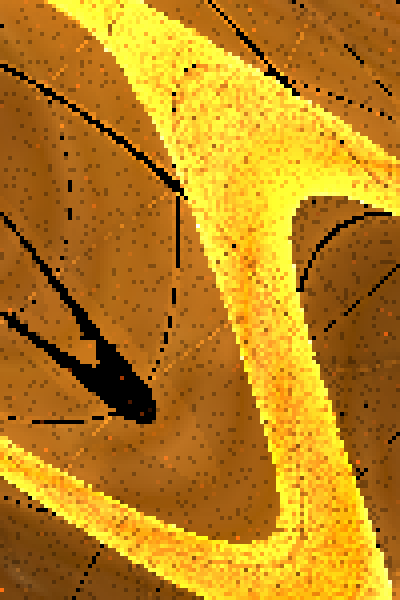} \\                       
    \end{tabular}}
     \caption{\label{fig:specular_nee}
        Equal-time comparison of rendering caustics \textbf{cast by displaced surfaces}. The top-left corner of each
        render shows the height map of its displaced surface. We compare our specular NEE query against
        the unbiased and biased variants of SMS~\cite{zeltner2020specular}. We report the mean squared error (MSE) and \FLIP{} scores
        with respect to reference images, which were rendered using unbiased SMS with a high sample count. Overall, our method 
        achieves a more than $10\times$ variance reduction without loss of energy.
    }
\end{figure*}

Our current implementation supports only specular paths with a single specular vertex. 
Extending it to handle consecutive specular interactions, for example by adapting the 
principles of Path Cuts~\cite{wang2020path} to tessellation-free displaced surfaces, 
is an interesting avenue for future work.
\section{Surface sampling}
\label{sec:measure-query}

\subsection{Area-guided hierarchical sampling}
The most basic form of surface sampling is to draw samples (approximately) in proportion to surface area. For an explicit 
triangle mesh, this is straightforward: one computes the cumulative distribution function (CDF) of the triangle areas and 
draws samples by inverting the CDF. However, this approach is not possible for a displaced surface that is implicitly described 
by its displacement map, unless the surface is tessellated. Inspired by Clarberg et al.~\shortcite{clarberg2005wavelet, clarberg2008practical},
we propose a hierarchical sampling scheme in which we stochastically descend the texture-space hierarchy. At each level, the decision 
is guided by the estimated displaced area covered by each node's footprint, which is made possible by the first-order 
statistics stored in the hierarchy. Our sampling method remains unbiased despite relying on estimated areas, and we show 
that it is efficient.

The overall sampling process consists of two steps: (1) sampling a base triangle and (2) sampling a point within that
triangle which is subject to displacement. For ease of exposition, we first describe the second step, assuming that a 
base triangle $T$ has already been sampled. To initiate the texture-space descent, $T$ is transformed to UV space and 
locate the smallest hierarchy node that covers it. The process then involves repeatedly descending from a parent node to 
one of its four children until a leaf node is reached, after which a point in its footprint is uniformly sampled. As 
hinted, the core operation is to weight each child node by 
the underlying displaced surface area covered by its footprint. This is computed by integrating the surface element over 
the texture-space footprint $\ndom$, or alternatively expressed by the mean Jacobian $\bar{J}_{\ndom}$ over the footprint:
\begin{equation} \label{eq:node_displaced_area}
    A_{\nd} = \int_{\ndom}  J(u, v) \duv = |\ndom| \frac{\int_{\ndom}  J(u, v) \duv}{|\ndom|} = |\ndom| \bar{J}_{\ndom}.
\end{equation}
In a simple case, if the base triangle has a constant normal, then $A_{\nd}$ of a leaf node
can be calculated exactly as the sum of areas of 
two displaced triangles. However, in the general case of interior nodes and interpolating normals (\autoref{eq:interpolated-fields}), 
exact $A_{\nd}$ is difficult to compute. A more efficient way is to instead estimate $\bar{J}_{\ndom}$ using the first-order 
information provided by our hierarchy. To see this, we examine the definition of $J(u, v)$ in world space, given a 
texture-to-world transform $\mathbf{A}$ (the linear part):
\begin{equation*} \label{eq:world_space_jacobian}
    J(u, v) = || (\mathbf{A}\partial_u \mathbf{S}) \times (\mathbf{A} \partial_v \mathbf{S}) || = || \,|\mathbf{A}| \mathbf{A}\invtrans (\partial_u \mathbf{S} \times \partial_v \mathbf{S}) ||.
\end{equation*} 
Expanding the cross product term by \autoref{eq:surface-derivatives} and rearranging yields 
\begin{align} 
    J(u, v) &= || \mathbf{K}_0 + \partial_u h \mathbf{K}_1 + \partial_v h \mathbf{K}_2 ||, \\ \nonumber
    \mathbf{K}_0 &= (|\mathbf{A}| \mathbf{A}\invtrans) \big( (\partial_u \mathbf{P} + h\partial_u \mathbf{N}) \times (\partial_v \mathbf{P} + h\partial_v \mathbf{N}) \big), \\ \nonumber
    \mathbf{K}_1 &= (|\mathbf{A}| \mathbf{A}\invtrans)\big( \Nbase \times (\partial_v \mathbf{P} + h \partial_v\Nbase) \big), \\ \nonumber
    \mathbf{K}_2 &= (|\mathbf{A}| \mathbf{A}\invtrans) \big( (\partial_u \mathbf{P} + h \partial_u\Nbase) \times \Nbase \big).
\end{align} 
Note that the partial derivatives of $\Pbase$ and $\Nbase$ only depend on the base triangle and can be obtained easily by the 
standard approach of solving $2\times2$ systems\cite{pharr2023pbrt}. Under our convention of uninterpolated 
normals (\autoref{eq:interpolated-fields}), the normal derivatives are constant over the base triangle. 
When integrating using the true $J(u, v)$, 
$\partial_u h$, and $\partial_v h$ all vary across the footprint $\ndom$. By replacing them with $h_n$, $g_{u,n}$, and $g_{v,n}$ that 
are stored in the node, we arrive at our estimated mean Jacobian
\begin{equation} \label{eq:estimated_mean_jacobian}
    \widehat{J}_{\ndom} = || \mathbf{K}_0(h_n) + g_{u,n} \mathbf{K}_1(h_n) + g_{v,n} \mathbf{K}_2(h_n) ||.
\end{equation}
\autoref{eq:estimated_mean_jacobian} is a first-order estimation and is exact for a flat (tilted) region. It can be 
further improved to have second-order accuracy if the slope $(g_{u,n}, g_{v,n})$ is replaced with the mean slope over the region. 
However, as we have already observed good accuracy in practice, we keep our choice in favor of a tight geometric node 
bound for other applications.

With the estimated mean Jacobians of child nodes, we stochastically select one of them to descend in proportion to 
their estimated areas: 
\begin{equation} \label{eq:area_sampling_select}
    p(n_i) = \frac{ |\uvdom_{n_i}| \widehat{J}_{\uvdom_{n_i}} }{\sum_{j=0}^3 |\uvdom_{n_j}| \widehat{J}_{\uvdom_{n_j}}}, \quad i=0, 1, 2, 3.
\end{equation}
$|\uvdom_{n_i}|$ is straightforward to obtain by computing the intersection area between the base triangle and the child node 
in texture space. 
To complete the final piece of the sampling algorithm, we return to the earlier question of how to choose a base 
triangle $T$. A straightforward approach is to sample each base triangle with probability proportional to its $A_n$, 
as estimated from its initial node. As we will see in \autoref{subsec:weighed_area_sampling}, however, this is not the only option.

\subsection{Weighted area sampling for displaced emitters} \label{subsec:weighed_area_sampling}
Many applications~\cite{keller1997instant, alliez2003anisotropic, li2010anisotropic} seek to draw surface samples based 
not only on surface area but also on other factors, such as light flux, surface curvature, or weight maps. Fortunately, the 
hierarchical sampling scheme makes it easy to inject additional weight factors on-the-fly during descent. This is achieved 
by extending \autoref{eq:area_sampling_select} by another node-specific weight:
\begin{equation} \label{eq:weighted_area_sampling_select}
    p(n_i) = \frac{ w(n_i) |\uvdom_{n_i}| \widehat{J}_{\uvdom_{n_i}} }{\sum_{j=0}^3 w(n_j) |\uvdom_{n_j}| \widehat{J}_{\uvdom_{n_j}}}, \quad i=0, 1, 2, 3,
\end{equation}
where $w(n_i)$ can be any non-negative function. It can be either computed procedurally or with the help of additional 
information stored alongside existing node attributes. Let $\mathbf{y}$ be a sampled point. The full PDF in surface area 
measure is the product of the selection probabilities along the descent to the leaf node $l$ and the uniform sampling 
density over the footprint of $l$:
\begin{equation} \label{eq:weighted_area_full_pdf}
    p(\mathbf{y}) = \Big( \prod_{n \in \text{path}(l)} p(n) \Big) \frac{1}{A_l},
\end{equation}
where $\text{path}(l)$ denotes the traversal path of the hierarchy.

We demonstrate our weighted area sampling technique on light sampling for Monte Carlo rendering. In this setting, we 
equip displaced surfaces with textured emission, transforming them into complex emitters that are challenging to sample 
efficiently. Let $\mathbf{x}$ be the shading point and $\mathbf{y}$ be a sampled point on the emitter. A good
\footnote{The ideal PDF should also incorporate visibility, but we do not expect it without any form of global visibility test or resampling.} 
sampling strategy should have the following sampling PDF in surface area measure:
\begin{equation} \label{eq:good_light_pdf}
    p(\mathbf{y} \mid \mathbf{x}) = \frac{L_e(\mathbf{y}) |\mathbf{n}_{\mathbf{y}} \cdot \omega_{\mathbf{x} \rightarrow \mathbf{y}}| |\mathbf{n}_{\mathbf{x}} \cdot \omega_{\mathbf{x} \rightarrow \mathbf{y}}|}{||\mathbf{x} - \mathbf{y}||^2}, 
\end{equation}
where $L_e$ is emitted radiance, $\mathbf{n}_{\ast}$ is surface normal, and $\omega_{\mathbf{x} \rightarrow \mathbf{y}}$ is the unit vector pointing from $\mathbf{x}$ to $\mathbf{y}$.
It is clear that emission strength should be incorporated into the sampling process. Therefore, we weight each node by the 
average emission luminance over its leaves $w(n) = \bar{E}_n$, which can be computed alongside hierarchy construction and 
stored per node. To compute the PDF of this weighted strategy, notice that the product term \autoref{eq:weighted_area_full_pdf} 
approximately collapses to emitted power (luminance) of $l$ divided by that of the entire surface. Thus, the PDF is approximately proportional 
to the emission (luminance): 
\begin{equation*} \label{eq:weighted_pdf_e}
    p_{\text{E}}(\mathbf{y}) \approx \frac{\Phi_l}{A_l} \frac{1}{\Phi} \propto E(\mathbf{y}),
\end{equation*}
whose error comes from \autoref{eq:estimated_mean_jacobian}. Although this strategy accounts for the emission term, 
it leaves the two foreshortening terms unaddressed and, more importantly, the inverse-square distance term, which becomes 
singular as $\mathbf{x}$ and $\mathbf{y}$ approach each other. To address them, we introduce a second, receiver-aware weight term
\begin{equation}
    R(\mathbf{x})=
    \begin{cases}
        \dfrac{\max\bigl(0,\ \cos(\max(0,\theta-\theta_b))\bigr)}{\max(\| \mathbf{x} - \mathbf{y} \|^2,\epsilon)}, & \|\mathbf{x} - \mathbf{y}\|>r_b,\\[8pt]
        \dfrac{1}{\max(r_b^2,\epsilon)}, & \text{otherwise,}
    \end{cases}
\end{equation}
where $\theta$ is the angle between $\mathbf{n}_\mathbf{x}$ and $\omega_{\mathbf{x} \rightarrow \mathbf{y}}$, $r_b$ is 
the radius of the node's world-space bounding sphere, $\theta_b = \arcsin (r_b / \|\mathbf{x} - \mathbf{y}\|)$, and $\epsilon$ 
is a small positive number. The design of $R(\mathbf{x})$ is inspired by Conty Estevez and Kulla~\shortcite{conty2018manylights}.
The PDF of this receiver-aware strategy, combined with emission weights, approximates
\begin{equation*} \label{eq:weighted_pdf_er}
    p_{\text{E+R}}(\mathbf{y} \mid \mathbf{x}) \appropto \frac{E(\mathbf{y}) |\mathbf{n}_{\mathbf{x}} \cdot \omega_{\mathbf{x} \rightarrow \mathbf{y}}| }{ \| \mathbf{x} - \mathbf{y} \|^2 }.
\end{equation*}
$R(\mathbf{x})$ currently does not account for emitter orientation because our hierarchy does not store information about 
displaced normals  
However, this can be supported by additionally storing normal distribution functions (NDFs) in the nodes~\cite{dupuy2013leadr, zhou2025appearance}.
Finally, our method is agnostic to the strategy used to sample base triangles. The emitted power of the base triangles, 
as estimated by our hierarchy, can be used to build either a power distribution table or a light BVH.

\subsection{Results} \label{subsec:light_sampling_results}
We integrate our hierarchical sampling method into our GPU renderer, allowing tessellation-free
displaced surfaces to act as both regular surfaces and emitters in path tracing. \autoref{fig:light_sampling} 
demonstrates this dual role in the \textsc{Volcanic Energy Station}, a scene lit \emph{entirely} by emissive displaced surfaces.
Both the volcano and the wall panels are extruded from planar base surfaces via high-resolution displacement maps 
combined with textured emission. We compare with two baselines: the first is the uniform ray casting sampling method by Ling et 
al.~\shortcite{ling2025uniform}. This is a generic sampling method applicable to any type of surface as long as an all-hit ray
casting subroutine is provided. By supplying our ray intersection query (\autoref{sec:ray-query}), we can uniformly 
sample displaced surfaces without tessellation. The second baseline is the light bounding volume hierarchy (BVH)~\cite{conty2018manylights} 
built on pre-tessellated geometry, which is widely adopted for many-light sampling. We refrain from comparing with resampling based 
approaches~\cite{bitterli2020spatiotemporal} because they are orthogonal: their candidate generation still requires a 
surface sampling method.

The equal-time comparison in \autoref{fig:light_sampling} provides rendering results and qualitative metrics (mean squared 
errors (MSE) and \FLIP scores~\cite{andersson2020flip}). Our method strikes a favorable balance between 
sampling efficiency and memory cost. While the uniform baseline does not require any additional memory, its efficiency 
is poor because it does not importance sample any term in \autoref{eq:good_light_pdf}, and it requires tracing an all-hit 
ray against the entire scene every draw. 
Both of our weighted strategies (E and E+R) are much more efficient as they perform importance sampling according to emission
strength. Of the two, the receiver-aware strategy (E+R) allocates more samples to nearby sources,
which further reduces variance in the regions surrounding strong emitters.
The pre-tessellated light BVH achieves the lowest errors but at the cost of heavy memory cost, as reported in 
\autoref{tab:light_sampling}. In fact, its efficiency advantage mostly comes from the raw speed advantage of hardware-accelerated 
ray tracing of pre-tessellated geometry rather than a better sampling distribution. Indeed, our method (E+R) can achieve 
comparable errors under an equal-sample setting. Finally, since our method readily tracks the PDF of each sample
(\autoref{eq:weighted_area_full_pdf}), it can be combined with multiple importance sampling (MIS) to further reduce 
variance (\autoref{tab:light_sampling}).

\definecolor{LightSampOrange}{HTML}{FF0280}
\definecolor{LightSampSkyBlue}{HTML}{00A7E1}
\colorlet{LightSampFirst}{red!25}
\colorlet{LightSampSecond}{orange!30}
\colorlet{LightSampThird}{yellow!35}
\newlength{\LightSampGap}
\newlength{\LightSampInsetW}
\newlength{\LightSampStackH}
\newlength{\LightSampFrameW}
\newlength{\LightSampOverviewH}
\newcommand{\LightSampMetric}[2]{%
  {\setlength{\fboxsep}{0pt}\colorbox{#1}{\makebox[\LightSampInsetW]{%
    \rule[-1.5pt]{0pt}{\dimexpr\fontcharht\font`0+1.5pt\relax}%
    \vphantom{$\downarrow$}#2}}}%
}
\newcommand{\LightSampNote}[1]{$^{#1}$\kern-1pt}
\newcommand{\LightSampInset}[2]{%
  \begin{tikzpicture}[baseline=(img.south)]
    \node[inner sep=0pt, outer sep=0pt] (img)
      {\includegraphics[width=\LightSampInsetW]{light_sampling/#2}};
    \draw[draw=#1, line width=\LightSampFrameW]
      ([shift={(.5\LightSampFrameW,.5\LightSampFrameW)}]img.south west) rectangle
      ([shift={(-.5\LightSampFrameW,-.5\LightSampFrameW)}]img.north east);
  \end{tikzpicture}%
}
\newcommand{\LightSampStack}[1]{%
  \vbox{\offinterlineskip
    \hbox{\LightSampInset{LightSampOrange}{insets_100x150+1000+600/#1}}%
    \vskip\LightSampGap
    \hbox{\LightSampInset{LightSampSkyBlue}{insets_100x150+1300+270/#1}}}%
}

\newcommand{\LightSampMain}[6]{%
  \begin{tikzpicture}[baseline=(img.south)]
    \node[inner sep=0pt, outer sep=0pt, anchor=south west] (img) at (0,0)
      {\includegraphics[height=\LightSampStackH]{light_sampling/#1}};
    \node[inner sep=0pt, outer sep=0pt, anchor=north west] (overview)
      at (img.north west)
      {\includegraphics[height=\LightSampOverviewH]{light_sampling/#6}};
    \draw[white, line width=\LightSampFrameW]
      ([shift={(-.5\LightSampFrameW,0)}]overview.north east) --
      ([shift={(-.5\LightSampFrameW,.5\LightSampFrameW)}]overview.south east) --
      ([shift={(0,.5\LightSampFrameW)}]overview.south west);
    \begin{scope}[x={(img.south east)}, y={(img.north west)}, shift={(img.south west)}]
      \draw[LightSampOrange, line width=\LightSampFrameW]
        ({#2/2100},{1-#3/900}) rectangle ({(#2+100)/2100},{1-(#3+150)/900});
      \draw[LightSampSkyBlue, line width=\LightSampFrameW]
        ({#4/2100},{1-#5/900}) rectangle ({(#4+100)/2100},{1-(#5+150)/900});
    \end{scope}
  \end{tikzpicture}%
}

\begin{figure*}[!t]
\setlength{\LightSampGap}{2pt}
\setlength{\LightSampInsetW}{\dimexpr(3\textwidth-22\LightSampGap)/36\relax}
\setlength{\LightSampStackH}{\dimexpr3\LightSampInsetW+\LightSampGap\relax}
\setlength{\LightSampFrameW}{1.2pt}
\setlength{\LightSampOverviewH}{0.4\LightSampStackH}
\renewcommand{\arraystretch}{0.5}
\centering
\begin{tabular}{@{}c*{5}{@{\hspace{\LightSampGap}}c}@{}}
    \textsc{\textbf{Volcanic Energy Station}} &
    \textbf{\small Reference} &
    \textbf{\small Ours} &
    \textbf{\small Ours} &
    \textbf{\small Uniform} &
    \textbf{\small Light BVH} \\ [1pt]

    $2100 \times 900$, 16 GPU seconds, Ours (E+R) shown&
    &
    \textbf{\small (E)} &
    \textbf{\small (E+R)} &
    \textbf{\small \cite{ling2025uniform}} &
    \textbf{\small (Pre-tess.)} \\ [1pt]

    &
    spp &
    136 &
    133 &
    64 &
    758 \\ [1pt]    

    &
    MSE$\downarrow$&
    \LightSampMetric{LightSampThird}{$7.5\text{e-}2$} &
    \LightSampMetric{LightSampSecond}{$3.9\text{e-}3$} &
    $2.4\text{e-}1$ &
    \LightSampMetric{LightSampFirst}{$1.3\text{e-}3$} \\

    &
    \FLIP{}$\downarrow$&
    \LightSampMetric{LightSampThird}{$0.359$} &
    \LightSampMetric{LightSampSecond}{$0.273$} &
    $0.645$ &
    \LightSampMetric{LightSampFirst}{$0.187$} \\ \noalign{\vskip\LightSampGap}

    \LightSampMain{receiver-aware_nee-only.png}{1000}{600}{1300}{270}
                  {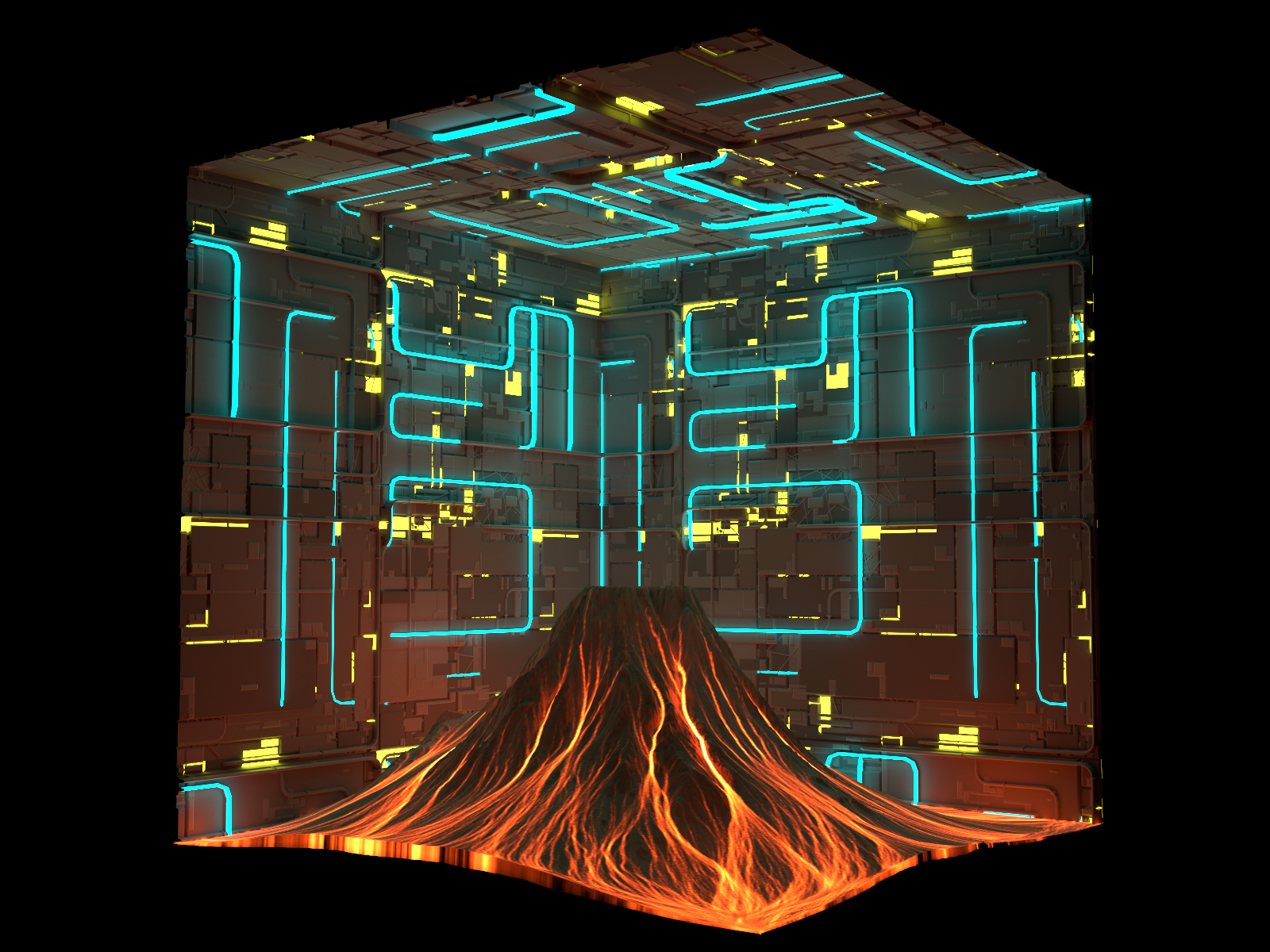} &
    \LightSampStack{reference.png} &
    \LightSampStack{weighted-area_nee-only.png} &
    \LightSampStack{receiver-aware_nee-only.png} &
    \LightSampStack{uniform-line_nee-only.png} &
    \LightSampStack{light-bvh_nee-only.png} \\
\end{tabular}
     \caption{\label{fig:light_sampling}
        Equal-time comparison of light sampling techniques. The scene is lit solely by emissive displaced surfaces (top left).
        In all renders, each technique is used for next-event estimation \textbf{without MIS}. Our method supports different
        area-weighting strategies, such as weighting by emission (E) and additionally by receiver proximity (E+R).
        The latter variant achieves sampling efficiency close to that of a light BVH but does not require heavy pre-tessellation.
        Please refer to \autoref{tab:light_sampling} for memory and additional statistics.
    }
\end{figure*}

\begin{table}
    \setlength{\tabcolsep}{3pt}%
	\centering
	\caption{ \label{tab:light_sampling}
        Additional statistics for the light sampling comparison shown in \autoref{fig:light_sampling}. \small{\LightSampNote{\ast}Includes
        displacement maps and all hierarchy channels. \LightSampNote{\dagger}Includes pre-tessellated vertex buffers, index buffers, and the 
        acceleration structure. \LightSampNote{\ddagger}Includes light BVH nodes, the light list, and bit trails.}
    }        
    \begin{tabular}{lll}
        \toprule
        Statistics & Ours (E+R) & Light BVH (Pre-tess.) \\
        \midrule
        Memory (MB)                             & 230.7$^{\ast}$ & 4901.8$^{\dagger}$ + 520.9$^{\ddagger}$ \\
        \midrule
        Per-spp time (ms)                       & 116.3 & 21.2 \\
        Ray intersection time $\%$              & 81$\%$ & 46$\%$ \\
        Light sampling time $\%$                & 16$\%$ & 37$\%$ \\
        \midrule
        64-spp MSE$\downarrow$                  & $1.3\text{e-}2$ & $9.3\text{e-}3$ \\
        64-spp \FLIP{}$\downarrow$              & 0.336 & 0.326 \\
        \midrule
        16-sec w/ MIS MSE$\downarrow$           & $7.4\text{e-}5$ & (not measured) \\
        16-sec w/ MIS \FLIP{}$\downarrow$       & 0.164 & (not measured) \\
        \bottomrule
    \end{tabular}
\end{table}
\section{Closest-point queries}
\label{sec:closest-query}
Finally, we consider closest-point queries, which are widely used in geometry processing~\cite{sharp2022spelunking}. The key idea to accelerate the search is that we can use the height planes and residuals from \autoref{sec:hierarchy} to derive lower bounds on the distance from the query point to each displaced surface region. These bounds guide a best-first branch-and-bound search~\cite{felten1989best}, allowing us to discard regions that cannot contain a closer point and query the surface without pre-tessellation. We demonstrate these queries through surface projection in Projected Walk on Spheres~\cite{sugimoto2024pwos}.

As in ray tracing, a coarse BVH organizes the displaced triangle faces, while our texture-space hierarchy guide the search within each face. We perform the search at two levels: the outer search selects promising faces, and the inner search refines texture regions within each selected face. The outer search supplies the distance $C_{\mathrm{out}}$ to its best surface candidate, or $+\infty$ if none is available. The inner search maintains its own candidate distance $U_T$ and uses $C=\min(U_T,C_{\mathrm{out}})$ to discard regions that cannot improve the best available candidate. \autoref{alg:closest-point} summarizes the search within one face.

\begin{algorithm}[!htb]
\caption{Closest-point search within one displaced face}
\label{alg:closest-point}
\small
\algrenewcommand{\algorithmicindent}{1em}
\definecolor{closestcomment}{HTML}{55A9D7}
\algrenewcommand{\algorithmiccomment}[1]{\hfill\textcolor{closestcomment}{$\triangleright$\,\textit{#1}}}
\begin{algorithmic}[1]
\Require Query point $\mathbf q$, face $T$, tolerance $\varepsilon_{\mathrm{cp}}$
\Statex Outer distance cutoff $C_{\mathrm{out}}$ ($+\infty$ if unavailable)
\State $\widehat{\mathbf y}_T\gets\Call{InitialCandidate}{\mathbf q,T}$ \Comment{\autoref{sec:closest-init}}
\State $U_T\gets\|\mathbf q-\widehat{\mathbf y}_T\|$; $L_{\mathrm{ret}}\gets+\infty$
\State Initialize $\mathcal Q$ with clipped root regions covering $\Omega_T$
\Statex \hspace{\algorithmicindent} keyed by \textsc{LowerBound} \Comment{\autoref{eq:closest-fo}}
\While{$\mathcal Q\ne\emptyset$} \Comment{Start the traversal of the priority queue}
  \State $L_T\gets\min(U_T,L_{\mathrm{ret}},\Call{TopSmallestBound}{\mathcal Q})$ \Comment{\autoref{sec:closest-traversal}}
  \State $C\gets\min(U_T,C_{\mathrm{out}})$
  \State \textbf{if} $U_T-L_T\le\varepsilon_{\mathrm{cp}}$ or $L_T>C$ \textbf{then break} \Comment{Converged or excluded}
  \State $(n,\ell_n)\gets\Call{PopSmallest}{\mathcal Q}$
  \If{$C-\ell_n\le\varepsilon_{\mathrm{cp}}$} \Comment{No improvement beyond tolerance}
      \State $L_{\mathrm{ret}}\gets\min(L_{\mathrm{ret}},\ell_n)$ \Comment{Retain closed or pruned bound}
  \ElsIf{$n$ is a texture leaf} \Comment{Refine within the texel}
      \State $(\mathbf y,\ell)\gets\Call{LeafSearch}{\mathbf q,n,C,\varepsilon_{\mathrm{cp}}}$ \Comment{\autoref{sec:closest-refine}}
      \State $L_{\mathrm{ret}}\gets\min(L_{\mathrm{ret}},\ell)$
      \State Update $\widehat{\mathbf y}_T,U_T$ if $\mathbf y$ is valid and closer
  \Else \Comment{Descend the hierarchy}
      \State Bound all nonempty clipped children \Comment{\autoref{eq:closest-fo}}
      \State Queue children with $C-\ell>\varepsilon_{\mathrm{cp}}$, keyed by $\ell$
      \State Retain remaining child bounds in $L_{\mathrm{ret}}$
  \EndIf
\EndWhile
\State $L_T\gets\min(U_T,L_{\mathrm{ret}},\Call{TopSmallestBound}{\mathcal Q})$
\State \Return $\widehat{\mathbf y}_T,L_T,(U_T-L_T\le\varepsilon_{\mathrm{cp}})$
\end{algorithmic}
\smallskip
\noindent\textsc{TopSmallestBound} returns $+\infty$ for an empty queue. $L_{\mathrm{ret}}$ retains the smallest bound from closed regions and completed leaf searches, including unresolved results. The returned Boolean indicates face-local convergence.
\end{algorithm}

\subsection{Initialization}
\label{sec:closest-init}
Before refining a face, we want to quickly find a surface point to establish a distance that the search must improve. Once this candidate is available, any region whose distance lower bound exceeds the candidate's distance can be discarded without searching inside it. We obtain the candidate through a small set of surface evaluations.
We cover the triangle's texture footprint with hierarchy cells and clip them to the footprint, obtaining the initial search regions. For each clipped polygon, we evaluate the displaced surface at its texture-coordinate vertices and at their arithmetic mean. The closest evaluated point becomes the face-local candidate $\widehat{\mathbf y}_T$ (\textsc{InitialCandidate}), with distance $U_T=\|\mathbf q-\widehat{\mathbf y}_T\|$. A closer initial candidate might allow more pruning, but any evaluated surface point suffices here as a valid upper bound. 
We then initialize a priority queue $\mathcal Q$ by ordering the distance lower bounds (\textsc{LowerBound}) of the clipped root regions.

\subsection{Distance bounds and traversal}
\label{sec:closest-traversal}

To discard an entire texture region, we need a lower bound on the distance to every surface point it contains. If this bound exceeds the current distance upper bound $C$,  no point in the region can improve the surface candidate found so far. 
Here we describe how do we derive \textsc{LowerBound} for a specific texture region/a node of our hierachy. 

Note that our hierarchy stores height planes and residual bounds in texture space and closest-point search requires bounds on world-space distance. For node $n$ on face $T$, we first restrict the node's rectangular texture region $\Omega_n$ to the triangle's texture footprint $\Omega_T$, giving $\Omega_{nT}=\Omega_n\cap\Omega_T$. We then combine the stored height plane with the base geometry and displacement directions over this clipped region. 

With a constant displacement direction $\mathbf N$, the height plane maps to a finite planar region, and the height residual scales by the direction's length. When directions vary, both the height plane and the direction field $\mathbf N(u,v)$ are affine, so their product introduces quadratic terms. We retain the constant and linear terms to obtain a planar approximation $\mathcal
A_n$. A deviation bound $R_n$ accounts for the omitted quadratic terms and the height residual scaled by the displacement direction. Subtracting $R_n$ from the distance to $\mathcal A_n$ gives a distance lower bound
(\textsc{LowerBound}):
\begingroup
\setlength{\abovedisplayskip}{2pt}
\setlength{\belowdisplayskip}{2pt}
\thinmuskip=2mu
\medmuskip=2mu
\thickmuskip=3mu
\begin{equation}
L_n^{\mathrm{FO}}(\mathbf q)=\max(0,\operatorname{dist}(\mathbf q,\mathcal A_n)-R_n).
\;\vcenter{\hbox{\includegraphics[width=0.27\columnwidth]{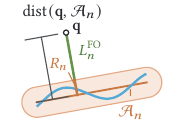}}}
\label{eq:closest-fo}
\end{equation}
\endgroup
Every surface point in the region lies within $R_n$ of its corresponding point in $\mathcal A_n$, so its distance to $\mathbf q$ cannot be smaller than $\operatorname{dist}(\mathbf q,\mathcal A_n)-R_n$.
Here, distance is measured to the finite polygon, including its edges and vertices, rather than its infinite supporting plane. For a constant unit displacement direction, $R_n=r_n$. The supplementary material gives the bound construction.

\begin{figure}[!htb]
 \centering
 \includegraphics[width=\columnwidth]{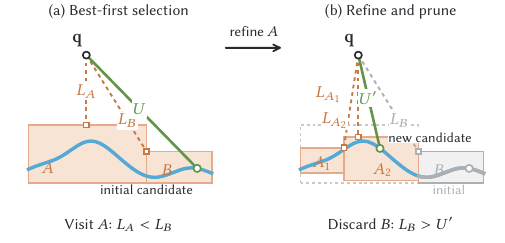}
 \caption{Best-first selection and pruning after finding a closer surface candidate. Rectangles illustrate regional enclosures, not the geometry of our first-order bounds.}
 \label{fig:closest-search}
\end{figure}

Next we describe how these lower bounds guide our traversal. We visit the queued region with the smallest bound first (\autoref{fig:closest-search}a). If its bound exceeds the current cutoff $C=\min(U_T,C_{\mathrm{out}})$, the region cannot improve the
best available candidate and we prune it. If its bound lies within $\varepsilon_{\mathrm{cp}}$ below $C$, any improvement is limited to the requested tolerance, so we may close the region. In both cases, we retain its lower bound for the termination test. Otherwise, we compute bounds for its children and queue those requiring further exploration, or refine the surface within a texel leaf (\textsc{LeafSearch}). Leaf refinement may find a closer candidate,
reducing $U_T$ and potentially $C$, allowing further pruning (\autoref{fig:closest-search}b). Our implementation evaluates surface candidates during initialization and leaf refinement, and we do not need to do any additional candidate evaluations at internal hierarchy nodes.

\subsection{Refinement and termination}
\label{sec:closest-refine}
Following the branch-and-bound descending, we may finally reach a texel leaf node, but this does not mean we have finished the search: a texel may still be too large to meet the requested distance tolerance and the closest point may lie anywhere within its displaced surface region.  We therefore subdivide its texture domain as needed, recompute distance lower bounds, and evaluate new surface candidates. This refinement returns the closest point it found (if any), and a lower bound on distance to the entire leaf region (\textsc{LeafSearch}). The supplementary material gives the leaf-refinement details.

To decide when to stop, we should account for the entire mesh becase another face may contain a closer point. Let $U_{\mathrm{obj}}$ be the distance to the closest surface candidate found so far. We obtain $L_{\mathrm{obj}}$ by taking the minimum of $U_{\mathrm{obj}}$ and the lower bounds covering all faces, retaining bounds from regions closed within tolerance. The query stops successfully when
\begin{equation}
L_{\mathrm{obj}}\le d^*(\mathbf q)\le U_{\mathrm{obj}},\qquad
U_{\mathrm{obj}}-L_{\mathrm{obj}}\le\varepsilon_{\mathrm{cp}}.
\label{eq:closest-certificate}
\end{equation}
With valid bounds, this ensures that the returned candidate's distance is within $\varepsilon_{\mathrm{cp}}$ of the minimum.

\subsection{Results}
\label{sec:eval-closest}
Our method enables closest-point queries directly on displaced surfaces without pre-tessellation. 
Here we evaluate the resulting memory savings at a common distance-error target by comparing against FCPW~\cite{FCPW} on tessellated surfaces. 
We then ablate the contribution of first-order bounds by replacing them with scalar min--max bounds in the same search. 
Finally, we demonstrate the use of these queries in Projected Walk on Spheres.
Comparisons and measurements were performed on CPU (see \autoref{subsec:specular_results} for machine specification).

\begin{figure}[!htb]
 \centering
 \includegraphics[width=\columnwidth]{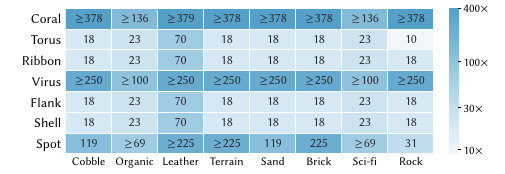}
 \caption{FCPW / ours query memory across 56 displaced surfaces. Entries marked $\ge$ are lower bounds where no tested tessellation meets the sampled distance-error target of $10^{-4}$.}
 \label{fig:app3-cross56}
\end{figure}

\paragraph{Comparison with pre-tessellation.}
We compare direct queries with FCPW~\cite{FCPW} on uniformly tessellated surfaces. We create a test set containing 56 displaced surfaces using seven base meshes and eight displacement maps. Our method reduces query memory by at least $55.4\times$ in geometric mean (\autoref{fig:app3-cross56}), counting all geometry and acceleration data retained after preprocessing. We time quries on 96 points at three surface-offset distances. Using 2,048 separate queries, we select tessellations with a maximum sampled distance error of $10^{-4}$, testing up to 64 subdivisions per base edge under a 50-million-triangle cap. 
Only 36 of the 56 displaced surfaces meet the distance-error target within the tessellation budget. The remaining 20 still fail at the finest feasible tessellation. We report relative memory usage for all 56 surfaces. 
For the 36 surfaces whose tessellations meet the target, FCPW provides $7.48\times$ faster queries in geometric mean. Our method instead prioritizes compact storage and direct queries without pre-tessellation.

\begin{figure}[!htb]
 \centering
 \includegraphics[width=\columnwidth]{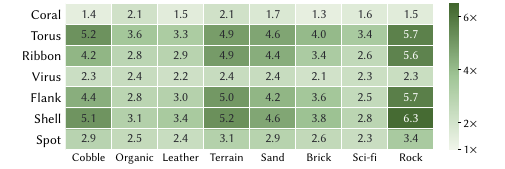}
 \caption{First-order query speedup over scalar min--max prism bounds, with other components fixed.}
 \label{fig:closest-bound-ablation}
\end{figure}

\paragraph{Ablation on the effect of first-order bounds.}
Within our direct-query algorithm, first-order bounds achieve a $3.06\times$ geometric-mean speedup over scalar min--max prism bounds, improving all test surfaces (\autoref{fig:closest-bound-ablation}).
The comparison holds traversal, the leaf solver fixed, isolating the effect of the regional bounds. 

\begin{figure*}[!t]
 \centering
 \includegraphics[width=\textwidth]{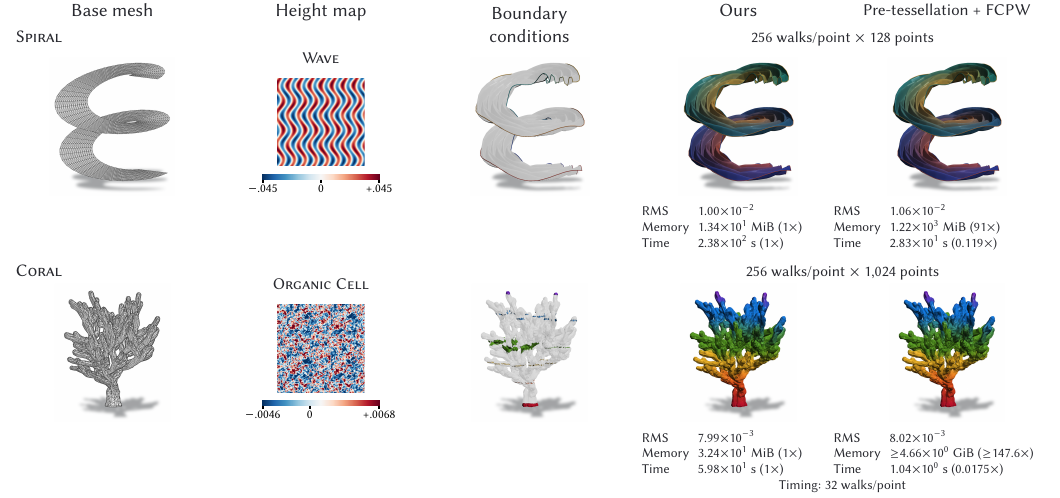}
 \caption{PWoS on displaced spiral and coral surfaces. Prescribed boundary colors define RGB harmonic fields on each surface. We estimate these fields at 128 points on spiral and 1,024 on coral, using 256 walks per point for each method, and interpolate the computed values to color the surfaces. RMS differences are measured against high-sample references with 4,096 walks per point for both examples. Counts exclude walk buffers and process overhead; coral FCPW memory is a lower bound that omits internal BVH nodes.}
 \label{fig:app3-pwos-helicoid}
\end{figure*}

\paragraph{Projected Walk on Sphere (PWoS) application.}
Projected Walk on Spheres~\cite{sugimoto2024pwos} is a Monte Carlo method for solving PDEs on surfaces that repeatedly projects walk positions onto the surface.
Now we use our closest-point algorihtm for these point projections, allowing the method to operate directly on displaced surfaces without pre-tessellation. 
Our compact representation reduces query memory by factors of $12.52$ and $91$ at sampled distance-error targets of $10^{-4}$ and $10^{-5}$, respectively. 
Since a tighter error target requires finer tessellation, our method provides more memory advantage in high-precision scenarios. 
In a separate ablation of complete walks, it is $3.19\times$ faster than the min--max AABB variant.
We also apply PWoS to a more complex displaced coral surface with locally prescribed boundary regions (\autoref{fig:app3-pwos-helicoid}) and use the resulting field to control the coral's emission in \autoref{fig:teaser}.

\section{Conclusion and future work}
\label{sec:conclusion}

In this work, we presented a first-order, texture-space hierarchy for tessellation-free displaced surfaces that supports a diverse set of geometric queries. Local height planes and residual bounds provide surface enclosures and area estimates, while additional slope bounds support specular-path search. We demonstrated our method across various applications including caustics rendering, sampling displaced emitters, and closest-point queries for surface PDE solvers, while retaining efficient direct ray tracing.

We believe our work opens up a fruitful avenue of research. Modern graphics, rendering, 
and geometry processing increasingly demand high fidelity and fine detail, which translate 
into massive amounts of geometry that often cannot fit into memory. Traditional geometry 
and scene representations were generally not designed for such memory-intensive workloads. 
Our work demonstrates that tessellation-free displaced surfaces are a promising candidate 
for a high-fidelity, general-purpose graphics primitive.

\section*{Acknowledgments}
The authors would like to thank Tzu-Mao Li and Ravi Ramamoorthi for their valuable input and proofreading, Cheng Chang for initial experiments on displacement map compression, and Anton Michels and Warren Hunt for many helpful early-stage discussions. 

\bibliographystyle{ACM-Reference-Format}
\bibliography{refs}

\end{document}